\documentclass{aa}  

\usepackage{graphicx}
\usepackage{txfonts}
\usepackage{amsmath}
\usepackage{booktabs}
\usepackage{hyperref}

\begin{document}

   \title{Candidate Young Stellar Objects in the S-cluster: The Kinematic Analysis of a Sub-population of the Low-mass G-objects close to Sgr~A*}

   \titlerunning{Dusty stellar objects of the S-cluster}

	\author{F. Pei{\ss}ker\inst{\ref{inst1},\ref{inst2}}
        \and M. Zaja\v{c}ek\inst{\ref{inst3},\ref{inst1}}
        \and M. Melamed\inst{\ref{inst1}}
        \and B. Ali\inst{\ref{inst1}}
        \and M. Singhal\inst{\ref{inst4}}
        \and T. Dassel\inst{\ref{inst1}}
        \and A. Eckart\inst{\ref{inst1},\ref{inst5}}
        \and V. Karas\inst{\ref{inst6}}
     } 

	\institute{I.Physikalisches Institut der Universit\"at zu K\"oln, Z\"ulpicher Str. 77, 50937 K\"oln, Germany\label{inst1} \and
		\email{peissker@ph1.uni-koeln.de}\label{inst2}
        \and Department of Theoretical Physics and Astrophysics, Faculty of Science, Masaryk University, Kotlá\v{r}ská 2, 611 37 Brno, Czech Republic\label{inst3}
		\and Astronomical Institute, Charles University, V Hole\v{s}ovi\v{c}k\'{a}ch 2, CZ-18000 Prague, Czech Republic\label{inst4}
        \and Max-Plank-Institut f\"ur Radioastronomie, Auf dem H\"ugel 69, 53121 Bonn, Germany\label{inst5}
        \and Astronomical Institute, Czech Academy of Sciences, Bo\v{c}n\'{i} II 1401, CZ-14100 Prague, Czech Republic\label{inst6}
     }

   \date{Received XXX; accepted XXX}

  \abstract
   {The observation of several $L$-band emission sources in the S cluster has led to a rich discussion of their nature. However, a definitive answer to the classification of the dusty objects requires an explanation for the detection of {compact} Doppler-shifted Br$\gamma$ emission. The {ionized hydrogen in combination with the observation} of mid-infrared $L$-band continuum emission suggests that most of these sources are embedded in a dusty envelope.
   {These} embedded sources are part of the S-cluster, and their relationship to the S-stars is still under debate. Until now, the question of the origin of these two populations is vague, although all explanations favor migration processes for the individual cluster members.}
   {This work revisits the S-cluster and its dusty members orbiting the supermassive black hole Sgr~A* on bound Keplerian orbits from a kinematic perspective. The aim is to explore the Keplerian parameters for patterns that might imply a non-random distribution of the sample. Additionally, various analytical aspects are considered to address the nature of the dusty sources.}
   {Based on the photometric analysis, we estimated the individual $H-K$ and $K-L$ colors for the source sample and compared the results to known cluster members. The classification revealed a noticeable contrast between the S-stars and the dusty sources. To fit the flux-density distribution, we utilized the radiative transfer code \texttt{HYPERION} and implemented a Young Stellar Object Class I model. We obtained the position angle from the Keplerian fit results, and additionally, we analyzed the distribution of the inclinations and the longitudes of the ascending node.}
   {The colors of the dusty sources suggest a stellar nature consistent with the spectral energy distribution in the near and mid-infrared domains. Furthermore, the evaporation timescales of dusty and gaseous clumps in the vicinity of Sgr~A* are much shorter ($\ll$ 2 years) than the epochs covered by the observations ($\approx$15 years). In addition to the strong evidence for the stellar classification of the D-sources, we also find a clear disk-like pattern following the arrangements of S-stars proposed in the literature. Furthermore, we find a global intrinsic inclination for all dusty sources of $60 \pm 20^{\circ}$, implying a common formation process.}
   {The pattern of the dusty sources manifested in the distribution of the position angles, inclinations, and the longitudes of the ascending node, strongly suggests two different scenarios: the main-sequence stars and the dusty stellar S-cluster sources share the common formation history or migrated with a similar formation channel in the vicinity of Sgr~A*. Alternatively, the gravitational influence of Sgr~A* in combination with a massive perturber, such as a putative IMBH in the IRS 13 cluster, forces the dusty objects and S-stars to follow a particular orbital arrangement.}


   \maketitle
%

\section{Introduction} 
\label{sec:1}

Almost three decades ago, \cite{Eckart1996Natur} found indications for a compact mass in the center of our Galaxy by identifying single stars in a crowded cluster that is located at the position of the compact and variable radio source Sgr~A* \citep[see e.g.][for reviews]{Eckart2017b,2022RvMP...94b0501G}. Subsequent observations of the same region by \cite{Ghez1998} and \cite{Eckart2002} suggested that Sgr~A* can indeed be associated with a compact mass of about $4\times10^6 M_{\odot}$ in agreement with predictions derived from the gas motion inside the minispiral \citep{Bhat2022} using the virial theorem \citep{Wollman1977}. Although the Nuclear Star Cluster (NSC) with a diameter of $\sim 1\,{\rm pc}$, often referred to as the {\it inner parsec}, offers various scientific opportunities that could be used as a template for other galaxies \citep{Schoedel2009, schoedel2020, Chen2023, Subrayan2023}, the S-cluster with a diameter of 40 mpc requires particular attention because of its members and the interplay with Sgr~A*. In this peculiar environment, the intriguing discovery of a gas cloud more than a decade ago \citep{Gillessen2012} shed light on near-infrared (NIR) line-emitting sources with a dust counterpart observable in the mid-infrared (MIR) \citep{Clenet2004b,Clenet2005b, Ghez2005}. The multi-wavelength colors and hence the temperature \citep{Eckart2013} of these objects challenged the hypothesis of coreless clouds that move on a Keplerian orbit around the supermassive black hole Sgr~A* in the center of our Milky Way. In this highly radiative and turbulent environment \citep{Eckart2017a, Witzel2021}, the formation process of coreless clouds with self-illuminating capabilities requires a theoretical model that is developed in a challenging framework. For example, some authors favor the idea of interacting stellar winds created by Wolf-Rayet stars that create clumps associated with the here-investigated dusty sources. While strong stellar winds in the close vicinity of Sgr~A* and the S-cluster exhibit typical velocities of a few hundred to a thousand km/s \citep{Krabbe1995}, the formation of clumps that move on a Keplerian orbit with an almost constant IR magnitude \citep{Witzel2014,Pfuhl2015,Plewa2017,peissker2021c} lacks a sufficiently satisfying theoretical background. The model of clumps formed within colliding winds, as proposed by \cite{Gillessen2012}, becomes untenable if one considers the radiative breaking point $r_{\rm _B}$ \citep[][]{Gayley1997, Owoki2002} that defines the flux of the objects that are associated with the investigated dusty objects as a function of the distance D to the wind sources, i.e., the S-stars. As is obvious from
\begin{equation}
   r_{\rm B}(D)\,=\,D/\left(1+\sqrt{\rm P_{\rm WR/\nu}}\right),
    \label{eq:breacking-point}
\end{equation}
where $r_{\rm B}(D)$ scales with the flux density of a putative dust blob, $D$ represents the distance to the stellar wind emitting stars, and $\rm P_{WR/\nu}$ is the mass-loss-dependent momentum flux to a Wolf-Rayet (WR) star. Until now, no configuration of S-stars nor their trajectory offers any possibility of a clump with a mass of $\rm \approx\,3M_{\bigoplus}$ moving on a Keplerian orbit around Sgr~A*. Furthermore, \cite{Luehrs1997} finds that the flux density of colliding wind created clumps declines quadratically with distance to the impact zone, which contradicts the proposed formation scenario of G2 and other similar sources in the S-cluster due to their almost constant intrinsic magnitude \citep{Gillessen2012,Ballone2013,peissker2021,Ciurlo2023}. Even more challenging are the timescales of clumps that can survive in the vicinity of Sgr~A* in contradiction of the observation of G2 over 15 years \citep{Calderon2016, Calderon2020_mnras}.\newline
As a result, several authors have proposed an alternative to the idea of coreless objects near Sgr~A* and are in favor of a stellar scenario \citep{Murray-Clay2012, Scoville2013, Eckart2013, Zajacek2014, Witzel2014, Witzel2017, Owen2023}. Not surprisingly, objects similar to G2 or the D sources in the Orion Nebula are classified as embedded young stars with protoplanetary envelopes and disks \citep{1967ApJ...147..799B,Mann2009a, Mann2009b, Kraus2009}.\newline 
Interestingly, \cite{Ciurlo2020} suggested that the dusty sources could be associated with the lower mass members of the S-cluster. If the S-cluster can be classified as a {\it traditional} cluster described by a standard initial mass function (IMF) \citep{Salpeter1955, Massey1998, Kroupa2001}, the question of the lower mass end of the stellar distribution should be addressed, since most stars are massive stars \citep{Habibi2017}. However, large-scale observations suggest a mixed picture: while \cite{Lu2013} is in favor of a top-heavy IMF, \cite{stephan2016, stephan2019} requires a Salpeter value to investigate the binary fraction of the NSC. Given the age of a few Myrs of the S-cluster \citep{Habibi2017} and the suggested starburst a few Gyrs ago \citep{schoedel2020, Chen2023}, it is challenging to find a comprehensive answer of how these different stellar habitats are related to each other. For example, it is ambiguous why the binary fraction of the S-cluster is much lower \citep{Gautam2019} compared to other galactic clusters with comparable age, such as RCW 108 \citep{Comeron2005, Comeron2007}. This fact is especially puzzling because it is well known that the evolution of massive stars is altered by low-mass siblings \citep{Sana2012}.\newline 

Hence, it is of great interest to obtain a complete picture of the stellar content of the S-cluster and to compare it to other sub-clusters in the NSC. Recently, \cite{peissker2023c} showed that the arrangement of dusty sources in the IRS 13 cluster follows a disk-like distribution as it is proposed for the main-sequence S-stars by \cite{Ali2020} or the NSC \citep{Paumard2006, Fellenberg2022}. Although theoretical models suggest such an organized gas distribution for inspiraling molecular clouds \citep{Hobbs2009} that could result in the disk-like pattern as observed by \cite{Paumard2006} and \cite{Fellenberg2022}, we will test this hypothesis for the dusty sources of the S-cluster. We will use the orbital elements of the dusty sources presented in \cite{Peissker2020b} as a starting point and expand our sample. Using these Keplerian orbital parameters will help us to constrain the exact configuration of our cluster member sample. We will further apply the radiative transfer model \texttt{HYPERION} \footnote{\texttt{HYPERION}: an open-source parallelized three-dimensional dust continuum radiative transfer code \citep{Robitaille2011}.} to estimate the mass of the dusty sources \citep{Zajacek2017}. Furthermore, we find a Doppler-shifted double peak for X8 that underlines its bipolar morphology \citep{Peissker2019}. 

This manuscript is structured as follows: In Sec. \ref{sec:analysis}, we briefly discuss the data and describe the used tools and methods. Section \ref{sec:results} presents the results, which is followed by the discussion outlined in Sec. \ref{sec:discuss}. Finally, we list our key findings in Sec. \ref{sec:conclusion}.

\section{Data and Tools} 
\label{sec:analysis}

\begin{figure*}[htbp!]
	\centering
	\includegraphics[width=.8\textwidth]{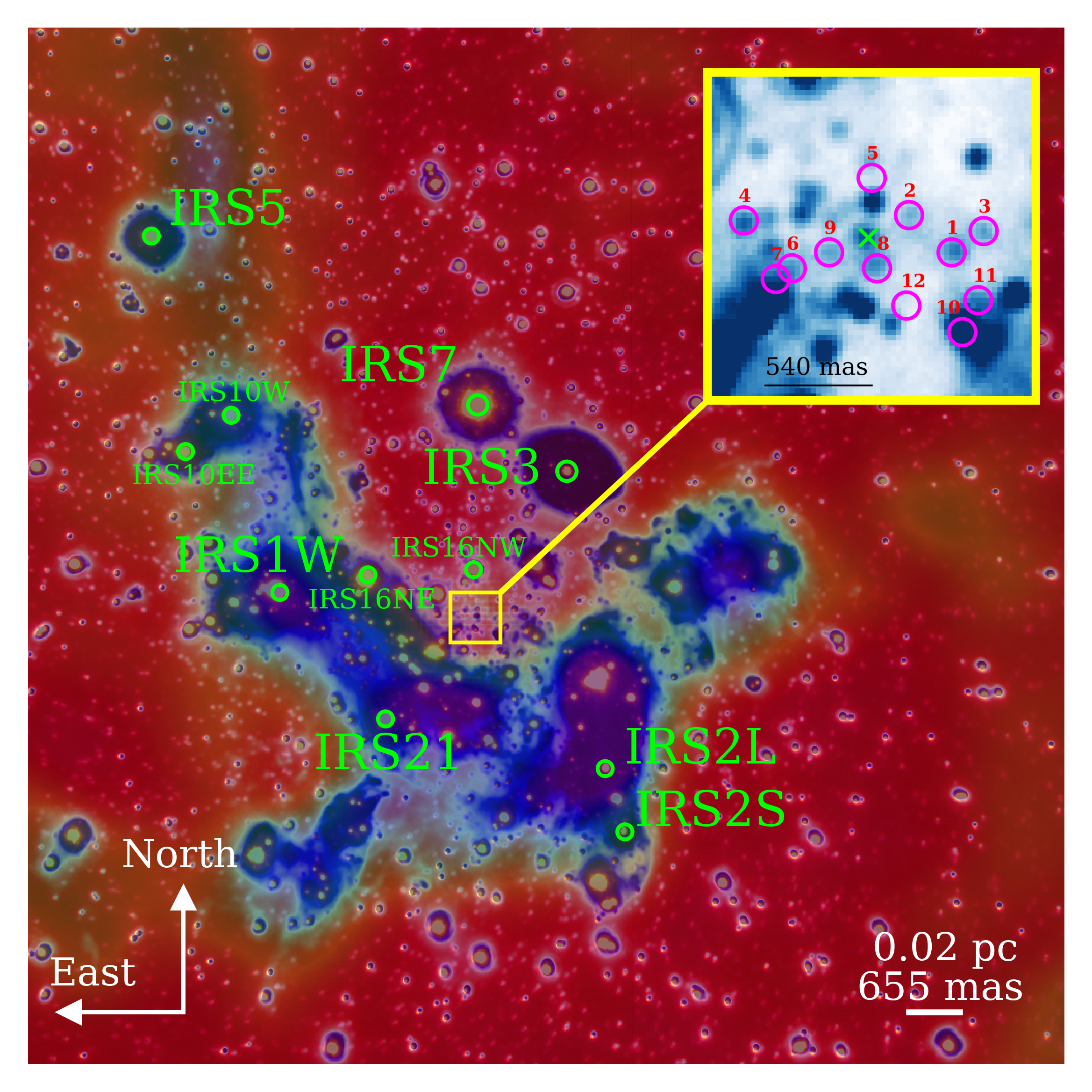}
	\caption{Multi-wavelength finding chart of the inner $\sim 0.4$pc of the Galactic center observed in the K (red) and L (blue) band observed with NACO (VLT). The green color represents 100 GHz observations carried out with ALMA and is based on the reduced data analyzed in \cite{Moser2017}. The brightest K-band (IRS7) and L-band source (IRS3) are indicated. The inset shows a zoomed view of the yellow box where we mark the position of Sgr~A* (lime-colored) with an $\times$. The positions of all dusty sources analyzed in this work are symbolized with a magenta-colored circle. North is up, east is to the left.}
\label{fig:finding_chart}
\end{figure*}

This section will briefly describe the instruments used and provide an overview of the applied tools. The data and telescopes/instruments are listed in Appendix \ref{ref:data_appendix}. To clarify the definition of the dust sources, we manifest a few key points in Table \ref{tab:dusty_source_def}. For example, all dusty sources are gravitationally bound to Sgr~A* which translates into a Keplerian orbit to describe their trajectory. All sources exhibit a Doppler-shifted Br$\gamma$ line accompanied by a NIR and MIR continuum magnitude that fulfills mag$_{NIR}\,\leq\,$mag$_{MIR}$.  
\begin{table}[htbp!]
        \centering
        \begin{tabular}{|c|cc|}
        \hline\hline
              & Doppler-shifted Br$\gamma$ & NIR/MIR continuum  \\  
         \hline  
       Emission & yes & mag$_{NIR}\,\leq\,$mag$_{MIR}$  \\
        \hline
       Remarks     &  \multicolumn{2}{c|}{Keplerian orbit}  \\     
        \hline
        \end{tabular}   
\caption{Definition of a dusty source. Because the majority of objects in the NSC can be classified as main-sequence stars, the photometric requirement mag$_{NIR}\,\leq\,$mag$_{MIR}$ is a necessary condition. Similarly, the detection of a Doppler-shifted Br$\gamma$ emission line is, despite the limitations imposed by the field of view of SINFONI, a necessary condition. As shown in \cite{Peissker2020b}, the presence of some additional emission lines, such as the forbidden iron line [FeIII], can vary for individual sources.}
\label{tab:dusty_source_def}
\end{table}
The identified dusty objects of the S-cluster are displayed in the multi-wavelength Fig. \ref{fig:finding_chart}. This finding chart shows an RGB image making use of ALMA and NACO data.
The NACO data is observed in the K- and L-band in 2008 and 2011 (large overview). To enhance the amount of dusty objects that are confused in some epochs, we use data from 2009 for the shown MIR inlet. The ALMA 100 GHz observations were executed in 2017 and taken from the data published along \cite{Moser2017}.

\subsection{SINFONI and NACO}

For the identification of the Doppler-shifted Br$\gamma$ (Table \ref{tab:dusty_source_def}) in the K-band, we use the Spectrograph for INtegral Field Observations in the Near Infrared (SINFONI). The data analyzed in this work are openly accessible through the ESO archive\footnote{\url{http://archive.eso.org/eso/eso_archive_main.html}} from which we incorporate all available data sets observed in the H+K band. As described in numerous publications covering the same region \citep[e.g.,][]{peissker2020a, peisser2021b, peissker2023b}, a sufficiently high data quality in combination with a sufficiently high on-source integration time, that exceeds at least two hours, is a strong requirement for observing the dusty sources. We refer the interested reader to \cite{peissker2021} and \cite{peissker2021c} for a detailed analysis of the impact of the integration time. The data reduction was performed applying the standard ESO pipeline whenever feasible. 
For the Nasmyth Adaptive Optics System (NAOS) - Near-Infrared Imager and Spectrograph (CONICA), abbreviated NACO, we use the data analyzed in \cite{Witzel2012}, \cite{Shahzamanian2017}, \cite{Parsa2017}, \cite{Peissker2020b}, \cite{Peissker2020c}, \cite{Peissker2020d}, \cite{peissker2021}, and \cite{peissker2021c}. Errors caused by missing keywords in the fits header were bypassed using DPuser \citep{Ott2013}, where we applied the standard reduction steps (DARK, FLAT, JITTER) to the raw data.

\subsection{Keplerian orbit and MCMC simulation}

For the orbital analysis of the dusty sources, we use a Keplerian fit to derive positions employing a least-square fit with respect to the location of Sgr~A*. The analysis benefits from the well-observed S2 orbit \citep{Schoedel2002, Parsa2017, gravity2018, Do2019S2}. From this orbit, we estimate the position of the SMBH, which is used as a reference source. The orbital elements for S2 are taken from \cite{Do2019S2} and match the outcome of the SiO {maser} identification discussed in \cite{Parsa2017}.
{The authors comment on the significance of connecting the radio emission with the infrared counterpart of the maser stars and determine the uncertainty range of 1-10 mas. This uncertainty range is comparable to the astrometric analysis of \cite{Plewa2018}, where the authors use single S-stars and flares to estimate the position of Sgr~A*.}
The resulting orbital elements are then associated with the priors for the MCMC simulations. We use these simulations to determine the uncertainties of the related Keplerian fit solutions due to the underestimation related to the minimized $\chi^2$ analysis that may not reflect the background variations of the S-cluster \citep{peissker2021c}.

\subsection{Line/Channel maps}

Some of the sources have been previously analyzed in \cite{Peissker2020b, peissker2021, peissker2021c}. In contrast, we present new or updated detections for D5, D9, X7, X7.1, and X8 using Doppler-shifted Br$\gamma$ maps. Since SINFONI data cubes consist of individual channels representing a specific wavelength, we isolate frequencies related to the Doppler-shifted Br$\gamma$ line that is used to identify the dusty sources. For the construction of line maps, we select the Doppler-shifted Br$\gamma$ peak and subtract the underlying continuum. Since the FOV of the resulting line map matches the continuum emission, where we already identified the location of Sgr~A*, we extract the related position of the investigated source and, hence, the distance to the SMBH.

\subsection{Radiative transfer model}

{For modeling the radiative transfer based on continuum data, we use \texttt{HYPERION} \citep{Robitaille2011, Robitaille2017}. The code uses optically thick cells to avoid trapped scattered photons in regions that are classified with high optical depth. Due to this principle, the code is a suitable choice for extreme conditions that are associated with the environment of an embedded YSO. Compared to other radiative transfer codes, modeling the surroundings of a pre-main sequence star or even a protostar is one of the strengths of \texttt{HYPERION} \citep{Steinacker2013}.}\newline
Due to the extensive available documentation of the 3-d dust continuum radiative transfer code \texttt{HYPERION} \citep{Robitaille2011, Robitaille2017}\footnote{See also \url{http://www.hyperion-rt.org/}.}, we will limit the description and focus on the basic features. The code utilizes a 4-element Mueller matrix for the calculation of random orientated grain scattering, which is defined as follows:\newline
$
\begin{pmatrix}
       I\\
       Q\\ 
       U\\ 
       V\\
\end{pmatrix}_{scattered}
=
\begin{pmatrix}
   S_{11} & S_{12} &    0   &    0    \\
   S_{12} & S_{11} &    0   &    0    \\
       0  &     0  & S_{33} & -S_{34} \\
       0  &     0  & S_{34} &  S_{33} \\
\end{pmatrix}
\times
\begin{pmatrix}
       I\\
       Q\\ 
       U\\ 
       V\\
\end{pmatrix}_{incident}
$\newline
Here, the vector ($I, Q, U, V$) defines the Stokes parameter where $I$ represents the total intensity, $Q$, $U$, and $V$ stand for the linearly/circularly polarized intensities. The efficient and accurate scattering process itself is realized by raytracing. As suggested by the name, individual (scattered) rays, including their respective radiation origins, are simulated which ensures non-isotropic scattering processes. Due to the nature of scattered rays, energy absorption becomes eminent for individual dust grains. Therefore, \texttt{HYPERION} uses an iterative approach to estimate the specific absorption rate $\dot{A}$. {This process is realized by defining cells that can be associated with resolution. The higher the cell number, the more computationally demanding the code.} The related local thermodynamic equilibrium (LTE) is defined as
\begin{equation}
    \dot{A}\,=\,4\pi\kappa_P(T)B(T),
\end{equation}
where $T$ describes the dust temperature, $B$ represents the integral of the Planck function, and $\kappa_P(T)$ is the mass absorption coefficient. 
All parameters describing a YSO are free and defined by user input. To some extent, the code produces SEDs to fit the spectral energy distribution. However, limitations of this process result in an insufficient output, i.e., \texttt{HYPERION} is not able to fit every flux density value to a given set of input parameters. This is evident for the SED of the cool carbon AGB star IRS 3 (Fig. \ref{fig:finding_chart}) which cannot be produced by \texttt{HYPERION} \citep{peissker2023c}.

\subsection{High-pass filtering}
\label{sec:smooth_subtract}
To increase the contrast but also reduce the imprint of destructive PSF wings, we use a high-pass filter to identify the dusty sources in the K-band. Although different implementations of the basic idea of filtering noise associated with low spatial frequencies exist, we will focus on the smooth-subtract algorithm. This algorithm ensures a robust astrometric analysis which we demonstrated in Fig. 2 in \cite{peissker2023c}. In the same publication, we shortly discuss the filtering technique which can be described as follows: The input image I$_{in}$ is smoothed to create I$_{sm}$. With I$_{in}$-I$_{sm}$, we get I$_{out}$. For further information, we refer to \cite{peissker2023b} and \cite{peissker2023c}. Since the majority of the dusty objects discussed here emit well above the detection limit, the usage of this filtering technique is not required for all sources. Whenever applied, we indicate its usage. 

\subsection{Photometric analysis}

Due to the eminent crowding of the S-cluster and the different field of view (FOV) of our mosaics observed with SINFONI ($\sim$1 arcsec) and NACO ($\sim$20 arcsec), it is sufficient to select various reference stars to enable flexibility for photometric analysis. To ensure an adequate precision of the reference stars in the smaller FOV of SINFONI, it is mandatory to select non-variable calibrator stars and compare the numerical values of all sources with the literature. Fortunately, the photometric analysis in this work benefits from long-time surveys of the Galactic center to classify variable and non-variable stars \citep{Rafelski2007, Pfuhl2014, Gautam2019}. From these surveys, it is known that IRS16NW (Fig. \ref{fig:finding_chart}) is a non-variable He star \citep{Krabbe1995, Blum1996, Ott1999}. The non-variable behavior of IRS16NW is reflected in the estimated K-band magnitudes of 10.0 mag from \cite{Blum1996} and \cite{Stolte2010} that precisely match the second decimal position. Furthermore, using the M-band magnitude of 5.5 mag of IRS2L from \cite{Viehmann2006} and \cite{Viehmann2007} results in 6.9 mag for IRS16NW in agreement with the numerical values from \cite{Moultaka2009}. For the calculation of the magnitude of the S-cluster stars S1, S2, and S4, we use the relation
\begin{equation}
    \rm mag_{source}\,=\,mag_{ref}\,-\,2.5\,\times\,\log_{10}(I_{source}/I_{ref})
\end{equation}
throughout this work, where $I$ refers to the intensity of the investigated source and the reference star. Comparing various literature values of several stars suggests that the reference magnitude values of IRS16NW and IRS2L listed in Table \ref{tab:ref_sources} are trustworthy\footnote{The zero point magnitudes for IRS16NW were taken from \cite{Frogel1987}, for IRS2L from \cite{Ott1999}.}. Furthermore, it is reassuring that the NIR and MIR magnitudes of S2 derived with the calibrator stars IRS16NW and IRS2L are in agreement with literature values listed in \cite{Clenet2004a},\cite{Schoedel2011}, \cite{Sabha2012}, and \cite{Witzel2014}.
\begin{table}[htbp!]
    \centering
    \begin{tabular}{|c|cccc|}
            \hline
            \hline
            Star  & H & K & L & M \\
            \hline
              S1       & 16.5$\pm$0.1 & 14.6$\pm$0.1 & 13.0$\pm$0.6 & 12.6$\pm$0.4\\
              S2       & 15.9$\pm$0.1 & 14.1$\pm$0.1 & 12.6$\pm$0.7 & 12.8$\pm$0.1\\ 
              S4       & 16.2$\pm$0.1 & 14.4$\pm$0.1 & 13.1$\pm$0.6 & 13.1$\pm$0.2\\
              \hline
              IRS16NW  & 12.0$\pm$0.02 & 10.0$\pm$0.04  & 8.4$\pm$0.1 & 6.9$\pm$0.1 \\
              IRS2L    & 14.2$\pm$0.2  & 10.6$\pm$0.2   & 6.4$\pm$0.1 & 5.5$\pm$0.1 \\
            \hline
    \end{tabular}
    \caption{Calibrator stars for the data set analyzed in this work. We adapt the listed magnitude values {for IRS16NW from \cite{Blum1996}, \cite{Stolte2010}, and \cite{Gautam2019}. The magnitudes for IRS2L are taken from \cite{Viehmann2006} and \cite{Viehmann2007}. The magnitudes for S1, S2, and S4 are estimated with these two IRS stars. All listed and independent derived magnitudes for the S-stars are in agreement with the literature (S1-S4, L/M-band: \cite{Clenet2004a}; S2, K-band: \cite{Sabha2012}; S1-S4, K-band: \cite{Habibi2017}}. With the listed reference stars, we estimate the NIR and MIR magnitudes of S2. Inter alia, we calculate the L-band magnitude of this star to $12.6\pm0.7$ in agreement with \cite{Witzel2014}. The magnitude of the standard deviation based uncertainty reflects the admissible fluctuation of S2 \citep{Hosseini2020}. The NIR values of S1 and S4 are calculated with the numerical magnitude results for S2, whereas IRS16NW is used for the MIR magnitude of the S-cluster star. {We refer to Appendix \ref{app_sec:photometry} for the complete list of all individual estimated magnitudes over a time of 12 years.}}
    \label{tab:ref_sources}
\end{table}
Since we established S2 as a proper reference star for the data set analyzed in this work, we calculate matching K-band magnitudes for S1 and S4 in agreement with \cite{Habibi2017}. Furthermore, we use IRS16NW to estimate the MIR magnitude of S1 and S4 to avoid the imprint of a possible variable behavior of S2 \citep{Hosseini2020}\footnote{We note that the reported variations of the L-band flux of S2 could be explained by local background fluctuations or the presence of interfering dusty sources.}. With the listed set of calibrator stars outlined in Table \ref{tab:ref_sources}, we adequately confront various crowding challenges for the NACO and SINFONI data set covering 2002-2019. Individual magnitude values of different epochs for the S-cluster stars S1, S2, and S4 are listed in Appendix \ref{app_sec:photometry}, Table \ref{tab:single_mag_epoch_photometry}.
For the final flux density values, we use average magnitude values of the individual dusty sources and the relation
\begin{equation}
    \rm F_{obj}\,=\,F_0 \times 10^{(-0.4(mag_{obj}-mag_{ref}))}
    \label{eq:flux}
\end{equation}
where $F_0$ is the zero flux of the reference source. To eliminate problems with the calibrators, we use two different stars and adopt the related values from \cite{Viehmann2006} and \cite{peissker2023b}. We list all numerical values used to estimate the flux density of the dusty sources with Eq. \ref{eq:flux} in Table \ref{tab:mag_fux_reference_values}.
\begin{table}[hbt!]
    \centering
    \begin{tabular}{|c|cccc|}
         \hline 
         \hline
           Filter & \multicolumn{2}{c}{IRS2L} & \multicolumn{2}{c|}{S2} \\
         \hline
                 & Magnitude  & Flux$_{\lambda}$  & Magnitude  & Flux$_{\lambda}$ \\
                 &  [mag] &  [Jy] & [mag] &  [Jy] \\
         \hline
         H-band  &  14.26   & 0.13 & 15.9  & 0.032  \\
         K-band  &  10.60   & 0.48 & 14.1  & 0.014  \\         
         L-band  &   6.4    & 2.98 & 12.6  & 0.009  \\         
         M-band  &   5.5    & 3.98 & 12.8  & 0.004  \\     
       \hline
    \end{tabular}
    \caption{Final magnitude and flux density values of the calibrator stars used in this work. All values are cross-checked with the literature.}
    \label{tab:mag_fux_reference_values}
\end{table}
As mentioned, the listed magnitude and flux values are consistent with the literature and provide a robust photometric analysis presented in this work and beyond that.

\section{Results}
\label{sec:results}

In this section, we present the results of the analysis of the infrared data observed at the VLT covering about two decades. The first part of this section covers the kinematic properties of the dusty sources in the S-cluster whereas the second part focuses on the photometric analysis, including the radiative transfer model.

\subsection{Identifying dusty objects}
\label{sec:ident_dusty_objects}
\begin{figure*}[htbp!]
	\centering
	\includegraphics[width=1.\textwidth]{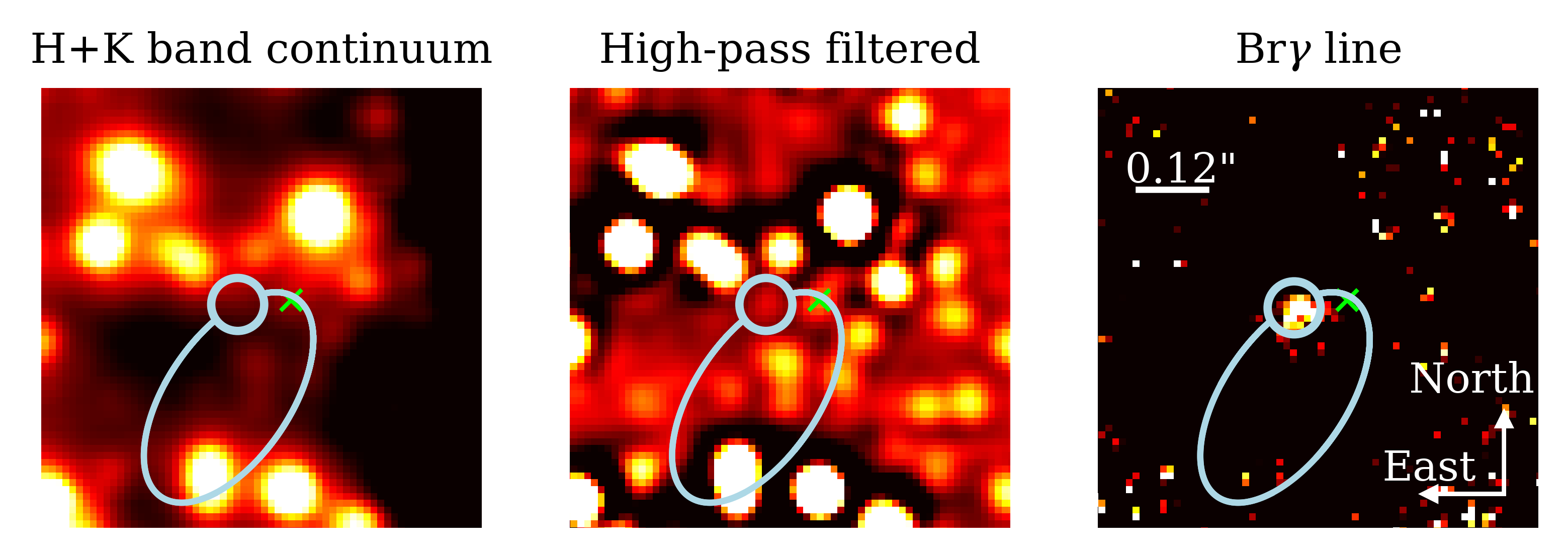}
	\caption{Identification of G2/DSO in the H+K continuum as proposed in \cite{Peissker2020b} and \cite{peissker2021c} observed with SINFONI in 2012. The left plot represents the H+K continuum view of a collapsed SINFONI data cube which is used as an input for the smooth subtract algorithm. The middle plot shows the result of smoothing the left image and subtracting it from itself. Further, the H+K-band continuum position of G2/DSO is consistent with the L-band detection shown in \cite{Gillessen2013b}. The right plot shows the Doppler-shifted line map detection of G2/DSO at about $2.18\mu m$. The orbit is adapted from \cite{peissker2021c} and the astrometric position of G2/DSO is consistent with the results presented \cite{Peissker2020b} where we used the Lucy-Richardson deconvolution algorithm to perform the high-pass filtering \citep{Lucy1974, Peissker2022}. The H+K-band continuum detection of G2/DSO demonstrates the superiority of this data processing method. Here, the north is up, and the east is to the left.}
\label{fig:dso_ident}
\end{figure*}

\begin{figure*}[htbp!]
	\centering
	\includegraphics[width=1.\textwidth]{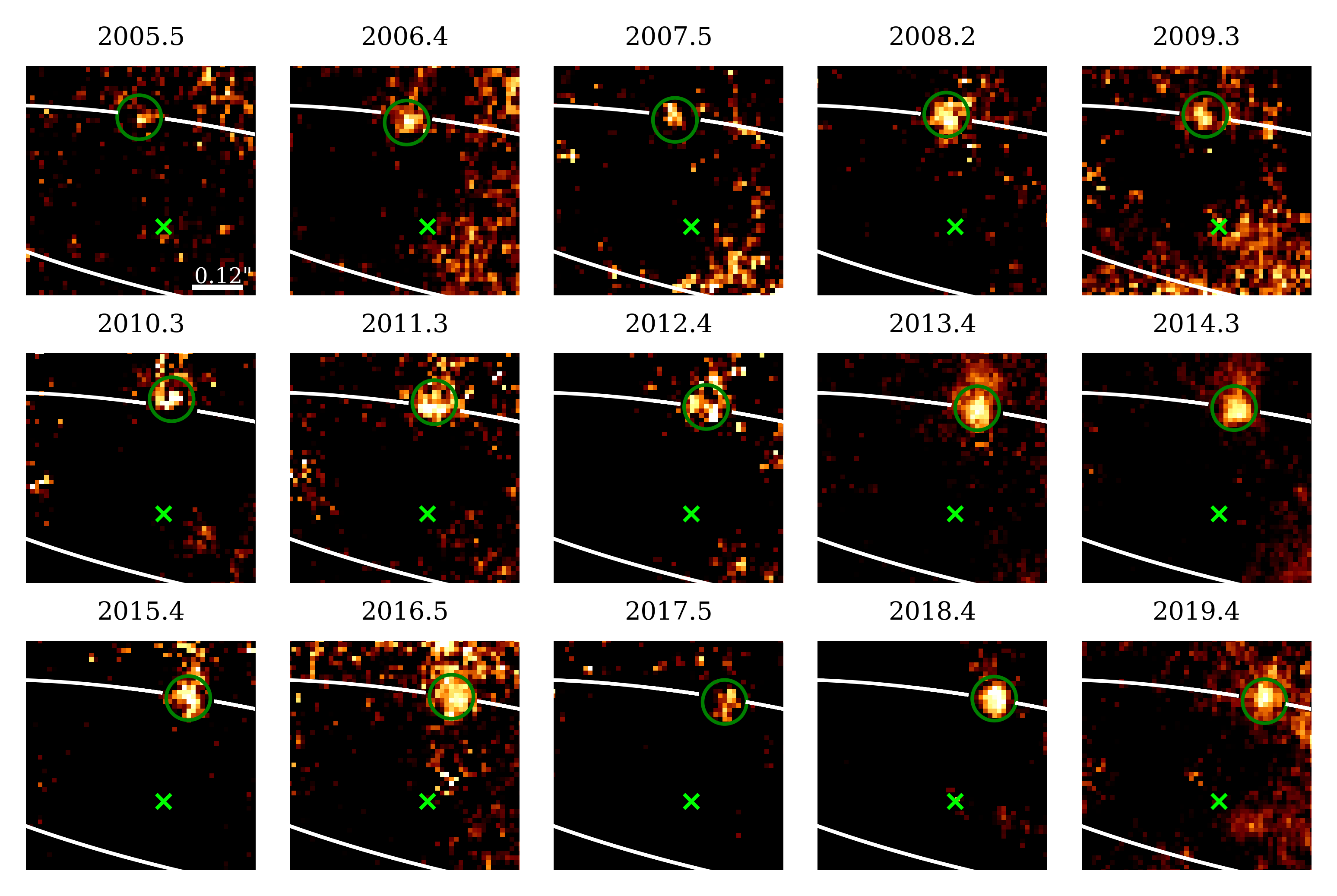}
	\caption{Br$\gamma$ line evolution of D9 between 2005 and 2019 observed with SINFONI. D9 can be identified in every epoch with a proper motion of about 250 km/s. The related distance of D9 and the LOS velocity is listed in Table \ref{tab:d9_positions}. Sgr~A* is marked with a lime-colored $\times$. Here, the north is up, and the east is to the left. For the epoch representing 2015 and 2016, we note remnants to the north of D9, which are related to insufficient instrument response of the observations.}
\label{fig:d9_ident}
\end{figure*}

Most dusty sources are identified in the Doppler-shifted Br$\gamma$ regime and in the L-band. Until now, every Br$\gamma$ line-emitting source is accompanied by a MIR counterpart suggesting a complex intrinsic composition that demands a temperature gradient. While the observation of the sources in the Doppler-shifted Br$\gamma$ regime and the MIR domain is not doubted, the high-pass filtered Lucy-Richardson \citep{Lucy1974} deconvolved NIR detection of some objects, such as G2/DSO, contradicts the missing K-band detection discussed by \cite{Gillessen2012}. Therefore, it is reassuring to identify G2/DSO with an independent filter such as the here discussed robust smooth subtract algorithm. In Fig. \ref{fig:dso_ident}, we show the detection of G2/DSO in the H+K band observed with SINFONI. For the shown detection, we collapse the data cube, which is constructed from observations carried out in 2012. All 2172 channels are used to create an H+K band image which is then treated with the smooth subtract algorithm described in Sec. \ref{sec:smooth_subtract}. With this technique, the astrometric precision can be enhanced by about 5-10$\%$.
As displayed in Fig. \ref{fig:dso_ident}, the majority of the identification of the dusty sources is accomplished using line maps. While we identified and analyzed in detail most of the sources shown in Fig. \ref{fig:dso_ident} in previous publications \citep{Peissker2019, Peissker2020b, peissker2021, peissker2021c}, we focus on the Br$\gamma$ line evolution of the inconspicuous object D9 between 2005 and 2019, as observed with SINFONI, in Fig. \ref{fig:d9_ident}.
This source is observed after its periapse and moves with an averaged LOS velocity of about 150 km/s (Table \ref{tab:d9_positions}) towards the projected west. To determine the positions of D9, we use a Gaussian fit which simultaneously provides a fitting error. However, this resulting uncertainty from the fit is in the range of 2-3$\%$ of a pixel. Due to the approximately constant distance of D9 from its current location on its orbit around Sgr~A*, we use the standard deviation to estimate a related uncertainty. Due to the large and approximately circular orbit with an eccentricity of $e\,=\,0.32\pm0.01$, we calculate a 3d distance for D9 with
\begin{equation}
    r\,\sim\,GM_{SgrA*}/v_{D9}^2,
\end{equation}
where we use the latest LOS velocity of $224.36\pm25.0$km/s observed in 2019 (see Table \ref{tab:d9_positions}) with the related proper motion of $249.43\pm5.01$km/s to estimate the total space velocity $\sim (v_{LOS}^2+v_{prop}^2)^{1/2}=335.49\,{\rm km/s}$. The uncertainties of the LOS are related to the spectral resolution of SINFONI, whereas the error for the proper motion stems from astrometric measurements. 
\begin{table}[htb]
\centering
\begin{tabular}{|c|ccc|}\hline 
 Epoch & Distance [mas]  & Br$\gamma$-line [$\mu m$] & Velocity [km/s] \\  
\hline 
2005.5 & 305.43$\pm$7.93 & 2.16521 & 123.26 \\
2006.4 & 289.21$\pm$7.93 & 2.16500 & 152.34  \\
2007.5 & 294.39$\pm$7.93 & 2.16492 & 163.42  \\
2008.2 & 306.44$\pm$7.93 & 2.16493 & 162.04 \\
2009.3 & 307.07$\pm$7.93 & 2.16550 & 83.09  \\
2010.3 & 313.37$\pm$7.93 & 2.16551 & 81.71  \\
2011.3 & 305.04$\pm$7.93 & 2.16492 & 163.42 \\ 
2012.4 & 294.12$\pm$7.93 & 2.16552 & 80.32  \\
2013.4 & 294.11$\pm$7.93 & 2.16551 & 81.71  \\
2014.5 & 291.37$\pm$7.93 & 2.16502 & 149.57 \\
2015.4 & 288.89$\pm$7.93 & 2.16497 & 156.50 \\
2016.5 & 291.98$\pm$7.93 & 2.16502 & 149.57 \\
2017.5 & 285.09$\pm$7.93 & 2.16451 & 220.21 \\
2018.4 & 298.90$\pm$7.93 & 2.16499 & 153.73 \\
2019.4 & 299.21$\pm$7.93 & 2.16448 & 224.36 \\
\hline
\end{tabular}
\caption{Distance with respect to Sgr~A* and LOS velocity of D9 between 2005 and 2009. The astrometric uncertainties that are based on the standard deviation reflect the noisy character of the investigated line maps. For the uncertainty of the velocity, we determine $\pm 47\,km/s$.}
\label{tab:d9_positions}
\end{table}
With a mass for Sgr~A* of $M_{SgrA*}\,=\,4.02\times10^6\,M_{\odot}$ \citep{Do2019S2,Peissker2022,eht2022}, we estimate a 3d distance for D9 of $\sim 0.15$ pc in 2019. Using the averaged LOS velocity for D9, the 3d distance increases to 0.19 pc, which we translate into an uncertainty for the 3d distance of D9 of 0.04 pc between 2005 and 2019. The final 3d distance for the investigated epochs is $0.17\pm0.02$pc.
With the parameters derived for the distance of D9 and the influence sphere of Sgr~A* of about 2 pc, the application of a Keplerian fit is justified, although we note a rather slow LOS velocity compared to, e.g., G2/DSO \citep{peissker2021c}.
In addition to D9, we furthermore revisited D5 \citep{Eckart2013,Peissker2020b}, X7 \citep{peissker2021}, X7.1 \citep{Peissker2020b, peissker2021}, and X8 \citep{Peissker2019} with the approach demonstrated here. All the measured astrometric analysis results used in this work can be found in Appendix \ref{app_sec:astrometric_measurments}.

\subsection{Keplerian orbit}

\begin{figure*}[htbp!]
	\centering
	\includegraphics[width=1.\textwidth]{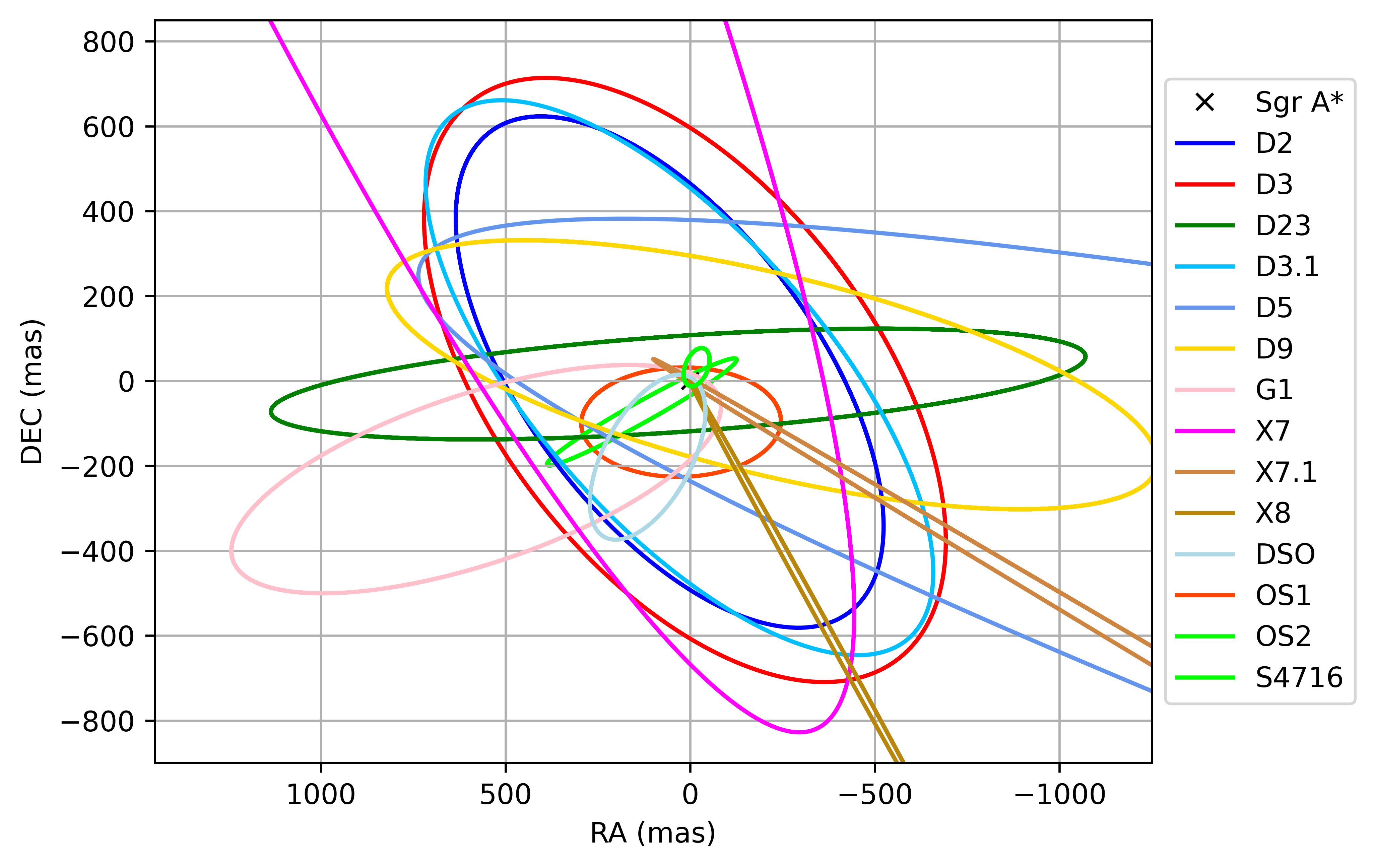}
	\caption{Keplerian orbits of all the investigated dusty objects in this work. The shown Keplerian solutions represent best-fit orbits which are limited by the data baseline. Therefore, more data results in a longer data baseline with increased accuracy. Here, Sgr~A* is located at (0,0). The orbit of S4716 is taken from \cite{Peissker2022}. North is up and east is to the left.}
\label{fig:all_orbits}
\end{figure*}

In addition to the pre-existing results that include D2, D23, D3, D3.1, G1, G2/DSO, OS1, and OS2 \citep{Peissker2020b, peissker2021c}, we list the Keplerian orbits of the revisited sources in Table \ref{tab:orbital_elements}. The uncertainties are related to the MCMC simulations displayed in Appendix \ref{app_sec:mcmc_simulations}.
\begin{table*}[htbp!]
    \centering
    \begin{tabular}{|ccccccccc|}
            \hline
            \hline
            ID  &Source & $a$ [mpc] & $e$ & $i$ [$^\circ$] & $\omega$ [$^\circ$] & $\Omega$ [$^\circ$] & $t_{\rm closest}$ [years]& PA [$^\circ$] \\
            \hline
            1 & D2     & 30.01 $\pm$ 0.04   &  0.15 $\pm$ 0.02  & 57.86 $\pm$ 1.71   & 44.11 $\pm$ 1.14   & 221.16 $\pm$ 1.43   & 2002.16 $\pm$ 0.03   & 129.9\\
            2 & D23    & 44.24 $\pm$ 0.03   &  0.06 $\pm$ 0.01  & 95.68 $\pm$ 1.14   & 93.39 $\pm$ 1.89   & 120.89 $\pm$ 3.15   & 2016.00 $\pm$ 0.03   & -84.2\\
            3& D3      & 35.20 $\pm$ 0.01   &  0.24 $\pm$ 0.02  & 50.99 $\pm$ 1.20   & 63.59 $\pm$ 2.29   & 206.26 $\pm$ 1.76   & 2002.23 $\pm$ 0.01   & 147.6\\
            3/1 & D3.1 & 31.46 $\pm$ 0.01   &  0.06 $\pm$ 0.02  & 63.02 $\pm$ 1.04   & 231.47 $\pm$ 1.18  &  47.89 $\pm$ 0.96   & 2004.37 $\pm$ 0.008  & -137.5\\
            4 & D5     & 2983.10 $\pm$ 2.11 &  0.991$\pm$ 0.001 & 100.84$\pm$ 4.58   & 343.20$\pm$ 8.02   &  73.33$\pm$ 3.89    &  2008.95$\pm$ 3.43   & 103.1\\
            5 & D9     & 44.00 $\pm$ 1.145  &  0.32 $\pm$ 0.01  & 102.55 $\pm$ 1.14  & 127.19 $\pm$ 8.02  & 257.25 $\pm$ 1.71   & 2309.13 $\pm$ 7.01   & -45.9\\
            6 & OS1    &16.53 $\pm$ 0.50    &0.758 $\pm$  0.075 & 107.71$\pm$ 20.68  &93.96 $\pm$  8.42   &  88.23 $\pm$ 16.50  &2020.68 $\pm$ 0.02    & -86.6\\
            7 & OS2    &12.54 $\pm$  0.19   &0.633 $\pm$ 0.019  & 95.11 $\pm$ 4.98   & 138.65 $\pm$  4.92 & 114.01 $\pm$ 2.17   &2029.41 $\pm$ 0.05    & -122.8\\
            8 & G1     & 48.58 $\pm$  0.05  &0.968 $\pm$ 0.009  &108.28 $\pm$ 1.79   &114.59 $\pm$  4.69  &  99.69 $\pm$ 4.89   &2001.08 $\pm$ 0.05    & -80.9\\
       9 & {G2/DSO}    &17.23 $\pm$  0.20   &0.963 $\pm$ 0.004  &120.32 $\pm$ 2.40   & 92.81 $\pm$  1.60  &  63.02 $\pm$ 1.37   & 2014.43 $\pm$ 0.01   & -59.4\\
            10 & X7    & 149.99 $\pm$ 0.36  & 0.76 $\pm$ 0.05   & 74.82 $\pm$ 7.84   & 190.22 $\pm$ 17.47 &  25.15 $\pm$ 10.77  & 1975.85 $\pm$ 0.30   & 178.7\\ 
            11 & X7.1  & 171.34 $\pm$ 5.27  & 0.97 $\pm$ 0.02   & 89.38 $\pm$ 3.72   & 28.64 $\pm$ 17.18  &  60.90 $\pm$ 8.59   & 2040.99 $\pm$ 15.41  & 117.7\\ 
            12 & X8    & 48.69 $\pm$ 1.12   & 0.99 $\pm$ 0.01   & 95.11 $\pm$ 5.61   & 57.24 $\pm$ 5.90   &  40.10 $\pm$ 4.52   & 1992.92 $\pm$ 10.54  & -33.4\\ 
            \hline
    \end{tabular}
    \caption{Best-fit orbital elements of the dusty sources. Please see Fig. \ref{fig:finding_chart} for the related ID of the dusty sources. The uncertainties are adapted from the MCMC simulations. The first column indicates the id as provided in the inset in Fig. \ref{fig:finding_chart}, where we mark the approximate position of the related dusty source (second column). As proposed in \cite{Peissker2020c}, D3.1 is most likely a dust and gas filament, such as the Br$\gamma$-bar \citep[see][]{Schoedel2011} with a weak radio/submm counterpart \citep{Yusef-Zadeh2017-ALMAVLA}.}
    \label{tab:orbital_elements}
\end{table*}
For G2/DSO, we adapt the updated orbital elements from \cite{peissker2023b}. Using the orbital elements indicated in Table \ref{tab:orbital_elements}, we find the distribution displayed in Fig. \ref{fig:all_orbits}.
This figure represents a projected on-sky view of the orbits of all of the investigated dusty sources analyzed in this work. While a projection does not necessarily reveal the 3d distribution, we investigate the position angle (PA) of the orbits. As shown in \cite{Ali2020}, the PA distribution enables us to inspect a cluster in terms of the patterns in the distribution of stars. 

\subsection{Position angle of the dusty sources}

\begin{figure}[htbp!]
	\centering
	\includegraphics[width=.5\textwidth]{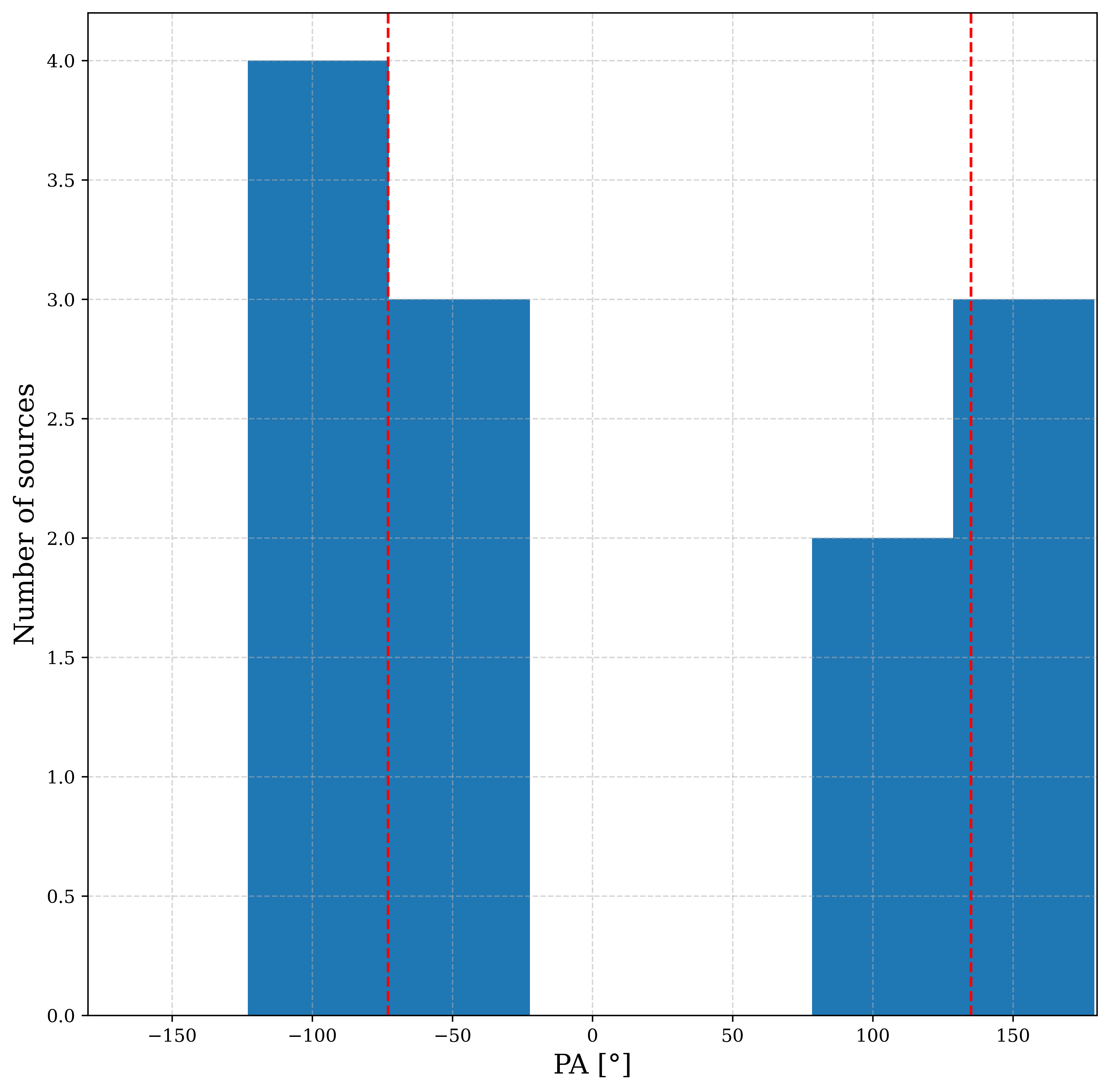}
	\caption{Position angle of the dusty sources. We find a clear bimodal distribution for the investigated sample. The red dashed lines are the average values of the related groups with $(-76.71\pm 25.60)^{\circ}$ and $(135.60\pm 26.23)^{\circ}$ where the uncertainties represent the standard deviation.}
\label{fig:pa}
\end{figure}

\begin{figure}[htbp!]
	\centering
	\includegraphics[width=.5\textwidth]{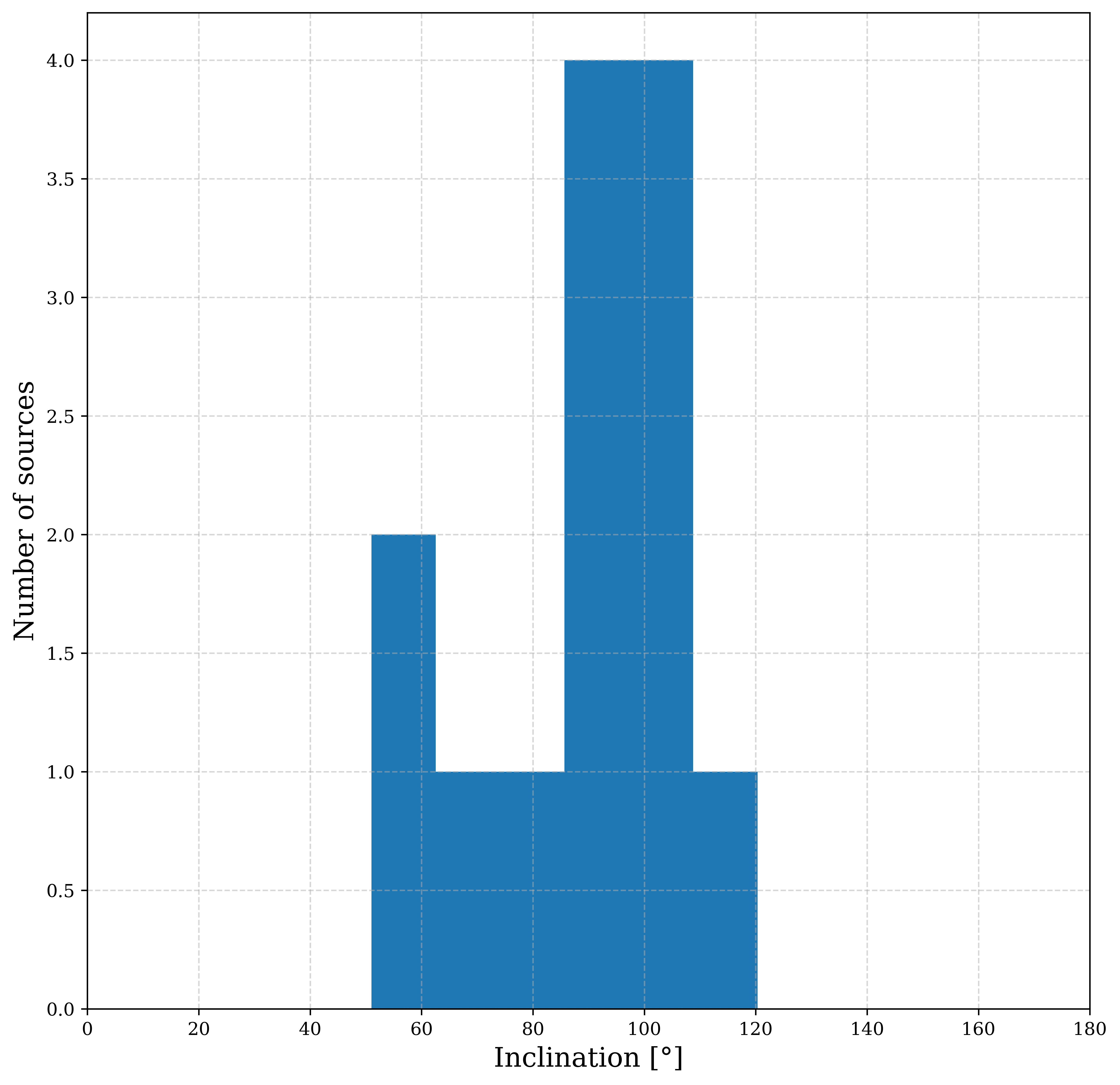}
	\caption{Inclination distribution of the dusty sources. The inclination is limited to angles between 0° and 180° and shows a strong correlation with the red disk distribution presented in \cite{Ali2020}.}
\label{fig:inc}
\end{figure}

\begin{figure}[htbp!]
	\centering
	\includegraphics[width=.5\textwidth]{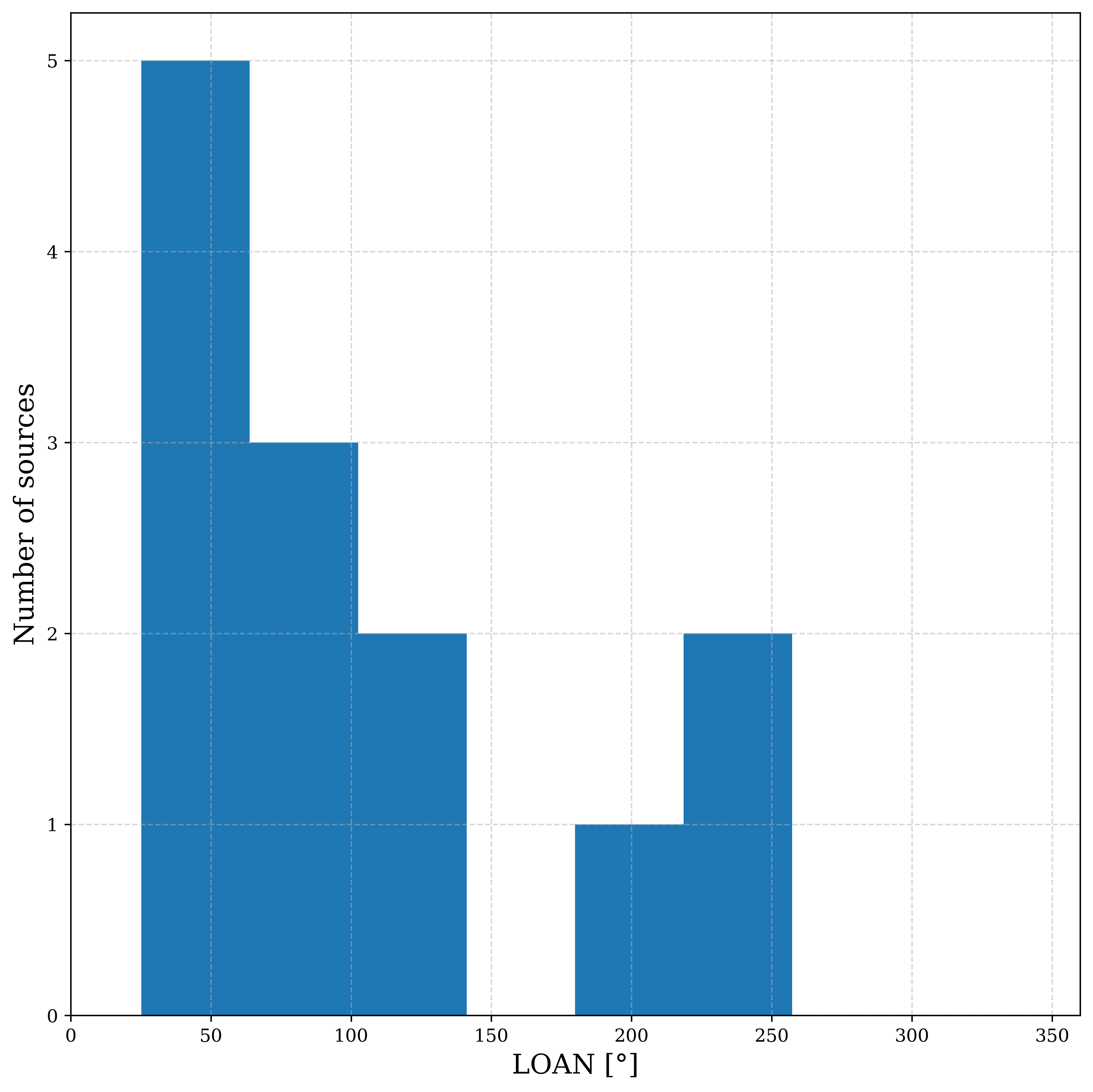}
	\caption{Distribution of the longitudes of the ascending node (LOAN) for the investigated dusty sources. As for the PA displayed in Fig. \ref{fig:pa}, we find a non-randomized distribution of the LOAN.}
\label{fig:loan}
\end{figure}

We determine the PA of the orbits of the cluster member sample with the approach presented in \cite{Ali2020}. The authors use the arctan2 function to estimate the PA because it ensures non-repetitive angles compared to the arctan function, i.e. one cannot distinguish between the location of the sources in individual quadrants. In other words, two quadrants of the arctan function will always feature the same set of angles inhibiting a correct PA analysis. Using the arctan2 function with given positions $x$ and $y$ that represent the orbital phase of the dusty sources results in
\begin{equation}
\theta=\arctan2(y,x)=
    \begin{cases}
        (-y,-x) & \text{if } \theta \in (-\pi,-\pi/2) \\
        (-y,x)  & \text{if } \theta \in (-\pi/2,0) \\
        (y,x)   & \text{if } \theta \in (0,\pi/2) \\
        (y,-x)  & \text{if } \theta \in (\pi/2,\pi) \\
    \end{cases}
\end{equation}
where $\theta$ represents the PA.
With the Keplerian fits listed in Table \ref{tab:orbital_elements}, we find an arranged structure of the orbits displayed in Fig. \ref{fig:pa}. This finding strongly suggests a non-randomized distribution of the dusty sources of the S-cluster in agreement with the arrangement of the S-stars found by \cite{Ali2020}.
By inspecting the related inclination of the sample sources listed in Table \ref{tab:orbital_elements}, we find a similar distribution as the S-stars in the red disk of the S-cluster \citep{Ali2020}. Using a bin size of half the sample number, we find the corresponding arrangement of the inclination displayed in Figure \ref{fig:inc}.
Intriguingly, we find a similar non-randomized arrangement for the LOAN as we do for the PA and the inclination. With the same bin size as in Fig. \ref{fig:inc}, we show the LOAN distribution for the investigated sources (Table \ref{tab:orbital_elements}) in Fig.~\ref{fig:loan}.
However, we will discuss this finding in Sec. \ref{sec:discuss}.

\subsection{Magnitude and flux}

\begin{figure*}[htbp!]
	\centering
	\includegraphics[width=1.\textwidth]{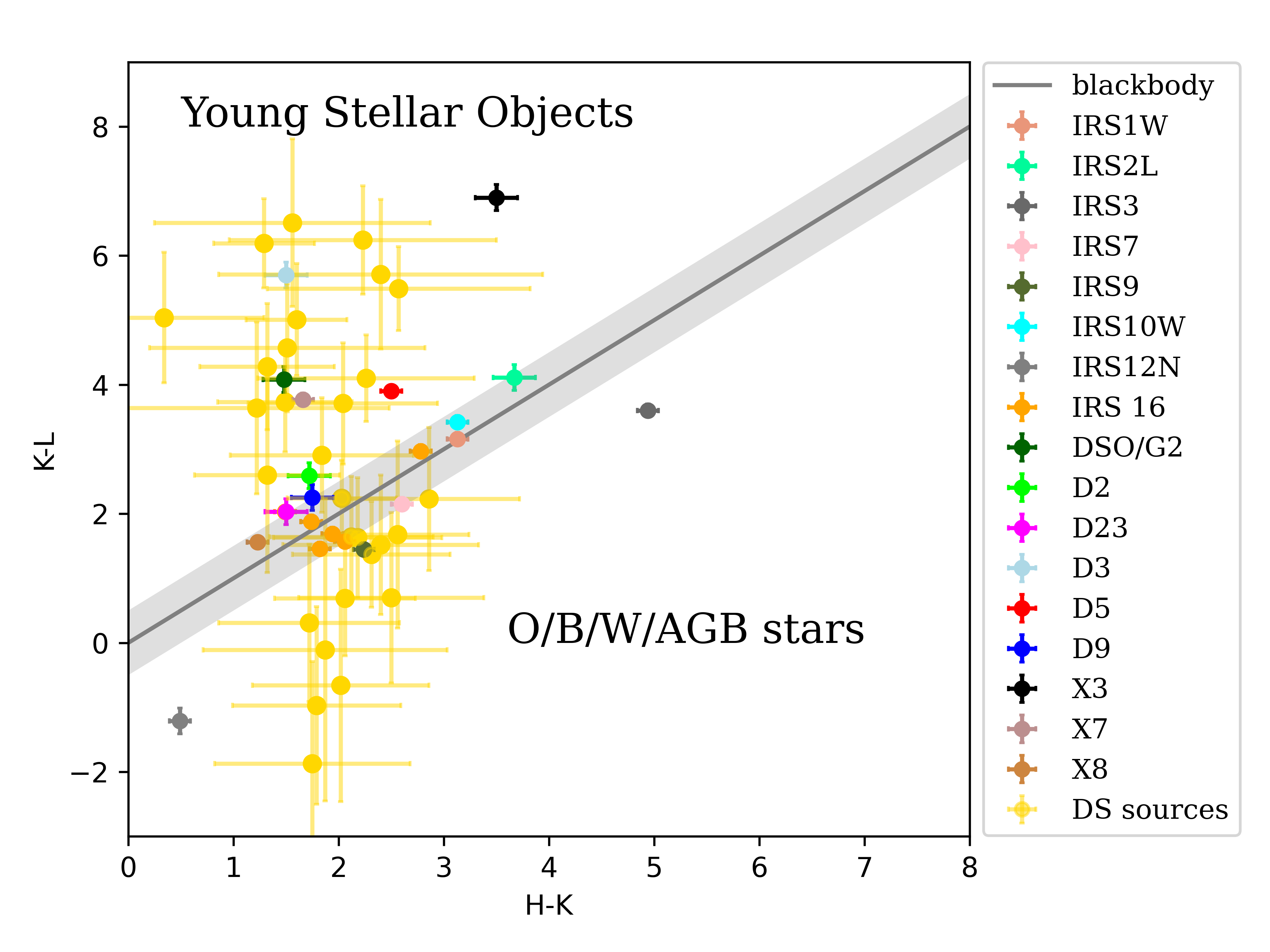}
	\caption{Color-color diagram of the dusty source of the S-cluster and IRS13 (gold-colored). Although the IRS13 sample number is slightly larger compared to the S-cluster objects investigated in this work, both groups of dusty sources show similarities implying a shared nature. Especially the dusty sources of the S-cluster seem to share a comparable photometric footprint. All uncertainties reflect the standard deviation of the related magnitude estimated for all available epochs between 2006 and 2019. The linear gray fit represents a one-component blackbody. Based on this classification, we conclude that the majority of dusty sources exhibit colors that are consistent with a YSO classification. As a comparison, we include the H-K and K-L colors for several main-sequence and AGB stars from \cite{Blum1996}, \cite{Ott1999},\cite{Pott2008}, and \cite{peissker2023b}.}
\label{fig:color_color_diagram}
\end{figure*}

The majority of the dusty sources exhibit Doppler-shifted Br$\gamma$ emission, which motivates the analysis of the SINFONI data. Due to the limited and crowded FOV, we used NACO data to estimate calibrator and reference sources in the S-cluster, namely S1, S2, and S4. With the magnitude values listed in Table \ref{tab:ref_sources}\footnote{Single epochs of the magnitude analysis are listed in Table \ref{tab:single_mag_epoch_photometry}, Appendix \ref{app_sec:photometry}.}, we inspect the H- and K-band of the SINFONI data cubes. To maintain a high level of consistency, we incorporate NACO L- and M-band data of the same epochs that are covered by the SINFONI observations. With this data set, we performed a photometric analysis of the four different bands (H, K, L, M) over a period of 14 years. For the individual numerical values listed in Tables \ref{tab:single_mag_epoch_photometry_dusty_sources_hband}-\ref{tab:single_mag_epoch_photometry_dusty_sources_mband}, we limit the analysis to the continuum data. In Table \ref{tab:mag_sll_sources}, we list the mean magnitudes and the H-K/K-L colors of the related dusty source.
\begin{table*}
\centering
\begin{tabular}{|c|cccccccccc|}
\hline
\hline
Source & \multicolumn{2}{c}{H-band} & \multicolumn{2}{c}{K-band} & \multicolumn{2}{c}{L-band} & \multicolumn{2}{c}{M-band} & K-L & H-K \\ 
\hline
   & [mag] &  [mJy] & [mag] &  [mJy] & [mag] &  [mJy] & [mag] &  [mJy] & & \\
\hline
D2      & 17.59$\pm$0.38  & 6.4$^{+0.3}_{-0.3}$  & 15.87$\pm$0.17  & 3.3$^{+0.4}_{-0.4}$ & 13.27$\pm$0.45 &  5.0$^{+0.2}_{-0.2}$ & 12.75$\pm$0.20 & 4.5$^{+0.4}_{-0.4}$ & 2.59      & 1.72      \\
D23     & 17.70$\pm$0.48  & 5.7$^{+0.3}_{-0.3}$  & 16.20$\pm$0.30  & 2.4$^{+0.3}_{-0.3}$ & 14.16$\pm$1.10 &  2.2$^{+0.1}_{-0.1}$ & 12.81$\pm$0.10 & 4.3$^{+0.3}_{-0.3}$ & 2.03      & 1.50      \\
D3      & 18.70$\pm$0.48* & 2.2$^{+1.1}_{-0.9}$  & 16.50$\pm$0.30* & 1.7$^{+0.7}_{-0.8}$ & 13.42$\pm$0.39 &  4.4$^{+0.2}_{-0.2}$ & 12.79$\pm$0.18 & 4.4$^{+0.3}_{-0.3}$ & 3.08*     & 2.19*     \\
D5      & 18.90$\pm$0.38* & 1.9$^{+0.7}_{-0.4}$  & 17.30$\pm$0.30* & 0.8$^{+0.1}_{-0.3}$ & 13.15$\pm$0.47 &  5.6$^{+0.2}_{-0.2}$ & 12.88$\pm$0.17 & 4.0$^{+0.3}_{-0.3}$ & 4.15*     & 1.59*     \\
D9      & 19.92$\pm$0.62  &0.7$^{+0.5}_{-0.5}$  & 18.17$\pm$0.21  & 0.3$^{+0.6}_{-0.6}$ & 15.92$\pm$1.32 &  0.4$^{+0.1}_{-0.1}$ &     -          &      -              & 2.25      & 1.75      \\
OS1     &      -          &      -               &     -           &  -                  & 13.86$\pm$0.63 &  2.9$^{+0.1}_{-0.1}$ &     -          &      -              &    -      &     -     \\
OS2     &      -          &      -               &     -           &  -                  & 13.97$\pm$0.42 &  2.6$^{+0.1}_{-0.1}$ &     -          &      -              &    -      &     -     \\
G1      &      -          &      -               &     -           &  -                  & 13.22$\pm$0.31 &  5.1$^{+0.2}_{-0.2}$ & 12.89$\pm$0.83 &      -              &    -      &     -     \\
G2/DSO  & 19.96$\pm$0.30* & 0.7$^{+0.3}_{-0.2}$  & 18.48$\pm$0.22* & 0.2$^{+0.2}_{-0.1}$ & 13.53$\pm$0.65 &  4.0$^{+0.1}_{-0.1}$ & 12.87$\pm$0.22 & 4.1$^{+0.4}_{-0.4}$ & 4.08*     & 1.48*     \\
X7      & 17.45$\pm$0.38  & 7.2$^{+0.3}_{-0.3}$  & 15.79$\pm$0.26  & 3.5$^{+0.4}_{-0.4}$ & 12.01$\pm$0.39 & 16.2$^{+0.7}_{-0.7}$ & 12.36$\pm$0.19 & 6.5$^{+0.5}_{-0.5}$ & 3.77      & 1.66      \\
X7.1    & 17.78$\pm$0.15 & 5.3$^{+0.2}_{-0.2}$  &     -           &  -                  & 13.30$\pm$0.44 &  4.9$^{+0.2}_{-0.2}$ & 12.76$\pm$0.21  & 4.5$^{+0.4}_{-0.4}$ &    -      &     -     \\
X8      & 17.05$\pm$0.34  &10.5$^{+0.5}_{-0.5}$  & 15.82$\pm$0.32  & 3.4$^{+0.4}_{-0.4}$ & 14.26$\pm$0.83 &2.0$^{+0.09}_{-0.09}$ & 12.87$\pm$0.22 & 4.1$^{+0.4}_{-0.4}$ & 1.56      & 1.23      \\
\hline
\end{tabular}%
\caption{Averaged magnitude and flux density values of the dusty sources of the S-cluster using H+K continuum data. Since we are not applying any filter to the data, the detection of some of the dusty sources is hindered due to the imprint of the PSF of bright nearby stellar sources. The asterisk-marked values are adopted magnitudes as well as H-K and K-L colors for G2/DSO, D3, and D5, which were taken from \cite{Eckart2013} and \cite{Peissker2020b, peissker2021c}. The SED for G1 will be analyzed in an upcoming publication (Melamed et al., in prep.).}
\label{tab:mag_sll_sources}
\end{table*}
The flux density values in Table \ref{tab:mag_sll_sources} are calculated using the calibration sources S2 and IRS2L (Table \ref{tab:mag_fux_reference_values}). It is reassuring that we find matching flux density values using both calibrators, demonstrating the robustness of the analysis. Minor variations between both calibrators are used to construct the indicated uncertainties of the related flux density values.
With the colors listed in Table \ref{tab:mag_sll_sources}, we construct a color-color diagram and adopt the analysis results of the dusty sources located in the IRS13 cluster from \cite{peissker2023c}. In Fig. \ref{fig:color_color_diagram}, we show that most of the investigated dusty sources exhibit an $H-K$ color of $<\,3$ and can be clearly distinguished from the high-mass YSO X3 \citep{peissker2023b}. 

\subsection{Spectral Energy Distribution}

\begin{figure*}[htbp!]
	\centering
	\includegraphics[width=1.\textwidth]{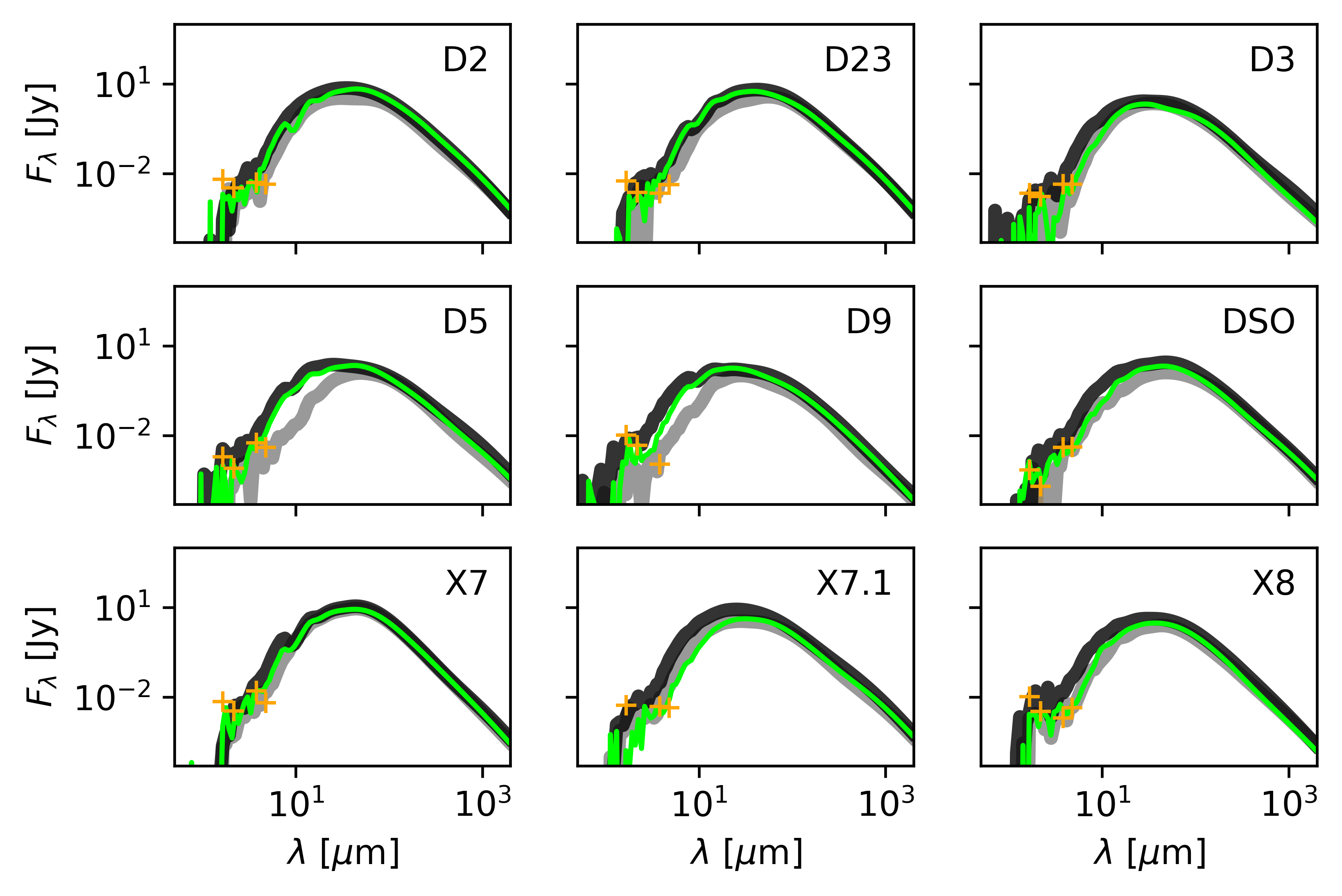}
	\caption{Best fit SEDs for some of the dusty sources discussed in this work derived with \texttt{HYPERION}. Due to the low spectral coverage because of the bright radio/submm counterpart of Sgr~A*, the analysis is limited to the NIR and MIR domain. We motivate the analysis with the color-color diagram shown in Fig. \ref{fig:color_color_diagram} and find a satisfying representation of the flux density distribution using a Class I model. The only exception marks the data point distribution for D9 which is hardly represented by the presented fit. See text for further details.}
\label{fig:SED}
\end{figure*}

\begin{figure}[htbp!]
	\centering
	\includegraphics[width=.5\textwidth]{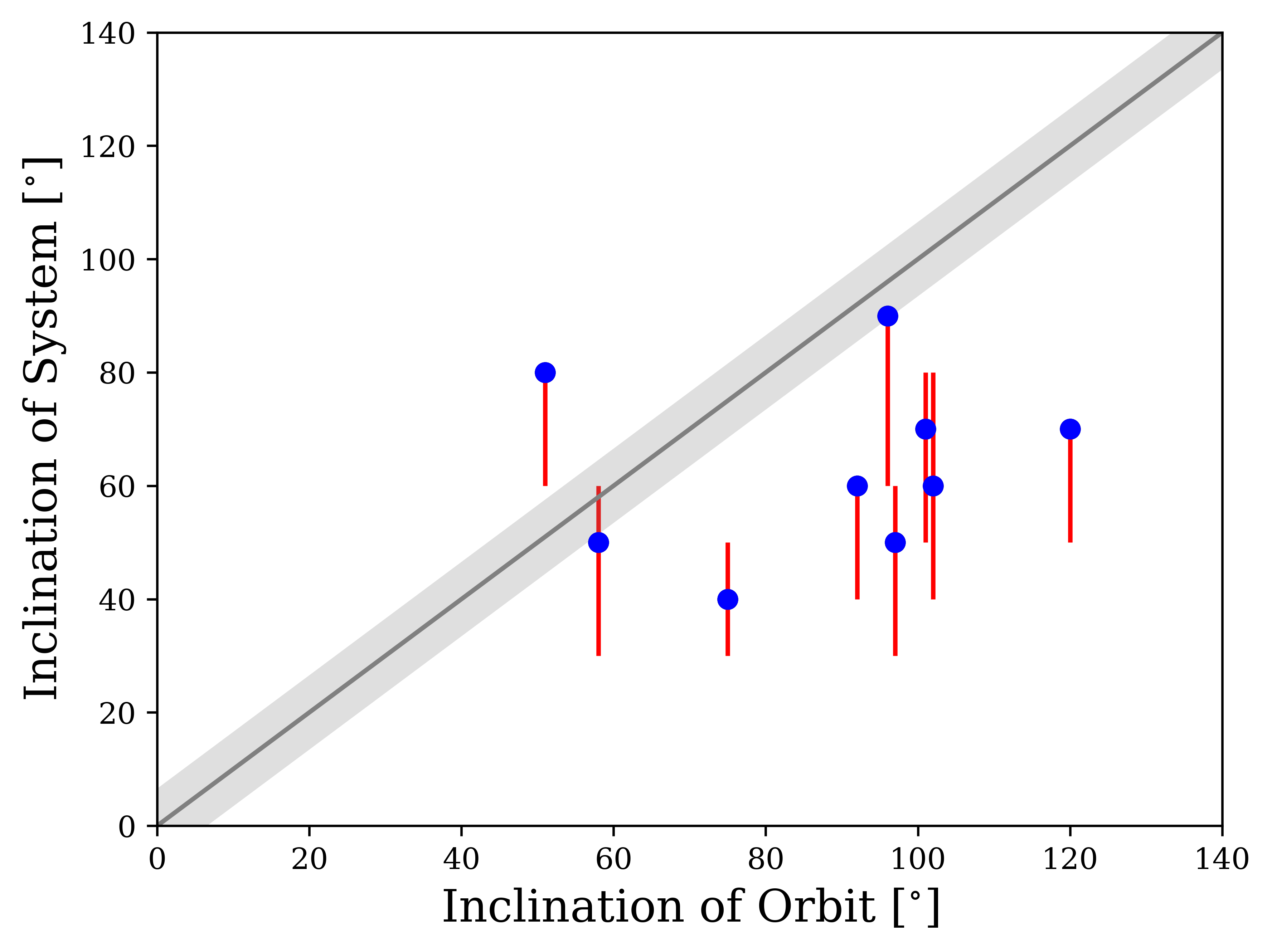}
	\caption{Correlation of the intrinsic inclination of the dusty sources listed in Table \ref{tab:sed_table} including its uncertainty range compared to the orbital parameter $i$. If the system is aligned with the orbit, we would expect a distribution close to the linear fit \citep{Monin2006}.}
\label{fig:inclination_correlation}
\end{figure}

To determine the properties of the dusty sources of the S-cluster, we inspect the color-color diagram presented in Fig. \ref{fig:color_color_diagram}. With the shown classification, we follow the suggested YSO nature by applying the radiative transfer code \texttt{HYPERION}. We use the flux density values listed in Table \ref{tab:mag_sll_sources} to define the input spectrum that is used to fit an analytical YSO model \citep{Robitaille2011, Robitaille2017} with individual parameters that describe the properties of the dusty sources. With these input quantities as a starting point, we conduct a comprehensive exploration to identify the best-fit parameters that represent the flux density values indicated in Table \ref{tab:mag_sll_sources}. The resulting best-fit solutions based on the radiative transfer model for the dusty sources are listed in Table \ref{tab:sed_table}.
\begin{table*}[tbh!]
\centering                                  
\begin{tabular}{|c|ccccc|}
\toprule
ID     & \multicolumn{1}{c}{Mass [$M_{\odot}$]} & \multicolumn{1}{c}{Luminosity [$10^3\times L_{\odot}$]} & \multicolumn{1}{c}{Radius [$R_{\odot}$]} & \multicolumn{1}{c}{Disk size [AU]} & Envelope size [AU] \\
\hline
D2        &   0.5$\pm$0.1  &  2000$^{+100}_{-200}$  &   2.09$^{+0.11}_{-0.19}$  & (6-125)$\pm$5 & (7-300)$\pm$20 \\
D23       &   0.5$\pm$0.1  &  2000$^{+200}_{-200}$  &   2.09$^{+0.10}_{-0.10}$  & (6-125)$\pm$5 & (7-300)$\pm$20 \\
D3        &   0.3$\pm$0.1  &  1500$^{+100}_{-100}$  &   2.09$^{+0.11}_{-0.09}$  & (4-75)$\pm$5  & (7-150)$\pm$10 \\
D5        &   0.3$\pm$0.1  &  1500$^{+100}_{-100}$  &   0.80$^{+0.10}_{-0.10}$  & (4-75)$\pm$5  & (7-150)$\pm$10 \\
G2/DSO    &   0.4$\pm$0.2  &  1300$^{+200}_{-200}$  &   0.30$^{+0.20}_{-0.20}$  & (4-70)$\pm$5  & (7-180)$\pm$20 \\
X7        &   2.0$\pm$0.5  &  2600$^{+200}_{-200}$  &   2.00$^{+0.50}_{-0.50}$  & (4-75)$\pm$5  & (7-400)$\pm$50 \\
X7.1      &   1.0$\pm$0.1  &  2000$^{+300}_{-300}$  &   1.00$^{+0.20}_{-0.20}$  & (4-100)$\pm$10& (7-200)$\pm$10 \\
X8        &   1.0$\pm$0.2  &  3000$^{+400}_{-400}$  &   1.00$^{+0.20}_{-0.20}$  & (4-80)$\pm$5  & (7-200)$\pm$20 \\
\hline
\end{tabular}
\caption{Best-fit properties of the candidate YSOs investigated in this work using \texttt{HYPERION} to reproduce their related SED. For these fit parameters, we derive the indicated uncertainty ranges, which are represented by three different SEDs displayed in Fig. \ref{fig:SED}. Due to the challenging observation of the OS sources close to the detection limit, we exclude them from this analysis. As mentioned, G1 will be analyzed in detail in Melamed et al. (in prep.). For all remaining sources, the radiative transfer code suggests a low-mass classification that was discussed in \cite{Ciurlo2020}.}
\label{tab:sed_table}
\end{table*}
As indicated, we confirm the presence of a stellar low-mass population in the S-cluster in agreement with \cite{Zajacek2017} and \cite{Ciurlo2020}. The related SEDs of these fits are shown in Fig. \ref{fig:SED} that are at the same time a verification of the significance of the method. We note that all the investigated sources are satisfyingly represented by a Class I YSO model except for D9. Since D9 exhibits a decreased MIR emission compared to all other sources in our sample, the approach used to find an SED with \texttt{HYPERION} that resembles the given flux density distribution seems to be challenging. However, the nature of D9 will be analyzed in detail in a forthcoming publication.   
Nevertheless, for the individual SEDs presented in Fig. \ref{fig:SED}, we use the best-fit and upper/lower values listed in Table \ref{tab:sed_table}. For every single SED, we select the best-fit inclination of the system. In Table~\ref{tab:inclination_table}, we list all the used inclinations for the related objects investigated in Fig.~\ref{fig:SED}.
\begin{table}[tbh!]
\centering                                  
\begin{tabular}{|c|cccc|}
\toprule
ID     & \multicolumn{1}{c}{Best-fit [$^{\circ}$]} & \multicolumn{1}{c}{Lower [$^{\circ}$]} & \multicolumn{1}{c}{Upper [$^{\circ}$]} & {Orbit  [$^{\circ}$]}  \\
\hline
D2        & 50 & 60 & 40 & 58 \\
D23       & 50 & 60 & 40 & 97 \\
D3        & 80 & 80 & 60 & 51 \\
D5        & 70 & 80 & 50 & 101\\
D9        & 60 & 80 & 40 & 102\\
G2/DSO    & 70 & 70 & 50 & 120\\
X7        & 40 & 50 & 30 & 75 \\
X7.1      & 60 & 60 & 40 & 92 \\
X8        & 90 & 90 & 60 & 96 \\
\hline
\end{tabular}
\caption{Intrinsic inclination values for the SEDs displayed in Fig. \ref{fig:SED}. We list the best-fit inclination and compare it with the lower and higher uncertainty range inclination values related to Table \ref{tab:sed_table}. In addition, we list the inclination of the orbit to explore the relationship between the Keplerian orbital orientation and the intrinsic arrangement of the associated system.}
\label{tab:inclination_table}
\end{table}
For visualization purposes, we show the relation listed in Table~\ref{tab:inclination_table} in Fig. \ref{fig:inclination_correlation}.
In this figure, we include the upper and lower limits for the system-related inclination and compare it with the orbital inclination $i$ adapted from Table \ref{tab:orbital_elements}. Furthermore, we include a linear fit that represents a state, where the system is aligned with the orbital plane. We will elaborate on these findings in Sec.~\ref{sec:discuss}.

\section{Discussion} 
\label{sec:discuss}
In this section, we will discuss our results and {analyze} the NIR and MIR detection of some of the sources. We will also provide a comprehensive perspective regarding the nature of the dusty sources.

\subsection{The nature of the dusty sources}

In the following, we will discuss each of the dusty sources investigated in this work individually and incorporate findings from the literature whenever possible. Furthermore, we will discard the {\it en vogue} idea to classify objects as coreless clouds in the high energetic radiation field of the supermassive black hole Sgr~A*. Due to the large amount of observations in various bands targeting the S-cluster over the last two decades, we will use G2/DSO as a case study to take a deeper look at the morphology of the dusty sources \citep{Gillessen2012, Valencia-S.2015, Shahzamanian2016, peissker2021c}. We will follow the argumentation of \cite{Scoville2013}, \cite{Eckart2013}, and \cite{Zajacek2017}, who classify G2/DSO as a young low-mass T-Tauri candidate. Furthermore, we strengthen the arguments for a stellar classification of the dusty sources by presenting a color-color diagram in Fig. \ref{fig:color_color_diagram} which covers stars from the entire inner parsec. 

\subsubsection{Electron temperature and gas temperature}
\label{sec:electron_temperature}

As mentioned by \cite{Gillessen2012}, the Br$\gamma$ electron temperature that is related to G2/DSO should be a few thousand Kelvin. In agreement with the NIR and MIR observation of the compact Br$\gamma$ and dust emission of G2/DSO, it is implied that the gas and dust are coupled. Here, the electron temperatures in the optically thin X-ray radiation-dominated environment of Sgr~A* are in the range of 5000-13000K \citep{Czerny2023}. As shown by \cite{Schutze1998} and \cite{Noor2015}, the electron and gas temperature is a function of pressure. While the electron and gas temperatures are identical in hot thermal plasma, this relationship does not exist in low-pressure cold plasma. In the case study of G2/DSO, the source is affected by the radiation field of Sgr~A*, which allows us to assume a gas temperature of about 9000 K \citep{gravitycollaboration2023, peissker2023b}. Moreover, the dust of the source is embedded in the gas of the proposed cloud, where it is expected that shocks and claimed drag forces \citep{Ballone2013, Gillessen2019} increase the gas temperature. Furthermore, it is known that dust located in hot gas suffers from evaporation, resulting in very low lifetimes for the grains. For a quantitative estimate of the usual lifetimes of grains, we follow the work of \cite{Draine1979} and \cite{Drozdov2021}. We adopt
\begin{equation}
    \eta_{S}\,\approx\,\rm 10^5yr(1+T_{Gas}^{-3})a_0 n^{-1},
    \label{eq:dust_timescale}
\end{equation}
that defines the characteristic evaporation timescale for dust located in hot gas \citep{Draine2011b} where $\rm T_{Gas}$ is associated with the gas temperature, $a_0$ the dust grain size, and $n_p$ the proton number density. Although there might be differences between $n_p$ and the electron number density $n_e$, we will use average numerical values, which include these variations \citep{Rowan2019}. We set $\rm n_p\,\approx\,n_e\,\approx\,10 \times 10^4\,cm^{-3}$, $a_0\,=\,0.1\,\mu m$, and $\rm T_{Gas}\,=\,10000\,K$ \citep{Drozdov2021, Czerny2023} and get $\eta_{S}=10\,yr$ in numerical agreement with \cite{Burkert2012} who estimated the gas evaporation timescales. This number becomes significantly lower and decreases to $\eta_{S}=2\,yr$ if we use $a_0\,=\,0.02\,\mu m$ as suggested by \cite{Gillessen2012}. The evaporation timescales here should be considered as an upper limit due to the slightly different approach of \cite{Draine2011b} and \cite{Drozdov2021}. Using the definition of Draine et al., given as
\begin{equation}
    \rm \eta_{S}\,=\,5.1\times 10^4 yr \left(\frac{n}{10^{5}cm^{-3}}\right)\left(\frac{R_{G2/DSO}}{0.0012 pc}\right)^2\left(\frac{T}{10^4 K}\right)^{-2.5},
    \label{eq:dust_timescale_draine}
\end{equation}
where $\rm \eta_{S} \approx 23\, s$. For the above equation, we use a cloud radius comparable to half of the SINFONI PSF, i.e., 60/2 mas. The different results in Eq. \ref{eq:dust_timescale} and Eq. \ref{eq:dust_timescale_draine} originate in the different evaporation mechanisms. While the cloud evaporates due to particle collision using the approach from \cite{Drozdov2021}, the object gets destroyed because of radiation transport between the hot (gas) and cold (dust) medium with the Ansatz suggested by \cite{Draine2011b}. {Assuming} a cloud radius of 100 times larger, i.e. R$\rm _{G2/DSO}\,=\,0.12$pc, Eq. \ref{eq:dust_timescale_draine} yields $\rm \eta_{S} \approx 2.7$days. {Although we used a Br$\gamma$ gas temperature of T$_{Gas}\,=\,10000\,K$ with Eq. \ref{eq:dust_timescale_draine}, we will explore the timescales for a dust temperature of 650 K \citep{Eckart2013}. Using Eq. \ref{eq:dust_timescale_draine}, we get $\approx$ 7 years under the assumption that no radiation transport in the ionized environment of Sgr~A* takes place or inefficient dust-gas coupling.}\newline
Regardless of the exact grain size or evaporation mechanism, we emphasize the magnitude of the timescale of this approach. If the components of the dusty sources were limited to ionized gas and dust, they would evaporate on observable timescales in the vicinity of Sgr~A*, which are covered by the SINFONI data set investigated here.

\subsubsection{Colliding winds}
\label{sec:colliding_winds}

\begin{figure}[htbp!]
	\centering
	\includegraphics[width=.5\textwidth]{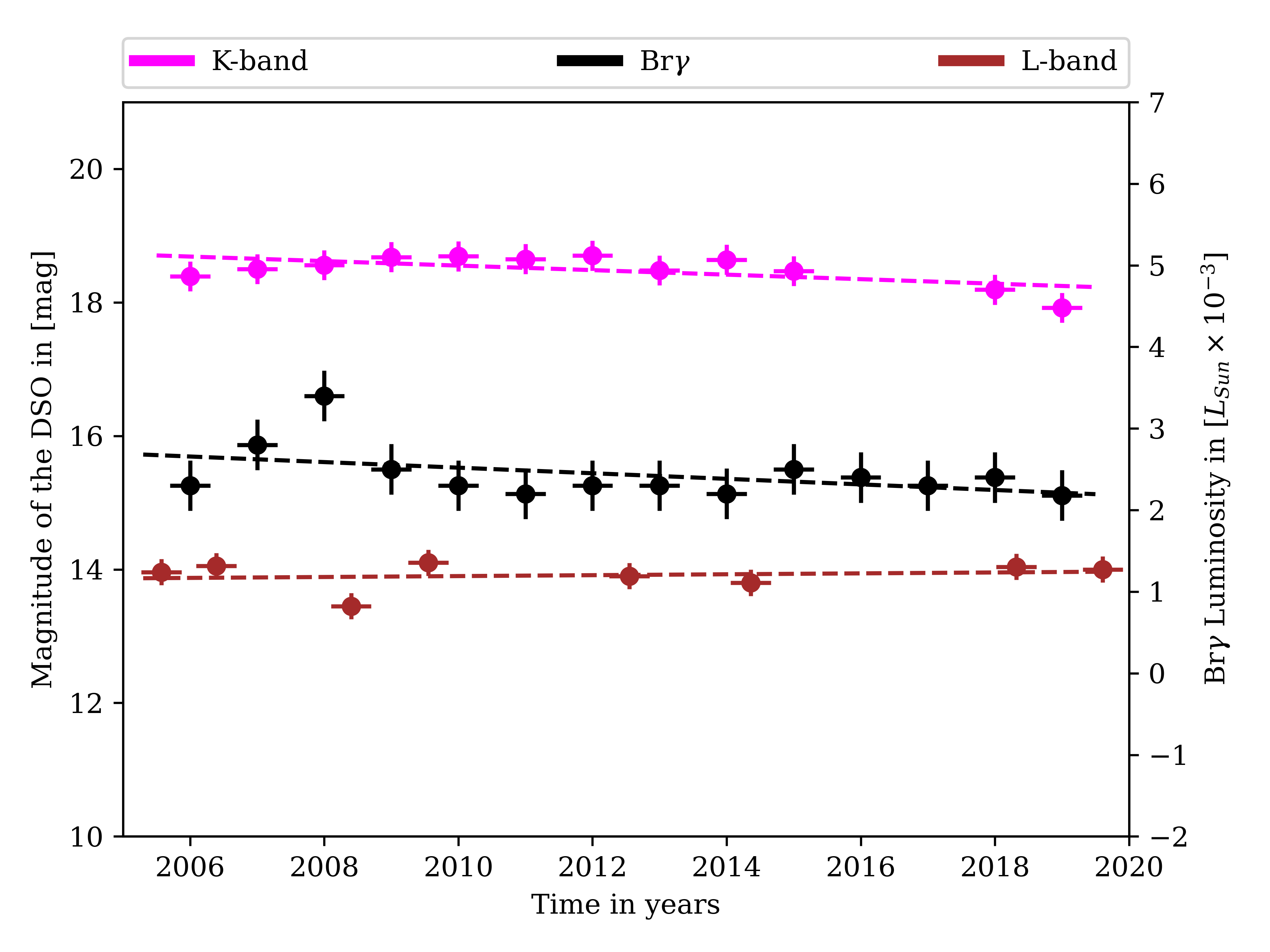}
	\caption{Magnitude of G2/DSO in different wavelengths. The magenta data points are adapted from \cite{peissker2021c}, whereas the black and brown data points are incorporated from \cite{Pfuhl2015} and \cite{Witzel2014}, respectively. Further, L-band data points from 2008 (NACO) and 2019 (NIRCAM2) are included. The Br$\gamma$ line luminosity after 2014 is based on the analysis presented in \cite{peissker2021c}. The dashed line for every data set represents a 1st-degree polynomial fit and underlines the almost constant magnitude/luminosity in strong contradiction to the expected variation discussed in \cite{Gillessen2012}, \cite{Ballone2013}, and \cite{Pfuhl2015}. Here, all the uncertainties are estimated from the standard deviation of the related data set.}
\label{fig:dso_kl_mag}
\end{figure}

In the former section, we have shown that a cold cloud with $T_{cloud}\ll T_{gas}$ immediately evaporates with the input parameter listed in \cite{Gillessen2012} using the relations discussed in \cite{Draine2011b}. We will now take a closer look at the origin of our case study G2/DSO, ignoring the evaporation timescales.\newline
As a possible formation-, and more importantly, radiation-process for G2/DSO, \cite{Gillessen2012}, \cite{Schartmann2012}, and \cite{Burkert2012} suggested the ultraviolet (UV) field of nearby massive stars. Excited filaments and structures have been found in numerous stellar wind-dominated habitats \citep{Koo2007, Artigau2011, Sundqvist2012, Herrera2017, Ramiaramanantsoa2018}, which justifies the assumed classification as a gas cloud from this particular point of view. Consequently, the origin of the UV radiation is related to the presence of stellar winds \citep{Prinja1989, Prinja1990}, which, in turn, are a reliable tool for determining several parameters of the star itself \citep{Kudritzki1995, Kudritzki2000, Vink2000}.\newline 
For this discussion, it is required that the stellar wind velocities do not exceed a few hundred km/s as shown by \cite{Cuadra2005, Cuadra2006}. It is important to emphasize that only slow stellar winds are considered to be the origin of clump formation because low temperatures are essential for forming clumps. This relation is expressed in
\begin{equation}
    t_{cool}\,\propto\,T^2/\Lambda(T),
    \label{eq:cooling_time}
\end{equation}
where $\Lambda(T)$ defines the optically thin cooling function and $T\,\propto\,v_{wind}$, i.e. clumps are formed after $t_{cool}$. This profound relation manifested in Eq. \ref{eq:cooling_time} directly excludes fast stellar winds of several 1000 km/s that could be accounted as a possible clump-formation origin, for example, produced by Wolf-Rayet stars \citep{Wang2020, Zhu2020}. This framework condition differs from the hydrodynamical simulations presented in \cite{Calderon2016, Calderon2020_mnras} who considered clump formation processes driven by colliding winds of Wolf-Rayet stars. Although the authors find comparable results as presented here, they conclude that migration scenarios of a clump with a mass of approximately $3 \times M_{\otimes}$ are highly unlikely due to evaporation timescales. Hence, we assume for this work, that G2/DSO did not migrate to the S-cluster but was formed by the S-stars.
This assumption is motivated by the calculations presented in the former section (Sec. \ref{sec:electron_temperature}), where we concluded that the required low stellar wind velocities are indeed essential because of the evaporation timescales of the dust.\newline
An established tool by \cite{Kudritzki1995}, \cite{Kudritzki2000}, and \cite{Vink2000} is the so-called wind-momentum luminosity relation (WLR), defined as
\begin{equation}
    \rm \log(\dot{M}v_{\infty} R^{1/2}_{\star})\approx \left(\frac{1}{\alpha}\right)\log(L)+const(z, sp.type),
    \label{eq:wlr}
\end{equation}
where $\dot{M}$ describes the mass loss rate, $z$ the metallicity, and sp.type refers to the spectral type. Typically, $\alpha$ is set to $\approx\frac{2}{3}$ \citep{Puls2000}. In the S-cluster, the brightest stellar K-band sources are found to be B0-B3 stars \citep{Habibi2017}. Given the derived luminosity range of the S-stars by Habibi et al., we can safely assume mass-loss rates of less than $10^{-8}\dot{M}/yr$ \citep{Nugis2000, Crowther2007}. Taking into account the analysis of \cite{Crowther2006}, we find $\log(\dot{M}v_{\infty} R^{1/2}_{\star})\,\approx\,26\pm0.5$ for the S-stars with the luminosities given in \cite{Habibi2017}. In contrast, we can use the mass-loss rate for young OB stars, i.e., $10^{-8}\dot{M}/yr$ and adapt an average stellar radius from Habibi et al. of $5R_{\odot}$ to calculate the WLR given in Eq. \ref{eq:wlr}. In agreement with the required low wind velocities by \cite{Cuadra2005, Cuadra2006}, we find rather low numerical values of a few 100 km/s for the OB stars of the S-cluster.\newline
These wind velocities are followed by the expected X-ray luminosity from the stellar winds, which can be calculated with
\begin{equation}
    \rm L_{x}\,\propto\,\frac{\dot{M}^2}{d},
    \label{eq:xray-luminosity}
\end{equation}
where d defines the distance of the interacting stars \citep{Wang2020}. This strong squared distance correlation of the observed luminosity is manifested in an expected L-band flux fluctuation \citep{Ballone2013}. For the compact dust blob, which is associated with G2/DSO, we expect with
\begin{equation}
    \rm T_{dust}\,\propto\,\frac{L_{\star}}{d^2},
\end{equation}
a comparable luminosity trend as stated in Eq. \ref{eq:xray-luminosity}. It is obvious that the temperature and the luminosity are a function of distance, which is even more true for the stellar wind-induced temperature $T_{W}$ \citep[see][]{Gayley1997, Owoki2002}. It is reasonable that we are allowed to define $T_{(d)}$ and $L_{(d)}$ in a broader context, that is, the distance relationship of the luminosity and temperature is not limited to a specific band. Lastly, we adopt the emissivity of the Case B Br$\gamma$ recombination line from \cite{Ballone2013}, which is given by
\begin{equation}
   \rm j_{Br\gamma}\,=\,3.44\times 10^{-27}(T_{(Br\gamma)}/10^4 K)^{-1.09} n_p n_e erg s^{-1} cm^{-3},
\end{equation}
where $T_{(Br\gamma)}$ is equal to the temperature of the ionized hydrogen gas. From \cite{Pfuhl2015}, we adapt a constant Br$\gamma$ luminosity of $\sim\,(2.5\pm 0.5)\times 10^{-3}L_{\odot}\,\approx\,10\times 10^{30} ergs^{-1}$ although \cite{Ballone2013} predicts an increase for G2/DSO by one magnitude close to Sgr~A*. In summary, we expect
\begin{enumerate}
    \item I$_{\rm dust}\,\simeq\,1/d^2$,
    \item I$_{\rm Xray}\,\simeq\,1/d$,
    \item I$_{\rm Br\gamma}\,\simeq\,d_{SgrA*}$.
\end{enumerate}
where I defines the observed flux density. For 1) and 2) in the above list, the emission decreases with the distance to the S-stars that are assumed as the origin of the emission of G2/DSO\footnote{We note that I$_{\rm Xray}$ becomes $\simeq\,d$ if the origin of the emission is not linked to the S-stars in contradiction to the statements in \cite{Gillessen2012}.}. In contrast, 3) increases towards Sgr~A* because \cite{Ballone2013} considered the X-ray radiation of Sgr~A* to be solely responsible for the Br$\gamma$ emission. While this scenario would require a dense and spherical distribution of ionized hydrogen around Sgr~A*, no such emission has been found at the position of the black hole \citep{Peissker2020c, Ciurlo2021}. Considering these arguments, a classification of G2/DSO as a gas and dust cloud contradicts any expected physical evolution of such a system close to a supermassive black hole. So far, there is a poor physical understanding of G2/DSO as a dust and gas cloud due to the lack of a comprehensive model that takes into account the challenging cornerstones listed above. Therefore, it is plausible to consider an intrinsic origin of the emission, which results in the observed constant magnitudes (Fig. \ref{fig:dso_kl_mag}), which emerges from a faint embedded source in the G2/DSO system. Consequently, an almost constant IR and line magnitude can only be realized by a stellar core associated with the embedded source inside the G2/DSO system, which excludes the gas cloud model.
In addition to these arguments, there is a general consensus that the dust temperature measures up to several hundred K inferred from a constant L-band magnitude \citep{Eckart2013, Witzel2014, Pfuhl2015} further underlining the stellar nature of G2/DSO and this discussion. With the classification of G2/DSO as a gas cloud with the mass of $3 \times M_{\otimes}$, it is expected to observe at least a marginal deposit of material to Sgr~A* during the periapse at about 1855 Schwarzschild radii \citep{peissker2021c, peissker2023b}. The observed flare in 2019 reported by \cite{Do2019} is not related to the periapsis of G2/DSO due to the cooling and evaporation timescales \citep[see Eq. \ref{eq:cooling_time} and][]{Ballone2013}. We conclude, that clumps, independent of their formation habitat, evaporate on accessible timescales in contradiction to the observations of G2/DSO \cite{Calderon2016, Calderon2020_mnras}.  

\subsubsection{Temperature response of coreless clouds}

Until now, it has been proposed that the nature of two objects, namely G2/DSO and X7, might be described by coreless clouds. We have shown in the preceding section that this classification is challenging considering the cooling timescales related to the stellar wind velocities. Magnitude variations are expected to occur in stark contrast to the constant magnitude/luminosity of G2/DSO displayed in Fig. \ref{fig:dso_kl_mag}. Although X7 does exhibit to some degree a variable L-band magnitude \citep{peissker2021}, the uncertainty range is well covered by the standard deviation with mag$\rm _{\rm X7, L-band}$=10.91$\pm$0.08 which is in the same uncertainty range as the  G2/DSO magnitude (mag$\rm _{\rm G2/DSO, L-band}$=13.91$\pm$0.19).
Since the magnitude is a temperature dependent quantity, we want to explore the expected behavior using our G2/DSO case study. Regardless of the specific band, a putative gas cloud depends on external excitation sources due to its nature. We can distinguish between possible source origins that are responsible for the magnitude related to G2/DSO:
\begin{enumerate}
    \item Stellar winds
    \item Sgr~A*
\end{enumerate}
Regarding 1), we remind the reader that slow stellar winds are required for the creation of G2/DSO in the first place due to cooling/evaporation timescales (please consider Sect. \ref{sec:colliding_winds}). These winds cannot produce the observed magnitude of G2/DSO, which is underlined by the work of \cite{Marle2009}. The authors investigate the possible luminosity input of WR stars with wind velocities of about 3000 km/s which exceeds the required numerical values for the creation of G2/DSO by a factor of ten. Even if wind velocities of about 3000 km/s are allowed in contrast to the work of \cite{Cuadra2005, Cuadra2006}, the luminosity impact of G2/DSO would be neglectable small \citep{Marle2009}. Therefore, we will focus on 2) where Sgr~A* is considered to be the origin of the emission that is responsible for the G2/DSO magnitude. In this scenario, Sgr~A* should be the dominant emission source for G2/DSO for a significant part of its orbit. We investigate the impact by adopting
\begin{equation}
    T(r)\,=\,1.2\times 10^8 K\left(\frac{1.4\times10^4R_S}{r}\right)^{\beta}
    \label{eq:temp_dso_sgrA}
\end{equation}
for the temperature of the surrounding ambient hot flow, see, e.g., \cite{Schartmann2012} and \cite{Ballone2013}, where $\beta$ is set to unity. In Eq. \ref{eq:temp_dso_sgrA}, $r$ defines the distance of G2/DSO from Sgr~A*, which is adapted from \cite{peissker2021c, peissker2023b}. In Eq. \ref{eq:temp_dso_sgrA}, $T(r)$ represents the temperature profile of an advection-dominated accretion flow (ADAF) to describe the atmosphere of Sgr~A*. The authors of Ballone et al. introduced a transformation factor\footnote{\cite{Ballone2013} used the term {\it passive tracer} to artificially stabilize the atmosphere of Sgr~A*.} to link the gas temperature of the ambient medium to G2/DSO. Since we are following a qualitative strain of arguments, a factor to explain the connection between the ambient medium and the gas temperature of a putative clump does not impact the global expectation. In other words, the closer G2/DSO gets to Sgr~A*, the hotter it should get, as shown and discussed in detail by \cite{Ballone2013} using the same theoretical framework.
\begin{figure}[htbp!]
	\centering
	\includegraphics[width=.5\textwidth]{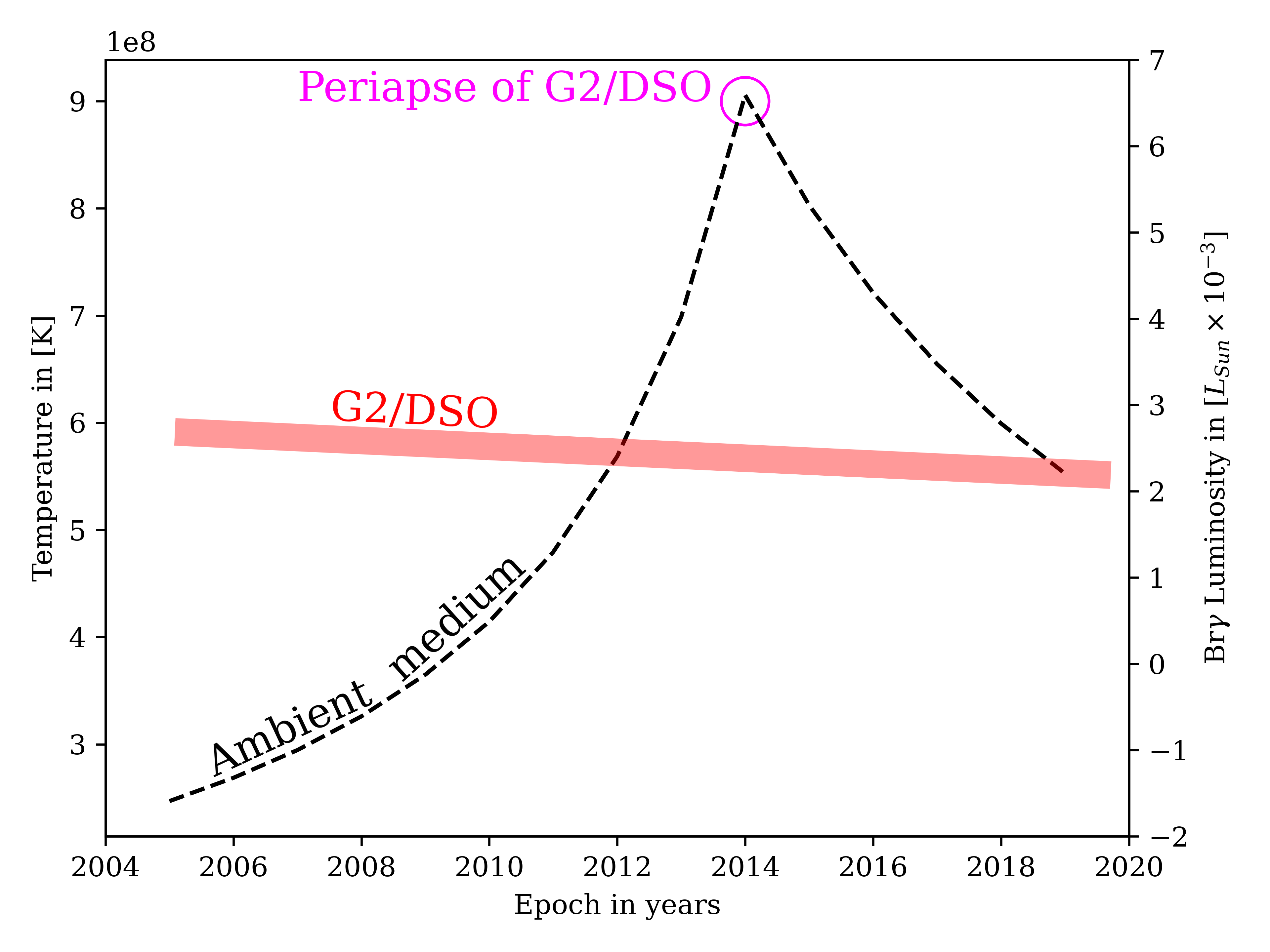}
	\caption{Comparison of the ambient medium temperature profile of Sgr~A* and the observed Br$\gamma$ luminosity of G2/DSO. The red line represents the fit of the Br$\gamma$ luminosity of G2/DSO displayed in Fig. \ref{fig:dso_kl_mag} whereas the black dashed line represents the expected temperature imprint of Sgr~A* (Eq. \ref{eq:temp_dso_sgrA}). In 2014, the distance of G2/DSO to Sgr~A* is about 1855 Schwarzschild radii. We strongly emphasize the irrelevance of the temperature magnitude (10$^8$K of the ambient medium against 10$^4$K of the Case B recombination) due to the missing transformation factor: the important quantity is the increasing and decreasing profile that should be imprinted in the luminosity of G2/DSO.}
\label{fig:expected_dso_mag}
\end{figure}
Combining the temperature of the ambient gas with the observed Br$\gamma$ luminosity of G2/DSO, we show a clear distance relation between G2/DSO and Sgr~A* as displayed in Fig. \ref{fig:expected_dso_mag} by the black dashed line. In strong contrast to the constant photometric behavior of G2/DSO displayed in Fig. \ref{fig:dso_kl_mag}, we find that the ambient medium and, consequently, Sgr~A is responsible for the temperature evolution of the object.\newline
Considering the data baseline covering 15 years, we find no indications that the nature of G2/DSO is consistent with a coreless gas cloud. Taking into account the multi-wavelength character of the dusty source in combination with the Br$\gamma$ line, which is associated with accretion processes \citep{Kraus2009, Valencia-S.2015, Zajacek2017}, the nature is sufficiently reflected by a proplyd-like classification as found in the Orion nebula \citep{Mann2009a, Mann2009b} or the Galactic center \citep{Zajacek2016, Yusef-Zadeh2015}.

\subsubsection{Compact Planetary Nebula}

\cite{Gillessen2012} argued that G2/DSO could be identified with a Compact Planetary Nebula (CPN) but rejected this idea due to the possibly destructive interaction with the dominant gravitational field of Sgr~A*. Consequently, the same argument should be applied to a coreless cloud of $\sim\,3\,M_{\oplus}$ in the harsh vicinity of an SMBH. 
Since we have presented compelling evidence and arguments regarding the nature of the dusty source, we are pursuing the exciting idea of the presence of a CPN. These objects show sub-structures \citep{Fang2018} that can be classified as the stellar source, jet-like outflows, bow-shocks, and dusty/gaseous envelopes. Most of the CPNs do exhibit a stellar mass of less than 3 M$_{\odot}$ and a related age between 100 Myrs and a few Gyrs after they leave the asymptotic giant branch (AGB) phase \citep{Salaris2014, Moreno2016}. For a short time of about 10\,000 years, CPNs exhibit a noticeable envelope that would be associated with the Br$\gamma$ emission in the case of G2/DSO. This envelope would explain the constant luminosity of the Br$\gamma$ emission (Fig. \ref{fig:dso_kl_mag}), but given the number of dusty sources with Doppler-shifted Br$\gamma$ emission in the S-cluster, it is highly unlikely to observe all these objects at this very specific evolutionary phase at once. Using the envelope lifetime $t_{\rm env}$ of approximately 10000 years and the age of CPNs of 100 Myrs (lower limit), we get a relative probability of 0.01 $\%$ to observe G2/DSO in the CPN stage. Since we find at least 12 Doppler-shifted Br$\gamma$ sources that are photometrically comparable to G2/DSO (Fig. \ref{fig:color_color_diagram}), the likelihood that all of them represent a CPN stage reduces to $10^{-46}\%$. Therefore, we rule out the possibility that the nature of the sample of sources examined can be associated with CPNs. In addition, \cite{Zajacek2020} suggested an explanation for the absence of red giants in the Galactic center. The authors argued that the outflow from the direction of Sgr~A* would have detached loose envelopes of AGB stars, which could explain the missing large red giants and AGB stars close to Sgr~A*. Taking this mechanism into account, the envelope of a CPN should have detached during periapsis due to the strong ram pressure of the ambient medium as well as the ADAF outflow. Hence, we would have expected a strong decrease of the Br$\gamma$ luminosity in contrast to the results presented in Fig. \ref{fig:dso_kl_mag}. 

\subsubsection{Binaries in the S-cluster}
\label{sec:be_star_d9}

Recently, \cite{Chu2023} presented evidence of a decreased binary fraction in the central 20 mpc around Sgr~A*. This is an important finding due to the fact that the evolution of (massive) stars is altered by their binary companions \citep{Sana2012}. Furthermore, it is known that most stars do form in multicomponent systems, as recently reviewed by \cite{Offner2023}. For over 60$\%$ of all early B-stars \citep{Abt1990,Duchene2013,Habibi2017}, it is suggested that these primaries are accompanied by a binary star which directly promotes the question, why most of the stellar content in the inner parsec appears to be arranged in a single component setup. Due to the missing binaries in the S-cluster, it is emphasized to consider alternative explanations such as the Hills mechanism proposed by \cite{Hills1988} or the eccentric Kozai-Lidov (KL) mechanism \citep{stephan2016}. The authors of Stephan et al. propose that the binary fraction for the S-stars decreases after $\approx$7 Myrs in line with the average stellar age of the brightest cluster members \citep{Habibi2017}. Although the KL mechanism cannot be ignored for the description of the multiplicity evolution of the S-stars, the decreased binary fraction in the S-cluster \citep{Chu2023} suggests that, at least for a fraction of cluster members, migration events took place for stars in the vicinity of Sgr~A*. This implication arises from basic star formation processes that demand a sufficiently low temperature and the absence of turbulence \citep{Jeans1902}. Therefore, other mechanisms should be explored, such as the mentioned Hills mechanism.
The author of \cite{Hills1988} proposes that a central potential, such as Sgr~A*, efficiently separates binary systems during their periapse. We already calculated the size of a possible stable binary system that orbits Sgr~A* to be of the order of a few AU \citep{peissker2021c}. Hence, only close and compact contact binary systems are able to resist the destruction by Sgr~A*. Although there is no doubt about the stellar nature of the dusty sources, some unresolved aspects might require a more comprehensive analysis. If the nature of the D-objects can be classified as binary mergers, as proposed by \cite{Ciurlo2020}, these objects impose a unique opportunity to study the history of the S-cluster from a more complex point of view. However, a detailed analysis of binarity tracers for the D-objects exceeds the scope of this work and should be explored separately. 

\subsubsection{Complementary aspects}

{The presence of this sub-population in the S-cluster increases the number of known cluster members by about 10$\%$ \citep{Ali2020}. Hence, a higher cluster member density results in an increased probability for violent events such as stellar encounters or disruption by Sgr~A*. However, \cite{Merrit2010} calculated the probability for such a disruption event to be in the order of $\approx 5\times10^{-4}\,yr^{-1}$. Given the age of the here discussed sub-population of a few $10^5$yrs, the main-sequence S-stars should be prone to stellar encounters or disruption events with a much higher probability. An exact quantitative statement remains challenging due to observational constraints. For example, the presence of fainter stars below the detection limit \citep{Sabha2012} may impact the collision probability. Since all known S-cluster members are on stable Keplerian orbits, the probability may be neglectable. Including the Kozai-Lidov mechanism \citep{stephan2016} but also Schwarzschild precession for stars with a short pericenter distance such as S2 \citep{Parsa2017, gravity2018, Do2019S2}, the orbits evolve over time which directly impacts the collision probability. An exact solution to this rather complex problem exceeds the scope of this work. However, to visualize one solution for the probability collision as a function of time, we model a cluster with 100 stars. In this model, we assume that close stellar encounters induce a random energy interaction that alters the orbit of both involved stars. We neglect Schwarzschild precession or the Kozai-Lidov mechanism.}
\begin{figure}[htbp!]
	\centering
	\includegraphics[width=.5\textwidth]{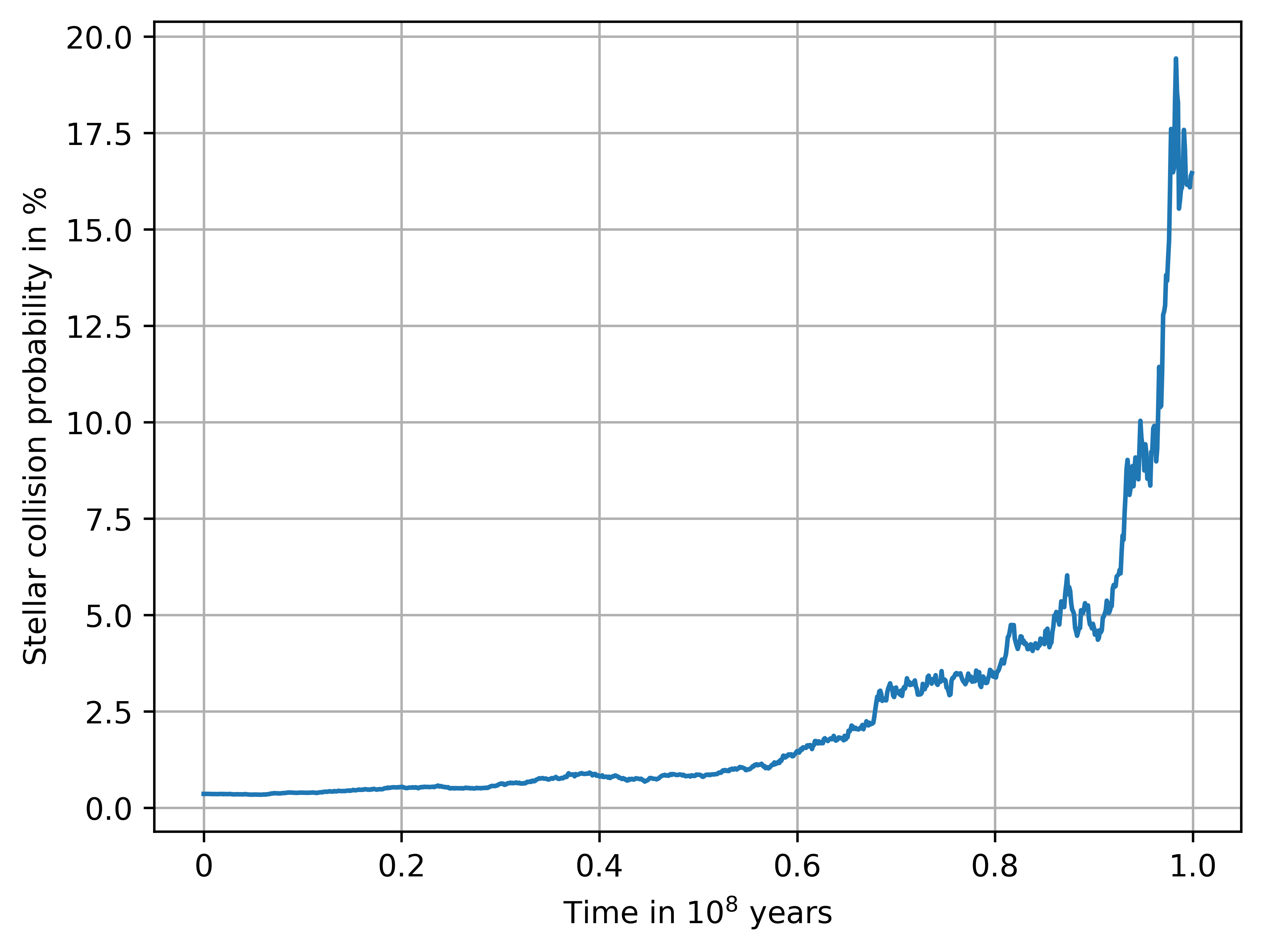}
	\caption{Monte Carlo simulation of the collision probability of a stellar cluster with 100 stars. To alter the orbits, we assume a random energy transfer for close stellar encounters. In this simplified simulation, we would expect a neglectable collision probability for the S-cluster with an age of almost $0.1\times 10^8$yrs.}
\label{fig:stellar_collison_probability}
\end{figure}
{In Fig. \ref{fig:stellar_collison_probability}, we show the results of the Monte Carlo simulations. We find that the stellar collision probability is comparable low for the first few million years. Given the age of the S-cluster of a few million years by considering the brightest S-stars \citep{Habibi2017}, we conclude that a collision probability of less than 1$\%$ is neglectable. Hence, we do not expect an increased collision probability for the dusty sources investigated in this work.}

\subsection{The individual dusty objects}

In the following, we will focus on the nature and properties of the individual dusty sources that are investigated in this work. Furthermore, we aim to include the results from the literature to provide a comprehensive picture of the origin and the connection of the dusty sources to the S-stars. {In addition, we discuss the limitations of the model used to fit the flux density distribution.}

\subsubsection{D2}

First discovered by \cite{Eckart2013} and confirmed by \cite{Meyer2014}, D2 is among the brightest L-band sources in the S-cluster. The source was later denoted as G3 by \cite{Ciurlo2020}. 
In addition to its NIR and MIR line flux well above the detection limit, it exhibits various emission lines \citep{Peissker2020b}. However, the strong Doppler-shifted Br$\gamma$ emission outshines additional lines such as [FeIII] and HeI by a factor of 2-3 as presented in \cite{Peissker2020b}. With the color-color analysis presented in Fig. \ref{fig:color_color_diagram} and the SED in Fig. \ref{fig:SED}, we conclude that D2 is a low-mass young stellar object with M$_{\rm D2}$=$0.5\pm 0.1 M_{\odot}$ that is embedded in a dusty envelope.

\subsubsection{D23}

So far, D23 is only reported in \cite{Peissker2020b}. It is named D23 due to its close distance to D2 and D3 in 2013 and the years thereafter. Likewise D2, it is a bright L-band source with NIR counterparts. Based on the photometric properties of D23, it shows similarities to D2. Taking into account the SED results listed in Table \ref{tab:sed_table}, we classify D23 as a low-mass candidate YSO. 

\subsubsection{D3}

Together with D2, D3 was already observed by \cite{Eckart2013}. It completes the trio of D2 and D23 since all three sources are located in projection nearby. While this can be explained by a chance association \citep{Martayan2016}, the photometric properties of all three sources are comparable (Table \ref{tab:mag_sll_sources}). Another similarity to D2 and D23 is the NIR detection of the source, which resulted, in combination with the MIR emission and its related colors, in a classification as a candidate YSO. As we have shown in Fig. \ref{fig:SED} and Table \ref{tab:sed_table}, the flux density of D3 follows the spectral energy distribution of a low-mass YSO.  

\subsubsection{D3.1}

We have included D3.1 only for consistency since it is most likely a large filament comparable to the Br$\gamma$ bar \citep{Schoedel2011, Peissker2020c, Ciurlo2021}. Nevertheless, the origin of this filament and its 3d distance towards Sgr~A* is unknown. Furthermore, a possible connection to the Br$\gamma$ bar cannot be excluded but raises, on the other hand, the question of the mechanism of its projected and 90$^{\circ}$ tilted location. We note that the alignment of this feature is reported for similar structures \citep{muzic2007}.

\subsubsection{D5}

The dusty source D5 is the only source that suffers from the limitations imposed by SINFONI. For the analysis, we have to rely on observations carried out in the epochs before 2008. Due to its location in the SINFONI data cubes at the edge of the FOV, we observe a hampered redshifted Br$\gamma$ emission with a LOS of $\sim$115 km/s \citep{Peissker2020b}. The source is prominent in the L-band and exhibits a proper motion that is directed towards the IRS 16 stars (Fig. \ref{fig:finding_chart}).

\subsubsection{D9}

In the sample of investigated cluster members, D9 stands out due to its low MIR emission. Compared to all other sources discussed in this work, we find a low L-band magnitude mag$_{\rm L-band}\approx$16 implying that the amount of emitting dust is limited. In part, this can be explained by the proximity of D9 to S2. In most of the epochs investigated in this work, the bright B2V star hinders a confusion-free detection of D9 which is why especially later epochs, i.e., 2017-2019, are relevant for the detection of the dusty source.
Considering a putative YSO association, this suggests that the object is close to the main-sequence phase (Fig. \ref{fig:color_color_diagram}). Due to its Dopplershifted Br$\gamma$ excess, it may be classified as a line emission star \citep{Kogure2007} or a Herbig Ae/Be star. A detailed analysis of the nature of D9 exceeds the scope of the work and will be covered in a forthcoming publication.

\subsubsection{G1}

G1 is historically significant in the analysis since it marks the beginning of the analysis of the dusty sources of the S-cluster population. Initially identified by \cite{Clenet2004a} and described in detail by \cite{Clenet2005a} and \cite{Clenet2005a}, G1 is also known as SgrA*-Flake and Extended Red. A long-term analysis was performed by \cite{Pfuhl2015} claiming that G1 is part of a gas streamer which will result in an in-spiral motion of the object towards Sgr~A*. Independently, this claim was not confirmed by \cite{Witzel2017} who further showed a bound Keplerian motion of G1. While a K-band counterpart is still under debate, we will focus on this particular source in Melamed et al. (in prep.). In the current work, we find in agreement with \cite{Witzel2017} a mean dereddened (L-M) color of 0.33 between 2006 and 2019. In contrast to Witzel et al., we do find a rather constant L-band magnitude using S2 and IRS2L as reference sources. For the moment, we do not have an explanation for this discrepancy which will be addressed in Melamed et al. (in prep.). One speculative explanation could be confused intensity measurements that arise from the lower resolution NACO observations compared to the Keck data. However, this contradicts the constant L-band magnitude listed in Appendix \ref{app_sec:photometry}.  

\subsubsection{G2/DSO}

Reported in \cite{Gillessen2012}, G2/DSO can be characterized by its prominent and strong Br$\gamma$ emission with a related average luminosity of $L_{G2/DSO}\,=\,(2.43\pm 0.30)\times 10^{-3}\,L_{\odot}$ where the uncertainty is based on the standard deviation of the data set observed between 2006 and 2019. This is about 60$\%$ higher than the luminosity of G1 estimated by \cite{Witzel2017} with $L_{G1}\,=\,(1.48\pm 0.17)\times 10^{-3}\,L_{\odot}$. The decreased Br$\gamma$ luminosity underlines the challenges related to the line map detection of G1 but emphasizes the chances of observing G2/DSO. Nevertheless, the multi-wavelength observations, the Keplerian orbit, and the theoretical framework discussed in this work strongly are in favor of a stellar classification for G2/DSO.
In agreement with the suggested classification of \cite{Ciurlo2020} for G2/DSO as a low-mass stellar source and the analysis of \cite{Zajacek2017}, we find a mass of $M_{G2/DSO}\,=\,0.4\pm 0.2\,M_{\odot}$. This estimate agrees with the analysis performed by \cite{Valencia-S.2015}, which also supports a low-mass stellar classification for G2/DSO \citep{Miralda2012,Murray-Clay2012,Scoville2013}. 

\subsubsection{X7}
\label{sec:x7_discuss}
Recently, \cite{Ciurlo2023} analyzed the bow-shock source X7 suggesting a non-stellar classification. As we have shown in \cite{peissker2021}, the S-cluster S50 exhibits an offset to the dust envelope, which becomes increasingly eminent after 2010. Due to the limitations of our data baseline, this offset between the tip of X7 and S50 seemed to have increased after 2018, which questions the connection between the S-star S50 and X7. In contrast, the distance between the tail and S50 seems rather constant. While the depletion of the envelope could be a reasonable explanation \citep{Zajacek2020}, we want to inspect the proposed nature of X7.\newline
Taking into account the classification for X7 of a coreless cloud, it is unexpected to observe a decreasing line-of-sight (LOS) velocity at the tip \citep[see Fig. 4 in][]{Ciurlo2023}. With a decreasing distance of X7 with respect to Sgr~A*, one would expect an increase of the tip LOS velocity for a coreless object as it was proposed for G2/DSO \citep{Gillessen2012}. This statement gets even stronger by considering the prominent and increasing elongation over two decades. If Sgr~A* is attracting a coreless cloud, we would expect an increasing LOS, especially at the tip because X7 is on the ascending part of the orbit. 
Furthermore, the authors of Ciurlo et al. demonstrate an L-band line profile that does not cover the complete bow-shock source. As shown in \cite{peissker2021} and \cite{Ciurlo2023}, the ionized hydrogen line intensity {exhibits} increased emission at the tip, which may be related to the bow-shock nature of the source or some outflow (supposedly originating at IRS16). In contrast, the dust emission traced by the L-band emission reveals a different and more complex picture. 

\begin{figure}[htbp!]
	\centering
	\includegraphics[width=.5\textwidth]{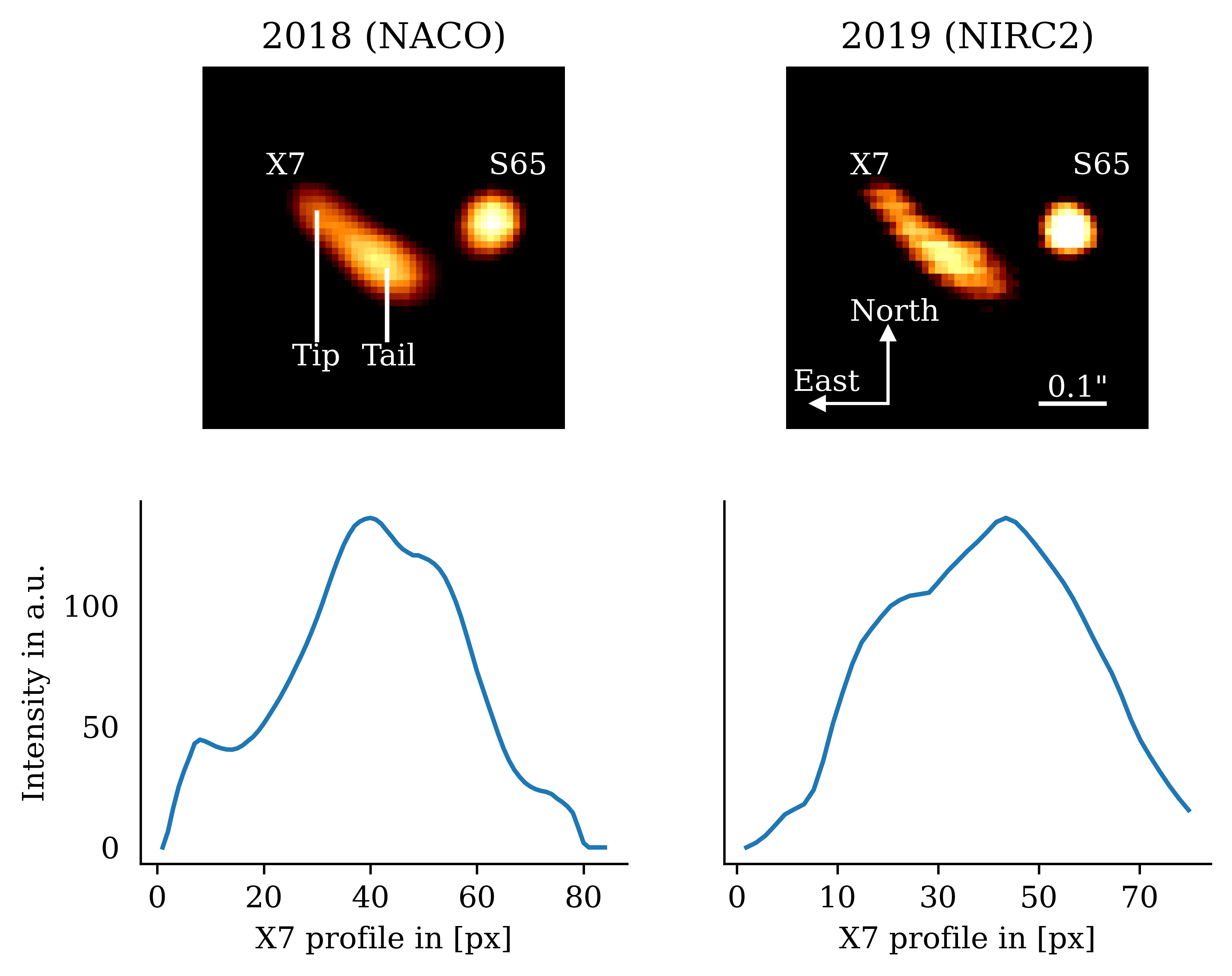}
	\caption{Radial profile of X7 from tip to tail in 2018 and 2019. We rebin the NACO data to match the pixel scale of the NIRC2 data, i.e., 10 mas. The profile is measured from the tip to the tail and shows a clear increased emission at the back of the bow-shock source. Please note that the distribution of the Doppler-shifted Br$\gamma$ line does not follow the here shown profile as we indicate in Fig. 5 and Fig. 11 in \cite{peissker2021}.}
\label{fig:x7_line_profile}
\end{figure}

As pointed out in \cite{peissker2021}, the dusty tail of X7 shows an increasing emission in strong contrast to the line profile shown in \cite{Ciurlo2023}. Hence, we revisited and analyzed publicly available L-band data to investigate the dust emission profile along the bow-shock source X7. As shown in Fig. \ref{fig:x7_line_profile}, we find in agreement with NACO observations presented in \cite{peissker2021} a clear increased emission located at the tail of X7 by including NIRC2 data observed with Keck in 2019. 
For a coreless dust feature that gravitationally interacts with Sgr~A* as proposed by Ciurlo et al., one would expect an increased dust emission at the tip location of X7. Considering the numerous publications related to G2/DSO discussing an apparent increase of the tip luminosity of a coreless elongated feature in the S-cluster approaching Sgr~A* \citep{Murray-Clay2012, Ballone2013, Pfuhl2015} brings the non-stellar classification of X7 into doubt. As implied by the offset of S50 to X7 shown in \cite{peissker2021} and confirmed by \cite{Ciurlo2023}, the S-star may not be the best NIR candidate, although it offers a solution to the increased thermal emission at the tail of the bow-shock source (Fig. \ref{fig:x7_line_profile}). 
Despite the relation between X7 and S50, the orbit was another strong argument that resulted in the classification as a coreless cloud. Due to the range of uncertainties for the astrometric data presented in \cite{Ciurlo2023}, the resulting Keplerian solution may suffer from limited significance. {Quantitative, the uncertainties in this work are ten times smaller for the orbital approximation compared to the Keplerian trajectory presented in \cite{Ciurlo2023}. This difference can not be explained by instrumental effects, especially considering the precise S2 orbit derived in \cite{Do2019S2}, which very well agrees with \cite{Parsa2017} and \cite{gravity2018}. Nevertheless, we} adopted the presented Keplerian solution from Ciurlo et al. and compared it with the one presented in Table \ref{tab:orbital_elements}.
\begin{figure}[htbp!]
	\centering
	\includegraphics[width=.5\textwidth]{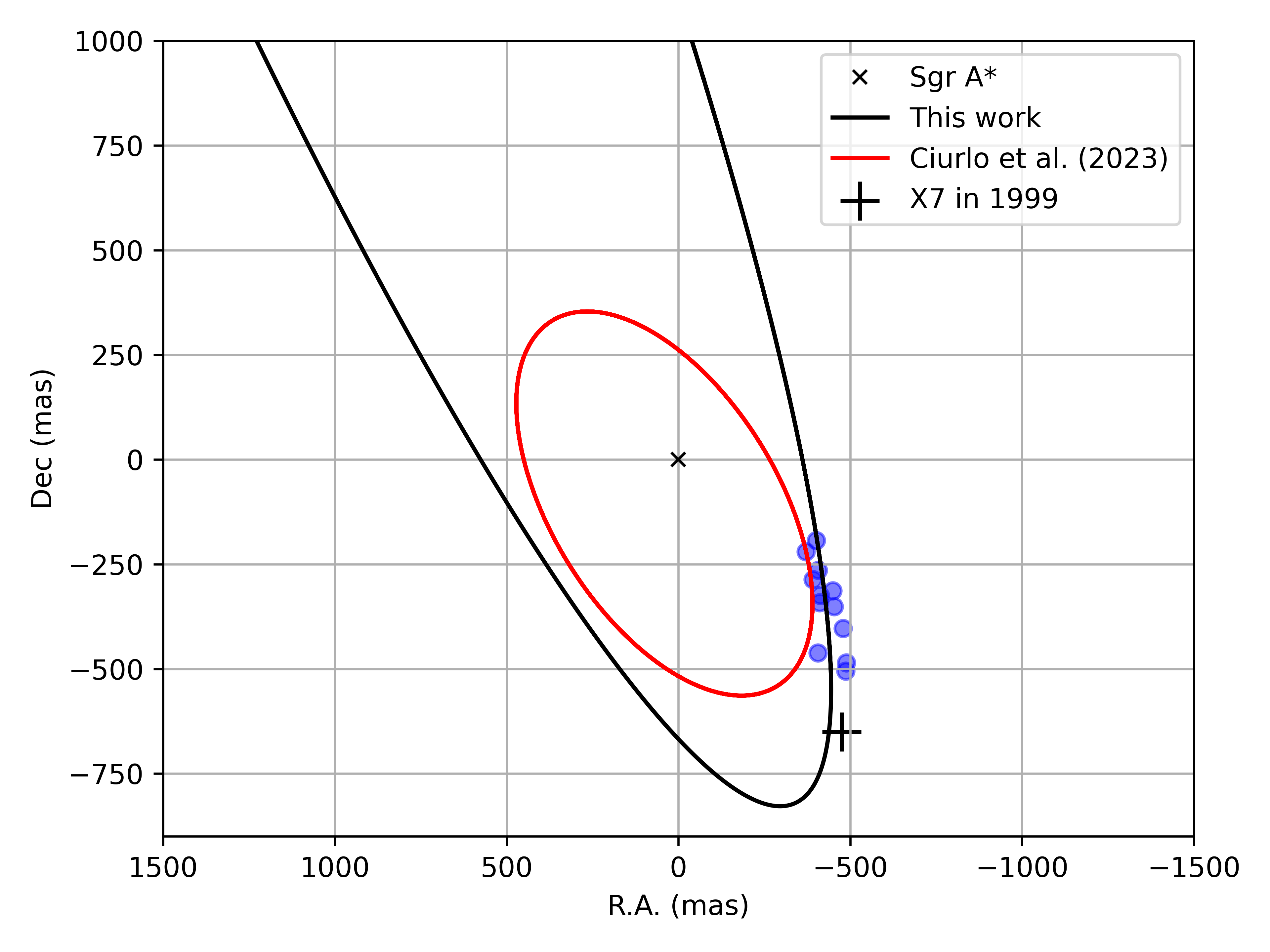}
	\caption{Comparison of the Keplerian solutions for the trajectory of X7 between \cite{Ciurlo2023} and this work. We implement the measured data points that are listed in the Appendix. In the upcoming epochs, we will be able to test the validity of the presented orbital solutions. However, L-band data from 1999 already reveals a strong tendency that favors the presented orbit in Table \ref{tab:orbital_elements}. This data point is represented by a black + sign in the figure above. See the text for details.}
\label{fig:x7_small_kepler}
\end{figure}
As shown in Fig. \ref{fig:x7_small_kepler}, both Keplerian solutions for X7 share a fraction of their related orbits, which demonstrates a general problem of the significance for individual objects in the S-cluster. As pointed out by \cite{PortegiesZwart2023}, close encounters could impact the orbit of the S-stars \citep{perets2007, Perets2008}. In addition, the presence of the so-called dark cusp, consisting of old faint stars that have migrated to the vicinity of Sgr~A* \citep{Morris1993, Alexander2014}, can potentially alter the trajectory of the cluster members. Hence, larger timelines decrease the expected astrometric uncertainty for objects in about 8 kpc distance. With the L-band data point of X7 in 1999 which is at $\Delta x\,=\,0.47''$ and $\Delta y\,=\,0.65''$ \citep{peissker2021}, we strongly favor the Keplerian solution presented in Table \ref{tab:orbital_elements}. Due to the significance of the astrometric measurements presented in \cite{Ciurlo2023}, we will explore upcoming high-resolution GC observations with ERIS (VLT), MIRI (JWST), and METIS (ELT) to enhance the {validity} of the presented orbital solutions.

\subsubsection{X7.1}

In \cite{Peissker2020b}, we reported the detection of a new dusty source very close ($\sim 20-30$mas) to X7. Due to its proximity, we called it X7.1. \cite{Ciurlo2023} confirmed this finding and reported in addition an additional source that the authors call X7.2. Because of the limitations of the SINFONI FOV and the lower resolution of the NACO data, we will explore this source with upcoming ERIS observations.
As pointed out in \cite{peissker2021}, X7.1 is independently identified as G5 in \cite{Ciurlo2020} where the authors present an orbit very similar to the one shown in Fig. \ref{fig:all_orbits}. In fact, both orbital solutions for X7.1 (Table \ref{tab:orbital_elements}) and G5 \citep{Ciurlo2023} agree reasonably well with the numerical values matching inside the uncertainty range. Regarding its (L-M) colors, we find 0.54 which implies a dust temperature close to 500 K \citep{Witzel2017}. Due to the missing K-band identification, we are not allowed to classify X7.1 using H-K and K-L colors. However, the radiative transfer model fit shows a satisfying agreement with the estimated flux density distribution (Table \ref{tab:mag_sll_sources}). Similarly to every other source covered in this work, X7.1 can be classified as a low-mass stellar object. Furthermore, the SED shown in Fig. \ref{fig:SED} suggests a classification as a candidate YSO.

\subsubsection{X8}

\begin{figure}[htbp!]
	\centering
	\includegraphics[width=.5\textwidth]{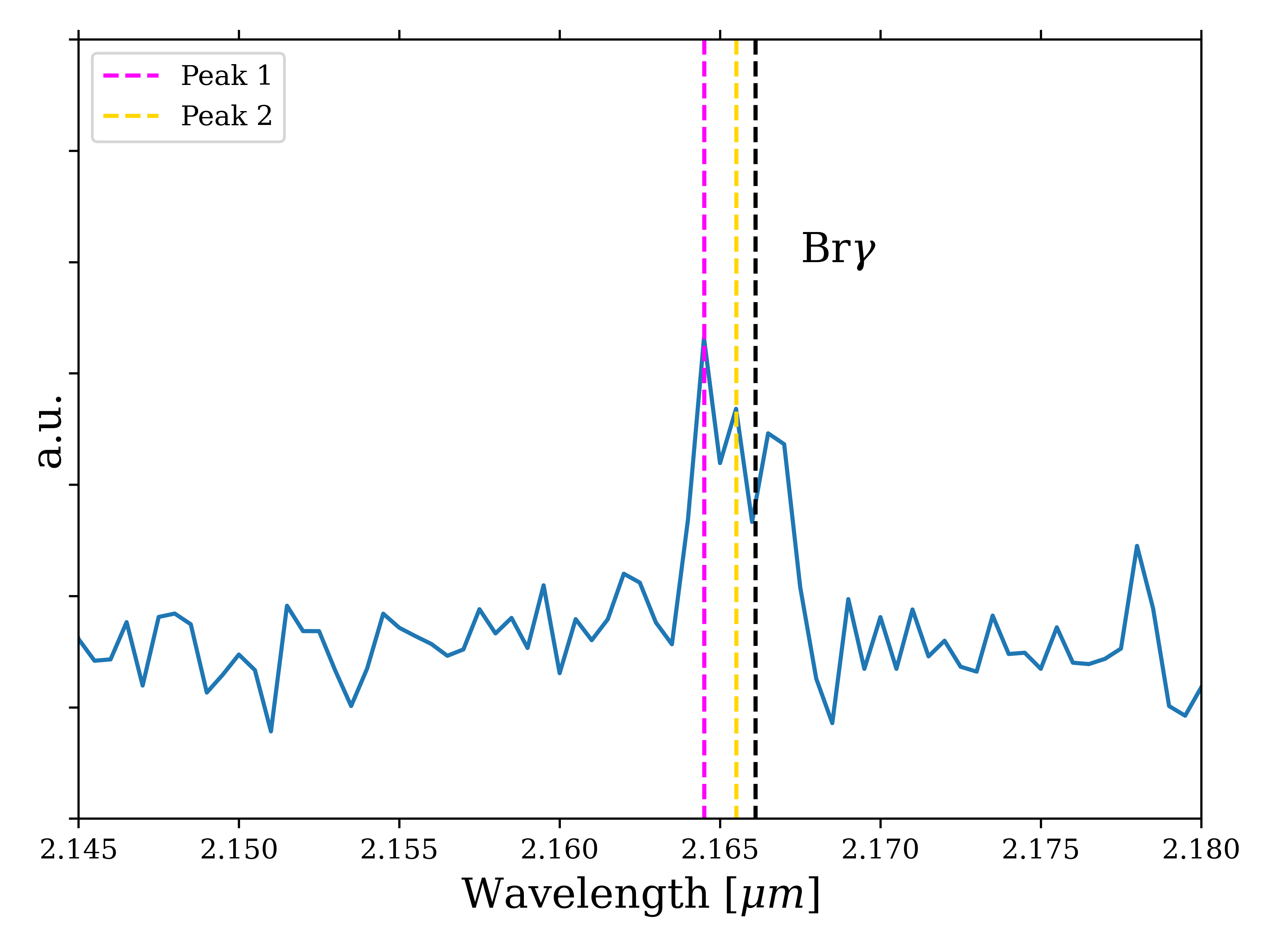}
	\caption{Spectrum of the bipolar outflow source X8 as observed with SINFONI in 2019. The dashed black line marks the Br$\gamma$ rest wavelength at 2.1661$\,\mu$m. The redshifted emission may be related to the Br$\gamma$-bar \citep{Peissker2020c} whereas the blueshifted line consists of two peaks at 2.1644$\,\mu$m (Peak 1, magenta) and 2.1654$\,\mu$m (Peak 2, gold). The related velocities of the two blueshifted peaks are 235.44 km/s and 96.94 km/s.}
\label{fig:x8_line}
\end{figure}

Already discovered in 2017, we reported the detection of a possible bow shock source that we called X8 \citep{Peissker2019} due to its proximity to X7 and its slightly elongated shape. While the classification of X8 as a bow shock source may be premature, the bipolar morphology of the source can be observed in various epochs. Using the latest observation of X8 with SINFONI of 2019, we detect a blueshifted double-line feature that is in strong agreement with the analysis presented in \cite{Peissker2019}. In Fig. \ref{fig:x8_line}, we show the blueshifted Br$\gamma$ line that exhibits the mentioned double-line feature which underlines the bipolar morphology of X8. Together with the findings of a population of low-mass bipolar outflow sources (BOS) shown in \cite{Yusef-Zadeh2017}, X8 marks the closest BOS toward Sgr~A*. Hence, it is reassuring that the color-color analysis suggests the classification of X8 as a candidate YSO. 
With the color-color classification, we utilize \texttt{HYPERION} and find that X8 is a low-mass candidate YSO with bipolar morphology. This result is an independent confirmation of the low-mass star formation population claimed in the inner parsec by \cite{Yusef-Zadeh2017}. 

\subsubsection{Limitations of the photometric analysis}

{Due to the prominent radio/submm emission of Sgr~A*, longer wavelengths hinder a detailed look into the S-cluster. Although MIRI (JWST) and especially Metis (ELT) will expand the photometric coverage, the inaccessible long wavelengths are an unavoidable obstacle. However, we want to emphasize that a stellar classification for the dusty sources, considering the evaporation timescales, is evident. It is, therefore, strongly suggested that the Br$\gamma$ emission is created intrinsically. In this framework, the presence of ionized gas directly points to stellar winds and their interplay with hot accretion disks where the Br$\gamma$ emission is produced \citep{Kraus2009, Seifried2012, Tanaka2016}. Considering the color-color analysis presented in Fig. \ref{fig:color_color_diagram}, systematic uncertainties would not impact the stellar classification by itself. Taking into account the consistent photometric results for the S-stars derived in this work compared with the literature, we conclude that any systematic uncertainties for the individual magnitudes are inside the given error range.}

\subsection{Lense-Thirring effect}
\label{sec:lense-thirring}

First proposed by \cite{Lense1918}, the Lense-Thirring (LT) effect describes the orbital precession that changes the argument of periapsis and inclination of a body that moves along a trajectory not aligned with the equatorial plane of a rotating central body, here Sgr~A*. LT precession is a genuine gravito-magnetic (Machian) effect of General Relativity \citep{1995mpfn.conf.....B,1995grin.book.....C,2003nlgd.conf.....R}; indeed, rotation (spin) of the central object stands behind LT precession and it combines with other influences, e.g. due to an extended distribution of nearby gravitating matter and/or higher multipoles caused by aspherical deformation (flattening) of the central source of the gravitational field, and the oblateness of the dense nuclear star-cluster \citep{2014A&A...566A..47S,2020A&ARv..28....4N}. Let us note that attempts to constrain the spin of Sgr~A* were already carried out in numerous independent works \citep{Melia2001,Iorio2011,Fragione2020}, however, a definitive answer about the spin orientation is still missing \citep{Shahzamanian2015_Msngr}. While the GRAVITY instrument helped set an upper limit on the extended mass \citep{Heissel2022, Peissker2022}, the EHT collaboration revealed the inclination and dimensions of the Schwarzschild radius of Sgr~A* \citep{eht2022}.

Even more, the same authors of \cite{eht2022b} confirmed the presence of a Kerr black hole with a mass of $4\,\times\,10^6\,M_{\odot}$ in numerical agreement with individual estimates based on single stellar orbits \citep{Do2019S2, Peissker2022}. Therefore, observing the imprint of the spin of Sgr~A* on the orbits of the S-cluster members would strengthen the significance of the findings listed above. Quantitatively, one would expect a randomized distribution of the S-stars under the influence of the rotating SMBH that alters the evolution of the orbits. The analysis of the S-stars presented in \cite{Ali2020} benefits from a larger data baseline that resulted in the finding of two distinct stellar disks, namely the red and black disks. Furthermore, the authors of \cite{Fragione2022}\footnote{The criticisms of \cite{Fragione2022} on the findings of the above-cited conclusions of \citet{Ali2020}, alleging that even \cite{Fellenberg2022} did not support the existence of a two-disk structure based on GRAVITY data, are not completely persuasive. This is because von Fellenberg et al. neither utilized GRAVITY data nor dismissed the possibility of a two-disk setup.} investigated Lense-Thirring precession with a focus on these S-cluster disks. More recently, \cite{Iorio2023} employed S4716 \citep{Peissker2022} to explore the imprints of Sgr~A* on the star. In light of these results, we investigate the influence of Sgr~A* on the dusty source population.

Using the results for PA indicated in Fig. \ref{fig:pa} and listed in Table \ref{tab:orbital_elements}, we find that the dusty sources are exclusively arranged in the red disk. Similar to \cite{Fragione2022}, we will use the angular momentum of the spin, $\mathcal{S}$ to define the dimensionless spin of Sgr~A* with
\begin{equation}
    \rm \chi_{SgrA*}\,=\,\frac{\lVert\mathcal{S}\rVert c}{GM_{SgrA*}^2}
\end{equation}
where we set the mass to $M_{SgrA*}\,=\,4\times 10^6 M_{\odot}$ \citep{Peissker2022}. Furthermore, we use
\begin{equation}
    \rm T_{\mathcal{S}}\,=\,\frac{c^3 a^3 (1-e^2)^{3/2}}{2\chi_{SgrA*}G^2M_{SgrA*}}
    \label{eq:lt}
\end{equation}
to calculate the Lense-Thirring timescales. In the above equation, we insert the individual orbital elements listed in Table \ref{tab:orbital_elements}.
\begin{figure}[htbp!]
	\centering
	\includegraphics[width=.5\textwidth]{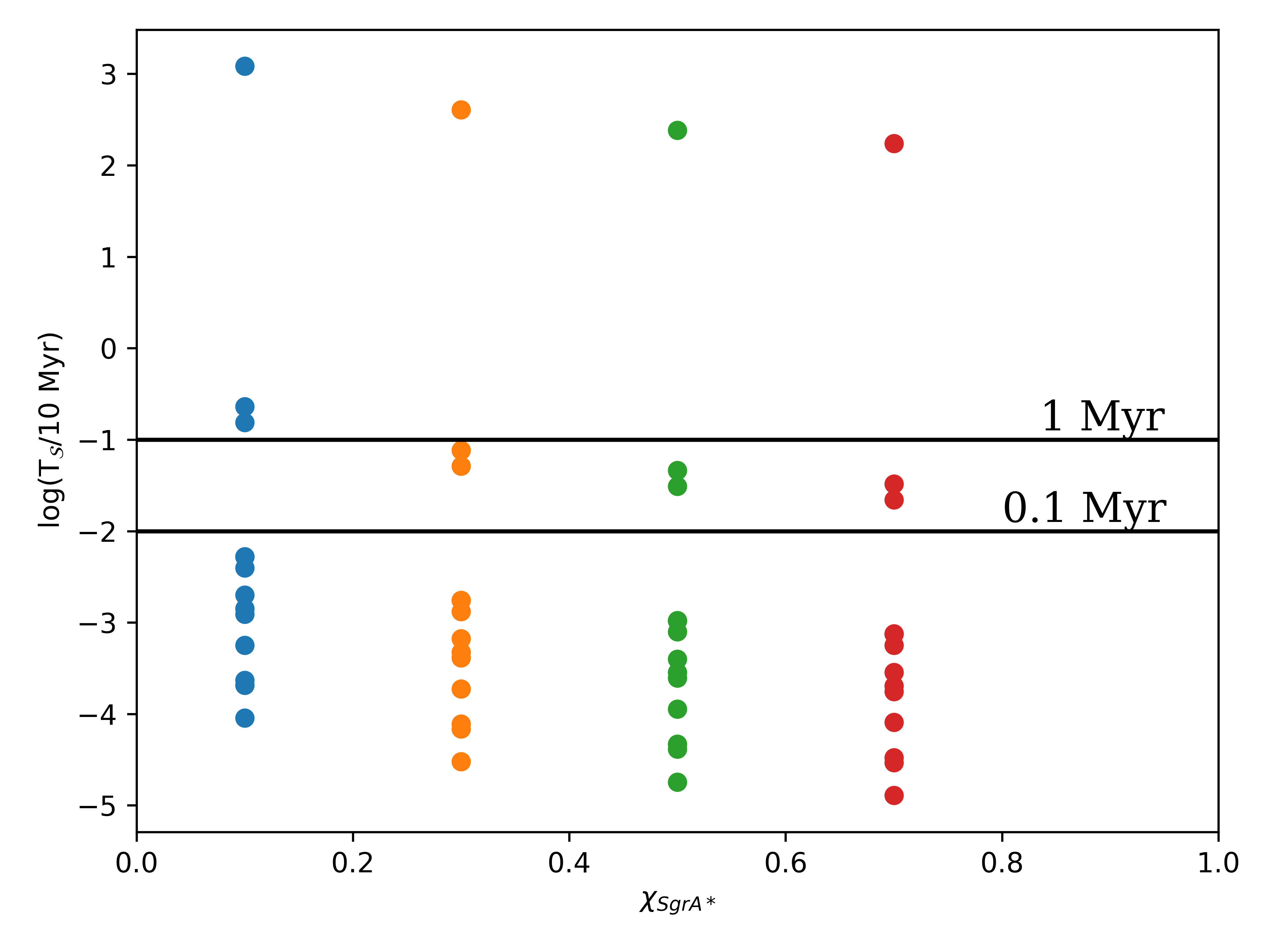}
	\caption{Frame-dragging timescales for the dusty sources for different spins $\chi$ of Sgr~A*. The solid lines mark the approximate age range for the dusty sources in this work. We assume an age of the S-cluster of 10 Myrs.}
\label{fig:lt}
\end{figure}
In Fig. \ref{fig:lt}, we show the results of Eq. \ref{eq:lt} by incorporating the semi-major axis $a$ and the eccentricity $e$ listed in Table \ref{tab:orbital_elements} as a function of different spin values with $\chi\in\{ 0.1,0.3,0.5,0.7\}$. Since we do find all dusty sources in a single disk, namely the red disk, which shows a mean inclination of about 90$^{\circ}$ roughly consistent with the plane of Sgr~A* \citep{gravity2018b}, we conclude that the dusty sources rapidly ($<$ 1 Myr) arranged in their current kinematic structure. This is consistent with the work of \cite{Iorio2023}, who concluded that a rearrangement due to the Lense-Thirring effect occurs on short timescales. Considering the age of the dusty source of $\lessapprox$ 1 Myrs, the imprint of the LT effect should have resulted in randomized orbits in contradiction to the observation. Therefore, it is tempting to raise the question about the origin of the observed non-randomized configuration.\newline
Speculative, the system constantly rearranges itself due to the interaction with massive disturbers such as IRS 13 \citep{Ali2020} or the CWD \citep{Paumard2006, Yelda2014}. As proposed by Ali et al., the reported two-disk structure is a natural outcome of this interaction (Singhal et al., submitted). The here-discussed dusty objects would obey the same dynamical influence as the main-sequence stars of the cluster.

\subsection{Origin of the dusty population}

As we have discussed in Sec. \ref{sec:lense-thirring}, the reorganization of the orbits of the D-objects of the S-cluster is possibly the imprint of a massive disturber given the LT timescales. This imprint is displayed in Fig. \ref{fig:pa}-\ref{fig:loan} and demonstrates the non-randomized character of the investigated orbits of the dusty sources. While we do not find a correlation between the orbit and system inclination, we notice a trend of $60\pm20$ degrees for the intrinsic arrangement of the dusty sources as shown in Fig. \ref{fig:inclination_correlation}. Despite the limited observed wavelength baseline that is used for the radiative transfer model to derive the intrinsic inclination of the systems (Table \ref{tab:sed_table}), the uncertainty of $\pm20$ degrees should compensate for the inaccessible bands, as the emission of the bright radio source Sgr~A* overshines all S-cluster members.
Due to the proposed nature of the dusty sources, we can safely assume that these cluster members migrated to the vicinity of the SMBH. It is irrelevant whether the evolution and nature of the dusty sources are categorized as binary mergers or tidally separated sources described by the Hills mechanism. Due to the comparable inclination of the sample investigated here, it is suggested that these sources have formed in a common process \citep{Wheelwrigth2011}, implying the possibility of star-formation channels due to infalling molecular clouds towards the inner parsec \cite{Bonnell2008, Hobbs2009, Jalali2014}. If the population of dusty sources has a common origin, it hints at a possible origin of the S-stars. Motivated by a rather global perspective on the NSC, we have recently identified another region that shows a similar age composition of the objects. The close-by and embedded massive cluster IRS 13 shows not only main-sequence stars but also candidate YSOs \citep{peissker2023c}. Although we cannot confirm nor neglect a putative evolutionary connection between the S-cluster and IRS 13, we can confidently conclude that the population of dusty objects did not originate within the S-cluster as depicted in Fig. \ref{fig:inclination_correlation} that shows the mismatch of the intrinsic system orientation and the orbital inclination\citep{Bonnell1992, Monin2006, Wheelwrigth2011}. 

\subsection{Potential relation to the S-stars}

In \cite{Ali2020}, we demonstrated that the S-stars do not show a randomized distribution and are organized into two distinct disks, namely the red and black disks. In agreement with the simulations of \cite{Bonnell2008}, it seems that the majority of stars in the inner parsec are arranged in disk-like structures. Recently, it has been independently shown in \cite{Burkert2024} that the distribution of the dusty objects follows the arrangement of the S-stars. Without a connection of the dusty sources to the two-disk setup of the S-stars, the authors of Burkert et al. concluded that all objects in the S-cluster formed in the same disk. Although this explanation seems tempting, it ignores the supposedly much younger age of the dusty objects compared to the S-stars as proposed by several authors such as \cite{Scoville2013}, \cite{Eckart2013}, \cite{Shahzamanian2016}, and \cite{Zajacek2017}. Taking into account the arrangement of the dusty sources that clearly follow the same distribution of the S-stars in the red disk using the PA, inclination, and LOAN, it is interesting to investigate the stellar properties of the black disk. Speculatively, the presence of an age gradient between the two disks could explain why the young dusty sources can be detected solely in the red disk. Although this exceeds the scope of this work, the origin of the different components may be a complex problem that demands theories that extend beyond an isolated star-formation origin in one disk. As proposed by \cite{Bonnell2008}, a migrating cloud that falls into the gravitational well of Sgr~A* favors an arrangement in one disk after the star formation processes took place. Therefore, it is plausible that multiple clouds could result in the formation of different disks. Although there is no theoretical basis for this claim yet, observations by \cite{Ali2020}, \cite{Fellenberg2022}, and \cite{Jia2023} favor a multi-disk arrangement of the stars of the inner parsec. A plausible origin of the initial clouds could be clumps in the CND that are formed under the influence of turbulent convection streams \citep{Dinh2021}.

\subsection{Astrometric robustness}

\begin{figure}[htbp!]
	\centering
	\includegraphics[width=.5\textwidth]{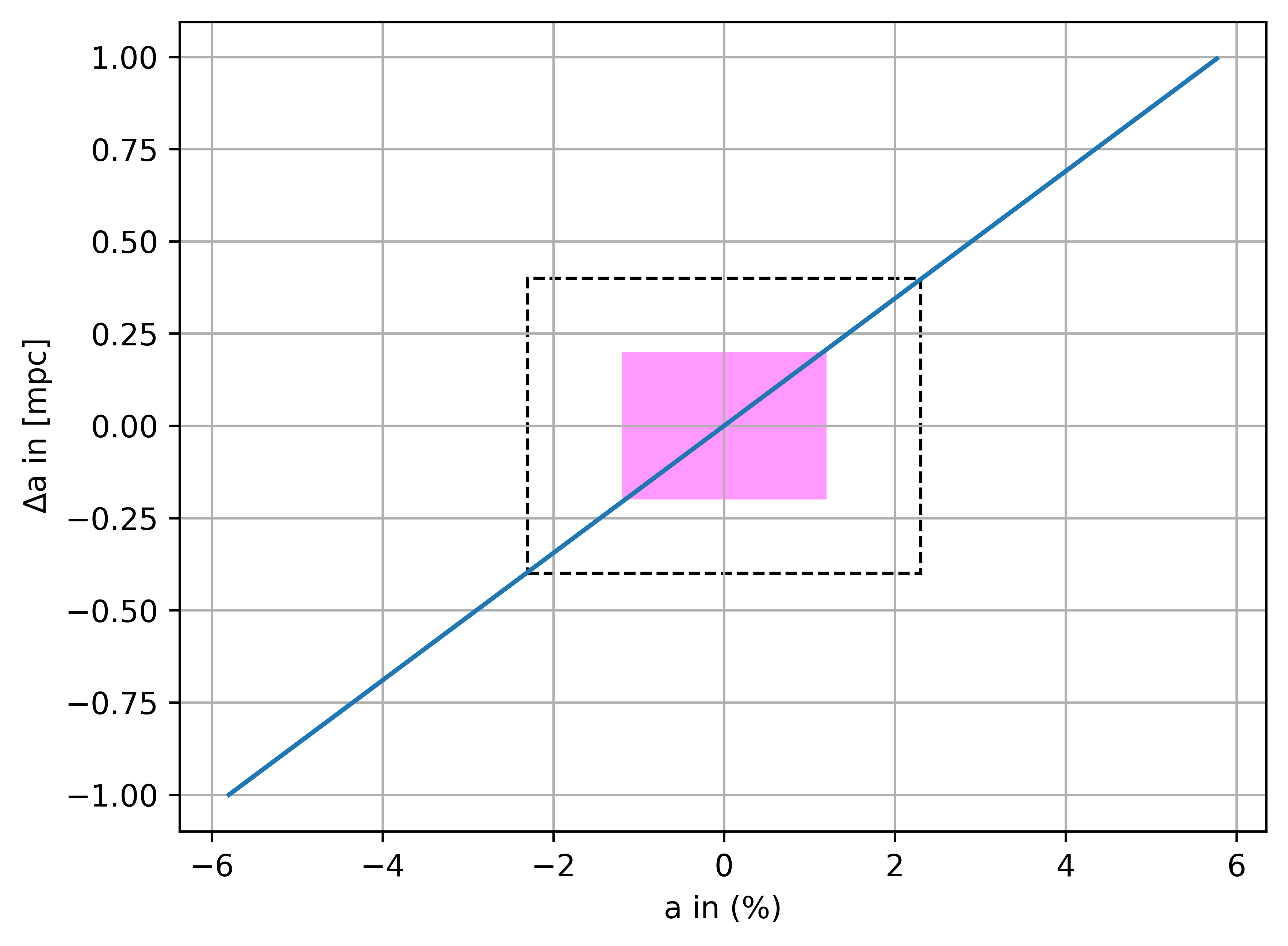}
	\caption{Impact of positional uncertainty $\Delta a$ on the semi-major axis of G2/DSO. The magenta patch represents the uncertainties derived by the MCMC simulations \citep[see Table \ref{tab:orbital_elements} and][]{peissker2021c} and are equivalent to 0.01 as or $\approx$ 1 pixel. The dashed box represents the uncertainty for 2 pixel or $\approx$ 0.02 as. See text for details.}
\label{fig:kepler_robustness}
\end{figure}

{The 3d arrangement of the dusty sources is based on the Keplerian analysis of single objects using SINFONI data. By focusing on one instrument and data set, systematic uncertainties would affect all measurements in the same way. Hence, we would expect the same outcome as discussed in this section. The uncertainties would be underestimated if our astrometric measurements solely relied on the Gaussian fitting error (Sec. \ref{sec:ident_dusty_objects}). Hence, we use a cumulative uncertainty constructed from the reference frame \citep[$\pm 2$ mas,][]{Parsa2017, Plewa2018}, distortion effects \citep[$\pm 4$ mas,][]{Eisenhauer2003, Peissker2020b}, and background noise in the Br$\gamma$ line maps \citep[$\pm 2-7$ mas,][]{Peissker2020c, peissker2021c}. Due to the variable background of the S-cluster, caused by the dark cusp \citep[Sec. \ref{sec:x7_discuss}][]{Alexander2014}, a qualitative uncertainty estimate of the magnitude remains challenging. Taking into account high-precision interferometric observations of S2 on $\mu as$ scales by \cite{gravity2018}, we may infer a correlation between the preexisting VLT (NACO, SINFONI) and VLTI (GRAVITY) data. Although the range and magnitude for all listed instruments may differ significantly, determining the position of S2 and vice versa, the position of Sgr~A*, results in robust outcomes. The high level of robustness is manifested in agreeing outcomes for estimating the Schwarzschild precession of S2. In \cite{Parsa2017}, the authors estimate a relativistic parameter of 0.88 using VLT data in numerical agreement with the analysis of \cite{Do2019_S2} where the authors use KECK observations. In addition, \cite{gravity2018} confirms these findings by deriving a relativistic parameter of 0.9 using GRAVITY. While these outcomes already suggest a high level of robustness, we want to demonstrate the advantages of an increased data baseline that intrinsically smooths any uncertainty fluctuations. 
In Fig. \ref{fig:kepler_robustness}, we show a variation of the semi-major axis of G2/DSO for an uncertainty of $\pm 1$ mpc as a function of the impact on the derived Keplerian elements listed in Table \ref{tab:orbital_elements}. In the same figure, we include the MCMC uncertainties from the same table and find that a variation by about 0.4 mpc (equivalent to 0.01 as and $\approx$ 1 pixel in the SINFONI data) correlates with an uncertainty of $\approx 2.5\%$ for the semi-major axis. Considering a larger positional uncertainty of 2 pixels (25 mas) in the investigated SINFONI data, we estimate an impact on the semi-major axis of $\approx 4.5\%$.}
{Although a 2-pixel uncertainty overestimates the astrometric precision, given the precise results from determining the relativistic parameter of S2, we conclude that our cumulative uncertainty range is well justified and covers fluctuations of the background and distortion effects introduced by the detector \citep{peissker2021c}. While the astrometric robustness is one (time-dependent) aspect, stellar confusion is another key by analyzing individual cluster members. Although the Doppler-shifted Br$\gamma$ line maps reduce the chance of confusion, we want to explore the probability of detecting a false cluster member on a similar Keplerian orbit as the source of interest. In Table \ref{tab:cluster_input}, we list the input parameter for the Monte Carlo simulation.}
\begin{table}[hbt!]
    \centering
    \begin{tabular}{|c|c|}
         \hline
           Parameter & Input \\
         \hline
         Cluster size     &  40 mpc   \\
         Number of stars  &  50-1000  \\         
         Resolution/FHWM  &   3 mpc   \\         
         Data baseline    &  15 years \\     
       \hline
    \end{tabular}
    \caption{Input parameter for the Monte Carlo simulations displayed in Fig. \ref{fig:cluster_statistics}.}
    \label{tab:cluster_input}
\end{table}
{Using the listed parameters, we compare the probability for false detections of stars on random Keplerian orbits in a cluster with an increasing number of members. Furthermore, we compare different data baselines to underline the significance of our long-term analysis.}
\begin{figure}[htbp!]
	\centering
	\includegraphics[width=.5\textwidth]{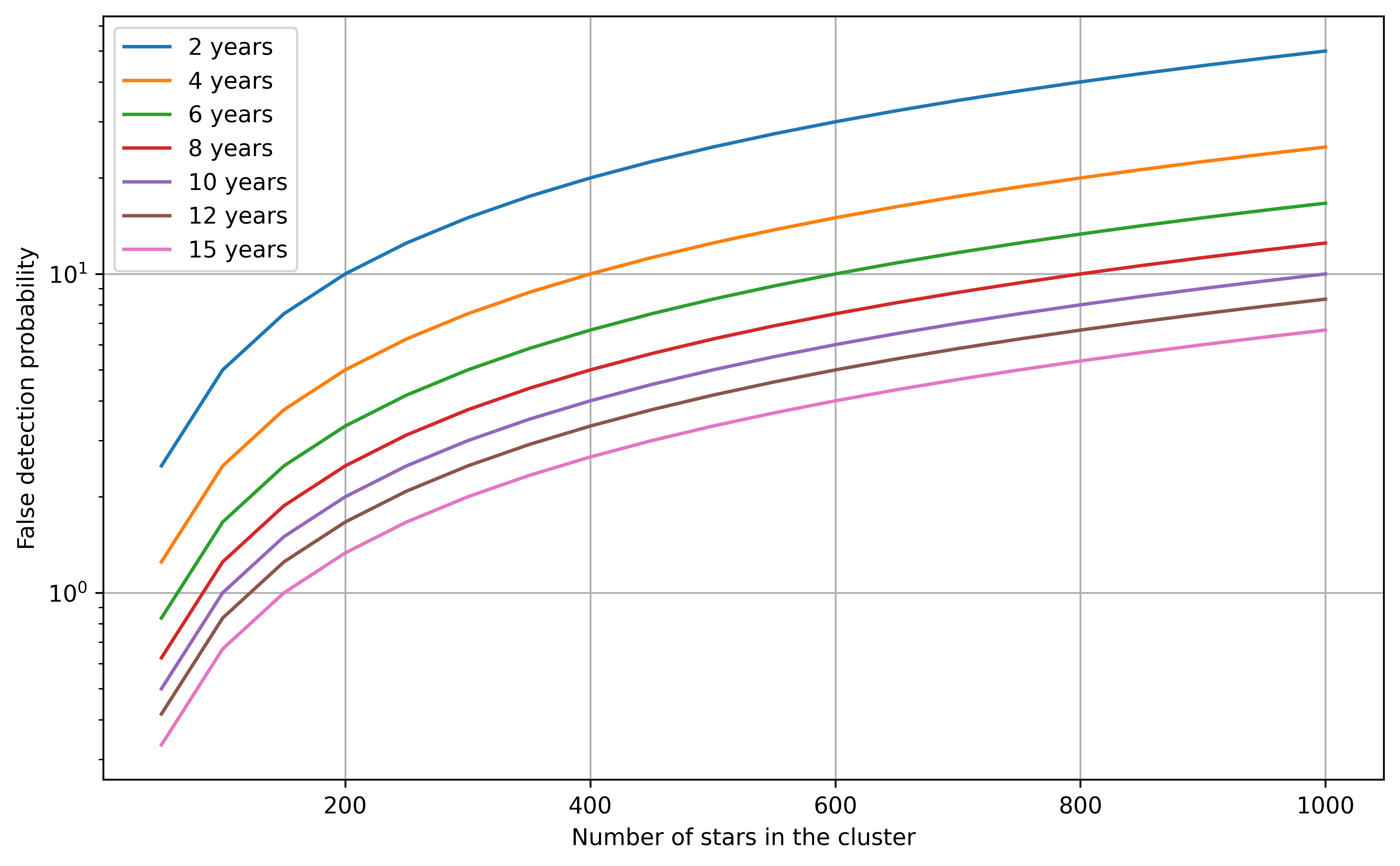}
	\caption{Monte Carlo simulations of the probability of detecting a confused star for different data baselines and cluster members. The y-axis is logarithmically scaled. For cluster sizes $\leq$100 stars and data baseline $\approx$15 years, the probability of detecting a confused star is $<1\%$.}
\label{fig:cluster_statistics}
\end{figure}
{Using randomly dispersed stars on Keplerian orbits in different-sized clusters, we find an expected probability that scales with the stellar density and is inversely proportional to the data baseline. For an S-cluster sized structure with 50-100 stars \citep{Ali2020}, we find a probability of $<1\%$ considering a data baseline of 10-15 years. The probability should decrease drastically if we take into account the magnitude of the source of interest.}\newline
{Based on the above discussion, we conclude that the here-derived results and interpretations are robust against outliers or systematic uncertainties. Hence, we expect the red disk arrangement (Fig. \ref{fig:pa}) of the dusty sources to be a statistically significant result in agreement with the non-randomized distribution stated in \cite{Burkert2024}.}

\section{Conclusion} 
\label{sec:conclusion}

In this work, we analyzed the population of dusty sources of the S-cluster using SINFONI and NACO data covering the epochs between 2002 and 2019. With the Keplerian orbital information, we are allowed to constrain a statement about the 3d orientation and the related PA of the individual object. We incorporated photometric measurements to estimate the flux density in various bands using several unique calibrator stars. The SED revealed stellar properties that underlined the idea that the dusty sources belong to an individual sub-population compared to the S-stars. In the following, we will list all key results of this work.
\begin{itemize}
    \item We estimate a Keplerian solution for the trajectory of 13 objects in the S-cluster,
    \item The PA of all sources shows that they are arranged in the red disk that was identified by \cite{Ali2020}, 
    \item Using the magnitude and consequently, the flux density values, we estimate the SED for the majority of objects,
    \item With \texttt{HYPERION}, we find that the flux density of the dusty sources follows the spectral energy distribution of low-mass YSOs in agreement with the works by \cite{Zajacek2017} and \cite{Ciurlo2020},
    \item The (dust) evaporation timescales of non-stellar low-mass clumps is $\ll$ 2 yrs,
    \item We find no indications of a non-stellar nature of the investigated sources in agreement with theoretical models,
    \item The multi-wavelength data favors a classification of the dusty sources as candidate YSOs with an age of $\lessapprox$ 10$^6$ yrs,
    \item Since the Lense-Thirring timescale is typically shorter than the lifetime of the dusty sources, an additional dynamical explanation is required to address the disk-like configuration,
    \item We find indications that the dusty sources, like the S-stars, undergo a dynamical interaction with massive structures, such as the IRS 13 cluster, that favors a disk-like arrangement,
    \item An age gradient for the cluster members suggests a different origin for the dusty sources and the S-stars,
    \item However, both populations might have migrated through a comparable star formation channel into the S-cluster and rearranged under the imprint of massive components of the NSC,
    \item Speculatively, the dusty sources will evolve into low-mass S-stars.
\end{itemize}
We note that continuous monitoring of the S-cluster with its dusty members is sufficient to establish a comprehensive picture of this subpopulation. It took G2/DSO almost a decade to exit the S-cluster. Therefore, it is plausible that we will find more G- and D-like objects in the future using instruments such as ERIS/VLT.

\section*{Acknowledgments}
FP gratefully acknowledges the Collaborative Research Center 1601 funded by the Deutsche Forschungsgemeinschaft (DFG, German Research Foundation) – SFB 1601 [sub-project A3] - 500700252. 
MZ is grateful to the GA\v{C}R JUNIOR STAR grant no. GM24-10599M for financial support. 
VK has been partially supported by the Collaboration Project (ref.\ GF23-04053L -- 2021/43/I/ST9/01352/OPUS 22). AP, SE, JC, and GB contributed useful points to the discussion. We also would like to thank the members of the SINFONI/NACO/VISIR and ESO's Paranal/Chile team for their support and collaboration. This paper makes use of the following ALMA data: ADS/JAO.ALMA$\#$2016.1.00870.S and ADS/JAO.ALMA$\#$2015.1.01080.S. ALMA is a partnership of ESO (representing its member states), NSF (USA) and NINS (Japan), together with NRC (Canada), MOST and ASIAA (Taiwan), and KASI (Republic of Korea), in cooperation with the Republic of Chile. The Joint ALMA Observatory is operated by ESO, AUI/NRAO and NAOJ.

\bibliography{bib}{}
\bibliographystyle{aasjournal}

\appendix

\section{Data}
\label{ref:data_appendix}

All data used and discussed in this work are listed in \cite{peissker2021} and \cite{peissker2021c}. For the finding chart displayed in Fig. \ref{fig:finding_chart}, we utilized NACO K- and L-band data and the final reduced 100 GHz ALMA observations published along \cite{Moser2017}. 
The SINFONI, NACO, and ALMA data that are observed between the epochs 2002 to 2019 are publically available in the ESO and ALMA archives\footnote{\url{www.eso.org}}. See Table \ref{tab:data} for an overview of the data and the related purpose.
\begin{table}[tbh!]
\setlength{\tabcolsep}{0.9pt}
\centering
\begin{tabular}{|cccc|}
\hline
Epoch & Telescope/Instrument & Band& Purpose\\
\hline
2005-2019 & VLT/SINFONI & H+K   & D, K, C\\
2002-2018 & VLT/NACO    & L     & FC, SED, Si, C\\
2011      & VLT/NACO    & K     & FC, SED\\
2012      & VLT/NACO    & M     & SED, C\\
2017      & ALMA        & 100 GHz & FC\\
\hline
\end{tabular}
\caption{Overview of used telescopes and data in this work. We list the individual epochs, telescopes/instruments, bands, and the related abbreviations: (C) Color analysis, (Si) Source identification, (SED) Spectral Energy Distribution, (K) Keplerian fit, (D) 3D distance.}
\label{tab:data}
\end{table}

\section{Astrometric measurements}
\label{app_sec:astrometric_measurments}

The Keplerian fits in Sec. \ref{sec:results} are based on the individual detection of the dusty sources primarily in the Doppler-shifted Br$\gamma$ line maps. We prefer this approach to reduce confusion effects due to the high stellar density of the S-cluster. Due to the limited FOV of SINFONI, D5 cannot be observed in the available data set after 2008. Hence, we used the NACO L-band observations to detect this particular source. In the same way, we could have implemented the continuum MIR NACO data showing X7 between 2002-2005 to increase the data baseline. However, we wanted to ensure a consistent analysis whenever possible and exclude possible distortion effects of the NACO instrument that may not be transferable to the SINFONI data set after 2005 \citep{Plewa2015}. Hence, every orbit derived and listed in this work is highly consistent within the selection of the data set. 
Since some of the sources are already covered in some previous publications \citep{Peissker2020b, peissker2021c, peissker2023a}, we limit this listing to uncovered objects in this series of papers that analyze the dusty sources of the inner parsec \citep{Peissker2019, peissker2021, peissker2023b, peissker2023c}.
\begin{table*}[htb]
\centering
\begin{tabular}{|c|cc|cc|cc|cc|cc|}\hline \hline 
      & \multicolumn{2}{c|}{D5} & \multicolumn{2}{c|}{D9} & \multicolumn{2}{c|}{X7} & \multicolumn{2}{c|}{X7.1} & \multicolumn{2}{c|}{X8}	\\ 
      \hline
        & x  & y &  x  & y & x  & y &  x  & y &  x  & y \\  
Epoch      & \multicolumn{2}{c|}{in [as]}& \multicolumn{2}{c|}{in [as]} & \multicolumn{2}{c|}{in [as]} & \multicolumn{2}{c|}{in [as]} & \multicolumn{2}{c|}{in [as]}	\\

\hline 
2005.5 & - & - & 0.066 & 0.298  & -0.486 & -0.505 & - & - & - & - \\
2006.4 & - & - & 0.056 & 0.283  & -0.406 & -0.462 & -0.698 & -0.352 & -0.200 & -0.300 \\
2007.5 & - & - & 0.044 & 0.290  & - & - & - & - & - & - \\
2008.2 & - & - & 0.024 & 0.305  & - & - & -0.644 & -0.383 & - & - \\
2008.4 & 0.621 & 0.067 & - & -  & - & - & - & - & - & - \\
2009.2 & 0.621 & 0.081 & - & -  & - & - & - & - & -0.210 & -0.330 \\
2009.3 & - & - & 0.038 & 0.304  & -0.489 & -0.485 & -0.589 & -0.339 & - & - \\
2010.3 & - & - & -0.021 & 0.312 & -0.478 & -0.403 & - & - & -0.230 & -0.360 \\
2011.3 & - & - & -0.018 & 0.304 & -0.411 & -0.341 & - & - & -0.250 & -0.360 \\                                
2012.4 & - & - & -0.040 & 0.291 & -0.412 & -0.324 & - & - & -0.230 & -0.370 \\
2013.3 & 0.679 & 0.108 & - & -  & -0.453 & -0.351 & - & - & - & - \\
2013.4 & - & - & -0.059 & 0.288 & - & - & -0.537 & -0.287 & -0.250 & -0.370 \\
2014.3 & - & - & - & - &  &  & - & - & -0.260 & -0.410 \\
2014.5 & - & - & -0.040 & 0.288 & -0.449 & -0.313 & -0.537 & -0.275 & - & - \\
2015.4 & - & - & -0.067 & 0.280 & -0.391 & -0.286 & -0.525 & -0.275 & -0.260 & -0.380 \\
2016.2 & 0.734 & 0.136 & - & -  & - & - & - & - & - & - \\
2016.5 & - & - & -0.065 & 0.284 & -0.406 & -0.264 & -0.475 & -0.262 & -0.275 & -0.400 \\
2017.5 & - & - & -0.090 & 0.270 & -0.371 & -0.264 & -0.450 & -0.262 & - & - \\
2018.3 & 0.742 & 0.189 & - & -  & - & - & - & - & - & - \\
2018.4 & - & - & -0.105 & 0.279 & -0.400 & -0.193 & -0.450 & -0.250 & - & - \\
2019.4 & - & - & -0.123 & 0.272 & - & - & - & - & - & - \\
\hline \hline
\end{tabular}
\caption{Astrometric positions with respect to Sgr~A*. These values are used as input parameters that result in the Keplerian orbital solution listed in Table \ref{tab:orbital_elements}. All other measurements used are listed in \cite{Peissker2019, Peissker2020b, peissker2021, peissker2023a}. Typical uncertainties are in the order of {$\pm\,6.5$} mas \citep{peissker2021c}.}
\label{tab:fit_positions}
\end{table*}

\section{Keplerian fit solution}
\label{sec:app_keplerian_fit_solution}

Using the astrometric measurements listed in Table \ref{tab:fit_positions}, we derive the Keplerian solution presented in Table \ref{tab:orbital_elements} for D5, D9, X7, X7.1, and X8. With this information, we plot all solutions to visualize the validity of the fit compared to the measured data points. 

\begin{figure*}[htbp!]
	\centering
	\includegraphics[width=1.\textwidth]{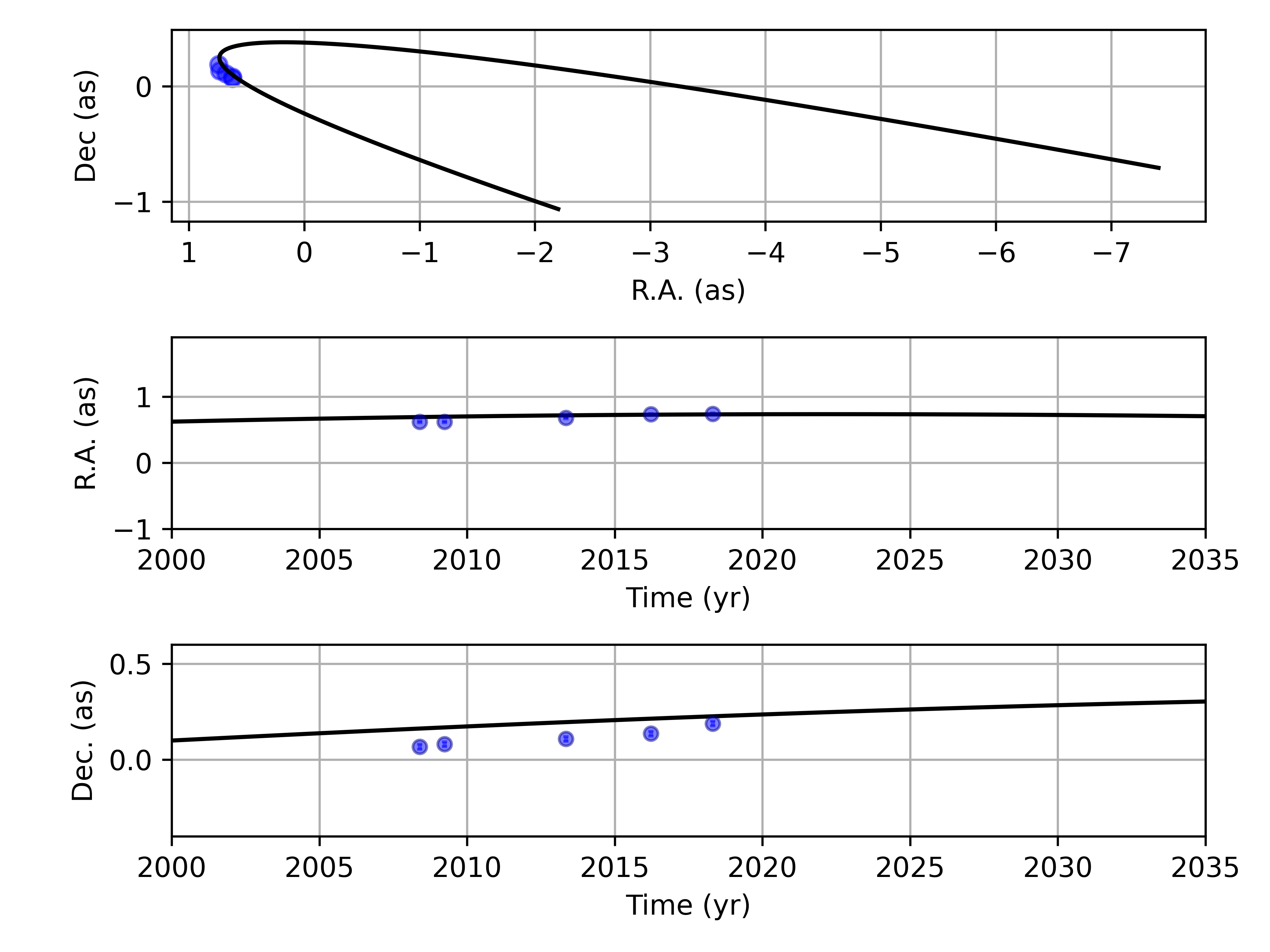}
	\caption{Keplerian solution for the dusty source D5. Due to its apoapsis proximity, the R.A. motion is expected to change in the forthcoming epochs.}
\label{fig:d5_keplerian_fit}
\end{figure*}

\begin{figure*}[htbp!]
	\centering
	\includegraphics[width=1.\textwidth]{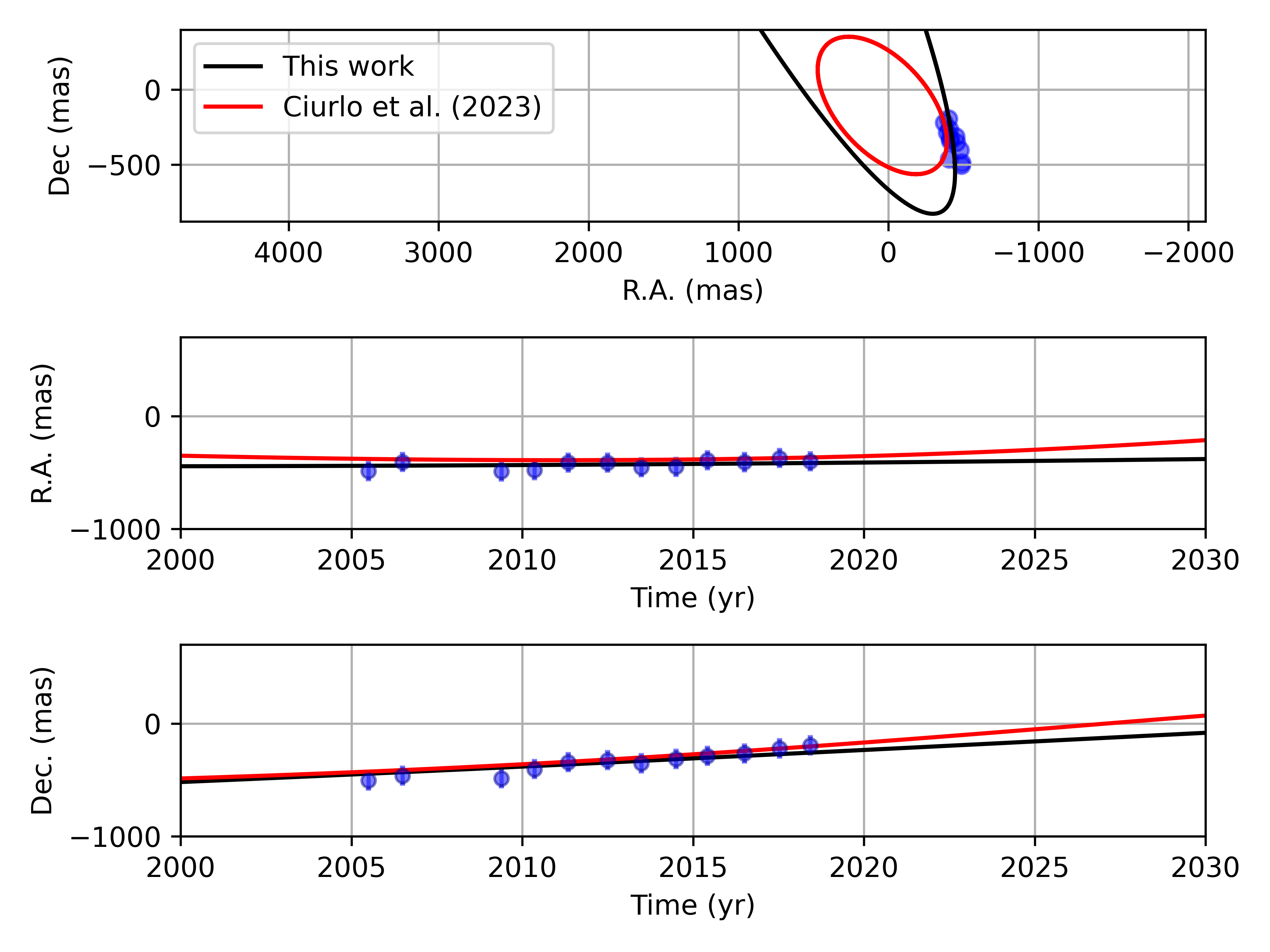}
	\caption{Best fit Keplerian solution for the astrometric measurements based on the Doppler-shifted Br$\gamma$ line observed with SINFONI. The red fit represents the orbital solution shown in \cite{Ciurlo2023}. While we find some overlap with our solution, the fit presented in Ciurlo et al.\ suffers from a low significance when we compare the astrometric measurements presented in this work.}
\label{fig:x7_keplerian_fit}
\end{figure*}

\begin{figure*}[htbp!]
	\centering
	\includegraphics[width=1.\textwidth]{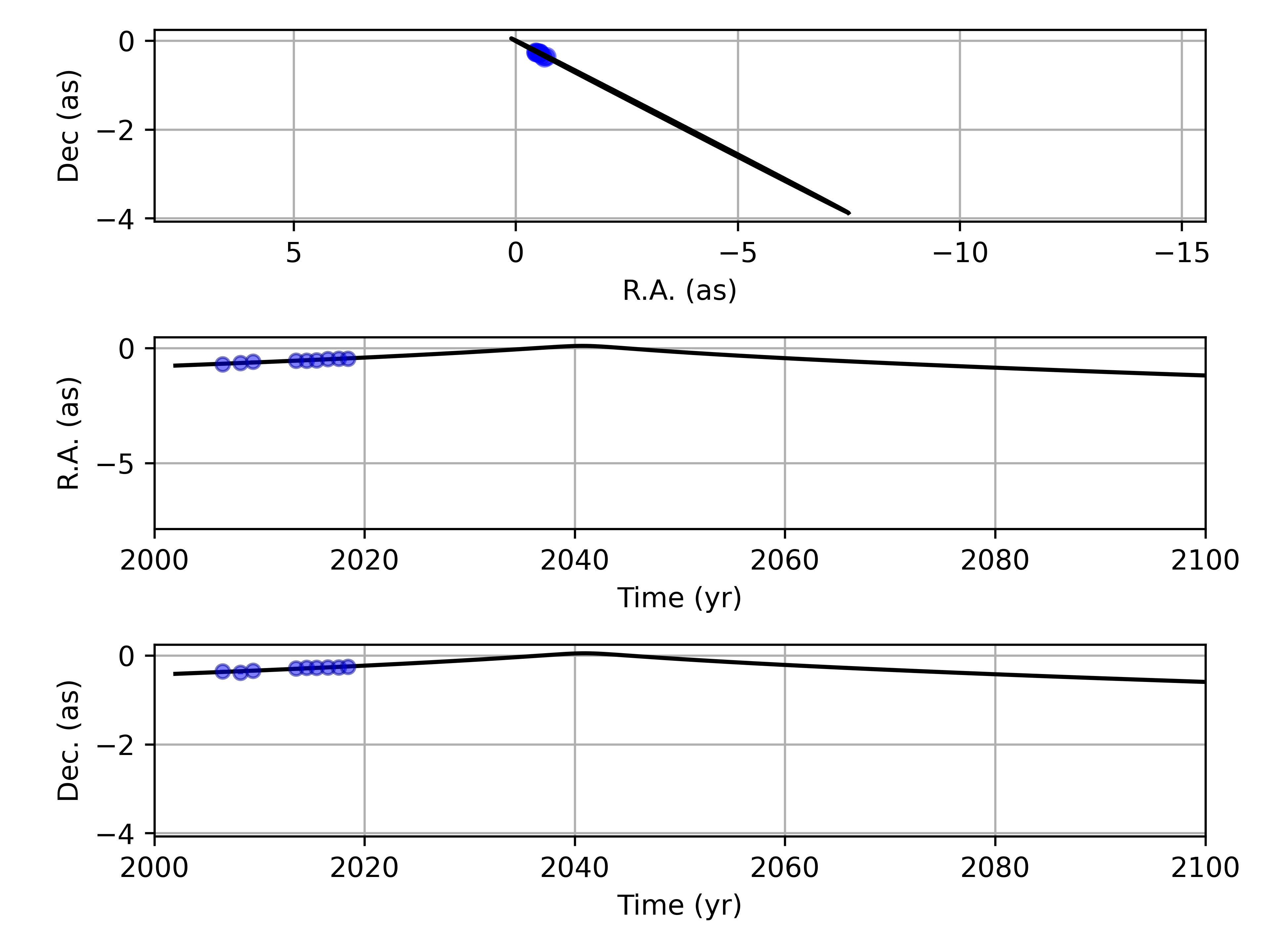}
	\caption{Keplerian best fit solution for the source X7.1 that moves close to X7 on a highly eccentric orbit.}
\label{fig:x71_keplerian_fit}
\end{figure*}

\begin{figure*}[htbp!]
	\centering
	\includegraphics[width=1.\textwidth]{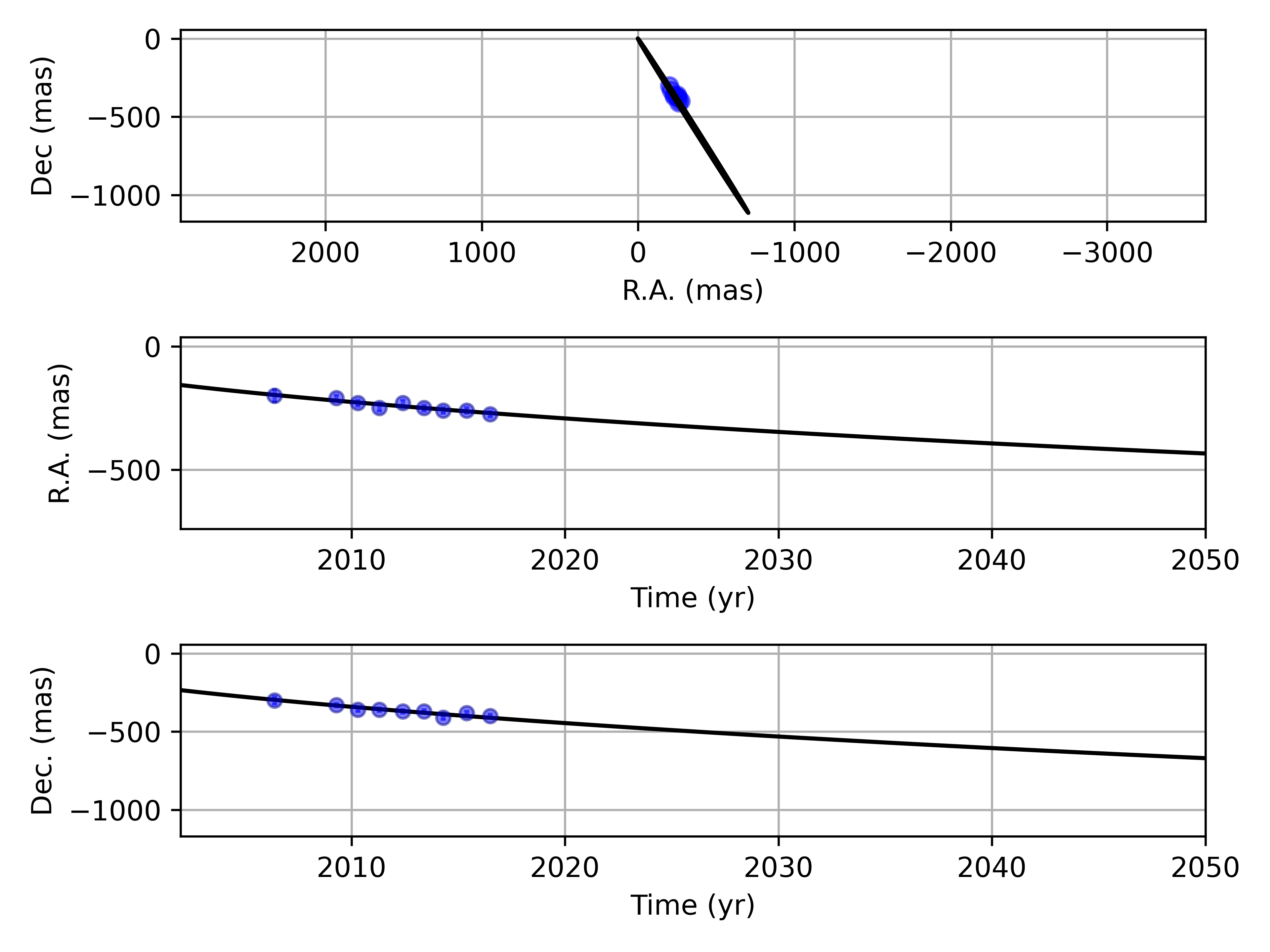}
	\caption{Orbit of X8. The sources moves, like X7.1, on a high eccentric orbit.}
\label{fig:x8_keplerian_fit}
\end{figure*}

In Fig. \ref{fig:d5_keplerian_fit}, we present the orbital solution for D5. Although a small offset in the declination is suggested, the overall projected orbit shows a strong correlation with the data points. The forthcoming JWST GTO observations will confirm or update the orbital solution.
The orbit shown in Fig. \ref{fig:x7_keplerian_fit} is consistent with the one displayed in Fig. \ref{fig:x7_small_kepler}. We plot the orbit in comparison with the solution presented in \cite{Ciurlo2023}. With the data points shown, we prefer the orbital solution derived in this work.
Close to X7, we detected another source that is discussed in \cite{Peissker2020b} and \cite{peissker2021}. In Fig. \ref{fig:x71_keplerian_fit}, we show the orbit solution for this source that we call X7.1 due to its proximity to X7. We find a highly eccentric orbit almost edge-on.
For X8, we find an even higher eccentricity of 0.99. We present this orbit in Fig. \ref{fig:x8_keplerian_fit}.

\section{Photometry}
\label{app_sec:photometry}

For the photometry, we focus on the H+K continuum data observed with SINFONI between 2006 and 2019. Since we use the Br$\gamma$ line to cross-check the individual positions of the objects, we limit the analysis of the NACO L-band to the epochs 2006-2018. Furthermore, this approach ensures a consistent data set where we can avoid various confusion/crowding challenges. 
For the calibrator stars, we use IRS16NW and IRS2L as our reference sources to estimate the H, K, L, and M magnitudes for S1, S2, and S4. With the analysis of \cite{Gillessen2009}, \cite{Sabha2010}, \cite{Stolte2010}, \cite{Witzel2014}, and \cite{peissker2023b}, we can cross-correlate the validity of the estimated magnitudes for the S-cluster stars. In Table \ref{tab:single_mag_epoch_photometry}, we outline the individual magnitude values for the investigated NIR and MIR bands.
\begin{table*}[htbp!]
    \centering
    \begin{tabular}{|c|cccc|cccc|cccc|}
            \hline
            \hline
            Epoch  &\multicolumn{4}{c|}{S1} & \multicolumn{4}{c|}{S2} & \multicolumn{4}{c|}{S4}  \\
                   & H & K & L & M & H & K & L & M & H & K & L & M  \\
            \hline       
             2004  & -  & - & 14.2 & - & 15.9 & 14.0 & 13.5    & -  & -  & -  & 13.9  & -   \\
             2006  &  16.4 & 14.6 & 12.6 & - & 15.8 & 13.9 & 12.4    & -  & 16.1  & 14.4  &  12.7 &  -     \\
             2008  &  16.5 & 14.6 & 12.5 & - & 15.6 & 14.1 & 12.3    & -  & 16.2  & 14.4  & 12.6  &  -    \\
             2010  &  16.4 & 14.6 & 13.1 & - & 15.9 & 14.2 & 13.4    & -  & 16.1  & 14.4  & 14.0  &  -    \\
             2012  &  16.5 & 14.6 & 12.7 & 12.6 & 16.1 & 14.1 & 11.4    & 12.8  & 16.2  & 14.4  & 12.6  & 13.1     \\
             2016  &  16.5 & 14.7 & - & - & -    &  -   & 12.4    &  - & 16.2  & 14.5  & -  &  -    \\
             \hline
             Mean  & 16.5  & 14.6 & 13.0 & 12.6 & 15.9 & 14.1 &  12.6   &  12.8 & 16.2  & 14.4  & 13.1  &  13.1    \\
             Std   &  0.1 & 0.1 & 0.6 & 0.4 & 0.1  & 0.1   &  0.7    &  0.1 & 0.1  & 0.1  & 0.6  &   0.2   \\
            \hline

            \hline
    \end{tabular}
    \caption{Calibrator stars for the S-cluster. For the K-band values, we use an extinction of $A_K\,=\,2.4$ adopted from \cite{Fritz2011} and a correction factor of 0.2 mag for the K' and K$_S$ magnitude \citep{Sabha2010}. For these values, we use the non-variable star IRS16NW and IRS2L as reference sources. For 2016, no H-band observations carried out with NACO are available. Due to corrupt calibration data for the K-band observations in 2016, we exclude this epoch from this list. The H- and K-band magnitudes for S1 and S4 are determined with S2 as the reference star and SINFONI H+K data between 2006 and 2016. As pointed out by \cite{Hosseini2020}, some magnitude variation of S2 is expected.}
    \label{tab:single_mag_epoch_photometry}
\end{table*}
With the mean magnitudes, we find consistent values with the listed literature references. All values listed in Table \ref{tab:single_mag_epoch_photometry} are dereddened where we use the zero flux information of IRS2L and IRS16NW. In the next step, we extract all counts for the individual objects in the FOV, including the reference stars. Finally, we estimate the magnitudes listed in Table \ref{tab:single_mag_epoch_photometry_dusty_sources_hband}-\ref{tab:single_mag_epoch_photometry_dusty_sources_mband}.
\begin{table*}[htbp!]
    \centering
    \begin{tabular}{|c|ccccccccccccccc|}
            \hline
            \hline
             Epoch & S1 & S2 & S4 & D2 & D23 & D3 & D5 & D9 & OS1 & OS2 & G1 & G2/DSO & X7 & X7.1 & X8  \\
            \hline     
            2006 &  16.34 & 15.9 & 16.09 &  -     & -     & - &   - &  -    &  - &   - &   - &  - &   -    &   -    &  17.40 \\
            2007 &  16.21 & 15.9 & 16.04 &  16.96 & 16.96 & - &   - &  -    &  - &   - &   - &  - &   -    &   -    &   - \\
            2008 &  16.56 & 15.9 & 16.21 &  18.15 & 18.59 & - &   - &  -    &  - &   - &   - &  - &   -    &   -    &  18.15 \\
            2009 &  16.18 & 15.9 & 16.18 &  17.49 &  -    & - &   - &  -    &  - &   - &   - &  - &   -    &   -    &   - \\
            2010 &  16.28 & 15.9 & 16.01 &  -     & 17.64 & - &   - &  -    &  - &   - &   - &  - &  16.89 &   -    &   - \\
            2011 &  16.28 & 15.9 & 16.01 &  -     &  -    & - &   - &  -    &  - &   - &   - &  - &   -    &   -    &   - \\
            2012 &  16.37 & 15.9 & 16.16 &  17.57 &  -    & - &   - &  -    &  - &   - &   - &  - &   -    &   -    &  17.57 \\
            2013 &  16.42 & 15.9 & 16.18 &  -     & 17.49 & - &   - &  -    &  - &   - &   - &  - &   -    &   -    &   - \\
            2014 &  16.24 & 15.9 & 16.11 &  -     & 17.75 & - &   - &  -    &  - &   - &   - &  - &  17.31 &   -    &   - \\
            2015 &  16.37 & 15.9 & 16.37 &  17.78 & 17.78 & - &   - &  -    &  - &   - &   - &  - &  17.78 &  17.78 &   - \\
            2016 &  16.55 & 15.9 & 16.24 &  -     &  -    & - &   - &  -    &  - &   - &   - &  - &   -    &   -    &   - \\
            2017 &  16.45 & 15.9 & 16.45 &  -     &  -    & - &   - &  -    &  - &   - &   - &  - &   -    &   -    &  17.20 \\
            2018 &  16.57 & 15.9 & 16.42 &  -     &  -    & - &   - &  -    &  - &   - &   - &  - &   -    &   -    &  17.93 \\
            2019 &  16.65 & 15.9 & 16.34 &  -     &  -    & - &   - & 19.92 &  - &   - &   - &  - &  17.84 &   -    &   -    \\ 
             \hline
            Mean  &16.39 & 15.9 & 16.20 & 17.59 & 17.70 & - & - & 19.92 & - & - & - & - & 17.45 & 17.78 & 17.65 \\
            Std   &0.14 & - & 0.14 & 0.38 & 0.48 & - & - & - & - & - & - & - & 0.38 & - & 0.34 \\

            \hline
            \hline
    \end{tabular}
    \caption{H-band magnitude of the dusty sources, including the reference stars S1, S2, and S4. The stars S1 and S4 are chosen to confirm the photometric analysis since their mean magnitude is consistent with the numerical values estimated from IRS2L and IRS16NW.}
    \label{tab:single_mag_epoch_photometry_dusty_sources_hband}
\end{table*}

\begin{table*}[htbp!]
    \centering
    \begin{tabular}{|c|ccccccccccccccc|}
            \hline
            \hline
             Epoch & S1 & S2 & S4 & D2 & D23 & D3 & D5 & D9 & OS1 & OS2 & G1 & G2/DSO & X7 & X7.1 & X8  \\
            \hline     
             2006  &  14.48 & 14.1 & 14.34 &   -   &   -   &  - &  - &   -    & - & - & - & - &   -    & - &  15.68  \\
             2007  &  14.49 & 14.1 & 14.36 & 15.75 & 15.75 &  - &  - &   -    & - & - & - & - &   -    & - &  16.31  \\
             2008  &  14.59 & 14.1 & 14.40 & 16.21 & 16.57 &  - &  - &  18.06 & - & - & - & - &   -    & - &  -  \\
             2009  &  14.35 & 14.1 & 14.39 & 15.81 &   -   &  - &  - &   -    & - & - & - & - &  15.81 & - &  -  \\
             2010  &  14.58 & 14.1 & 14.28 & 15.64 &   -   &  - &  - &   -    & - & - & - & - &  15.20 & - &  -  \\
             2011  &  14.54 & 14.1 & 14.29 &   -   &   -   &  - &  - &   -    & - & - & - & - &   -    & - &  15.43  \\
             2012  &  14.47 & 14.1 & 14.43 &   -   &   -   &  - &  - &   -    & - & - & - & - &   -    & - &  -  \\
             2013  &  14.61 & 14.1 & 14.45 & 15.91 & 16.11 &  - &  - &   -    & - & - & - & - &  15.60 & - &  -  \\
             2014  &  14.37 & 14.1 & 14.42 &   -   &   -   &  - &  - &   -    & - & - & - & - &  15.73 & - &  -  \\
             2015  &  14.56 & 14.1 & 14.21 &   -   & 16.35 &  - &  - &   -    & - & - & - & - &  15.99 & - &  -  \\
             2016  &  14.62 & 14.1 & 14.49 & 15.89 &   -   &  - &  - &   -    & - & - & - & - &   -    & - &  -  \\
             2017  &  14.77 & 14.1 & 14.62 &   -   &   -   &  - &  - &  18.48 & - & - & - & - &  15.89 & - &  15.89  \\
             2018  &  14.78 & 14.1 & 14.61 &   -   &   -   &  - &  - &   -    & - & - & - & - &  16.11 & - &  -  \\
             2019  &  14.70 & 14.1 & 14.52 &   -   &   -   &  - &  - &  17.97 & - & - & - & - &  15.97 & - &  -  \\
             \hline
             Mean  &14.56 & 14.1 & 14.42 & 15.87 & 16.20 & - & - & 18.17 & - & - & - & - & 15.79 & - & 15.82 \\
             Std   &0.12 & - & 0.11 & 0.17 & 0.30 & - & - & 0.22 & - & - & - & - & 0.26 & - & 0.32 \\
            \hline
            \hline
    \end{tabular}
    \caption{K-band magnitude of the dusty sources, including the reference stars S1, S2, and S4.}
    \label{tab:single_mag_epoch_photometry_dusty_sources_kband}
\end{table*}

\begin{table*}[htbp!]
    \setlength{\tabcolsep}{5.0pt}
    \centering
    \begin{tabular}{|c|ccccccccccccccc|}
            \hline
            \hline
             Epoch & S1 & S2 & S4 & D2 & D23 & D3 & D5 & D9 & OS1 & OS2 & G1 & G2/DSO & X7 & X7.1 & X8  \\
            \hline     
             2006 & 12.83 & 12.6 & 12.95 & 13.25 & 13.89 & 13.46 &  13.10 & -      &  13.76 &  13.77 &  13.22 &  13.64 &  12.06 &  13.37 &  14.30 \\
             2007 & 13.13 & 12.6 & 13.25 & 13.48 & 13.76 & 13.85 &  13.37 & -      &  14.41 &  14.84 &  13.83 &  14.20 &  11.98 &  13.76 &  15.59 \\
             2008 & 13.03 & 12.6 & 12.86 & 13.03 & 13.80 & 13.25 &  12.97 & -      &  13.44 &  13.40 & -      &  13.27 &  12.14 &  13.03 &  13.76 \\
             2009 & 12.68 & 12.6 & 12.57 & 12.93 & 16.69 & 13.35 &  12.73 & -      &  13.99 &  14.25 &  13.39 &  13.48 &  11.26 &  12.92 &    -   \\
             2010 & 12.45 & 12.6 & 13.32 & 14.42 & -     & -     &  13.23 & -      &  15.05 &  13.89 & -      &  14.83 &  11.27 &  12.74 &    -   \\
             2011 & 12.55 & 12.6 & 12.61 & 12.81 & -     & 12.84 &  12.50 & -      &  12.81 & -      &  12.84 &  12.82 &  12.18 &   -    &    -   \\
             2012 & 12.57 & 12.6 & 12.65 & 12.79 & 12.91 & 12.85 &  12.65 & -      & -      & -      & 12.80      &  12.75 &  12.31 &   -    &    -   \\
             2013 & 12.65 & 12.6 & 12.89 & 13.18 & -     & 13.37 &  13.12 & -      &  13.47 & -      &  13.27 &  13.23 &  12.28 &   -    &    -   \\
             2014 & -     & -    & -     & -     & -     & -     & -      & -      & -      & -      & -      & -      & -      &   -    &    -   \\
             2015 & -     & -    & -     & -     & -     & -     & -      & -      & -      & -      & -      & -      & -      &   -    &    -   \\
             2016 & 12.98 & 12.6 & 13.05 & 13.27 & 13.75 & 13.70 &  13.63 & -      & -      &  13.82 & -      & -      &  12.09 &   -    &  13.38 \\
             2017 & -     & -    & -     & -     & -     & -     & -      & -      & -      & -      & -      & -      & -      &   -    &    -   \\
             2018 & 13.43 & 12.6 & 13.30 & 13.52 & 14.31 & 14.09 &  14.19 & -      &  13.95 &  13.83 & -      & -      &  12.52 &  13.99 &    -   \\
             2019 & -     & -    & -     & -     & -     & -     & -      & 15.92  & -      & -      & -      & -      & -      & -      &    -   \\
             \hline
             Mean  &12.83 & 12.6 & 12.94 & 13.27 & 14.16 & 13.42 & 13.15 & 15.92 & 13.86 & 13.97 & 13.22 & 13.53 & 12.01 & 13.30 & 14.26 \\
             Std   &0.29 & - & 0.26 & 0.45 & 1.10 & 0.39 & 0.47 & - & 0.63 & 0.42 & 0.31 & 0.65 & 0.39 & 0.44 & 0.83 \\
            \hline
            \hline
    \end{tabular}
    \caption{L-band magnitude of the dusty sources, including the reference stars S1, S2, and S4.}
    \label{tab:single_mag_epoch_photometry_dusty_sources_lband}
\end{table*}

\begin{table*}[htbp!]
    \centering
    \begin{tabular}{|c|ccccccccccccccc|}
            \hline
            \hline
             Epoch & S1 & S2 & S4 & D2 & D23 & D3 & D5 & D9 & OS1 & OS2 & G1 & G2/DSO & X7 & X7.1 & X8  \\
            \hline     
             2012  & 12.84 & 12.8 & 12.89 & 12.75 & 12.81 & 12.79 & 12.88 & - & - & - & - & 12.87 & 12.36 & 12.76 & 12.87 \\

            \hline
            \hline
    \end{tabular}
    \caption{M-band magnitude of the dusty sources, including the reference stars S1, S2, and S4.}
    \label{tab:single_mag_epoch_photometry_dusty_sources_mband}
\end{table*}
We only utilize S2 in this analysis step to validate the independently estimated magnitudes for S1 and S4 through the use of IRS2L and IRS16NW. It is vital to note that this consistency check aligns with the literature and serves as a reliable estimate within the data set, as we obtain matching outcomes for all reference stars.

\section{L-band magnitude of G1}

As discussed in this article, we have opted to forgo the use of a filtering technique in order to ensure enhanced comparability with existing literature whenever feasible. Furthermore, the use of non-filtered data can be used as a consistency check with pre-existing results that were obtained by applying a high-pass filter. Due to confusion and crowding effects in dense clusters, this approach could tend to produce magnitude values that may be brighter compared to filtered data. Although we tested the impact of dense and crowded clusters on the photometric approach \citep{peissker2023c}, we will explore the literature L-band magnitude values of G1 listed in \cite{Witzel2017}.

In Table \ref{tab:g1_comp}, we list the results of the L-band comparison between the smooth-subtracted NACO data and the Keck observations presented in \cite{Witzel2017}. For this, we use the technique explained in Sec. \ref{sec:smooth_subtract} and apply a 104 mas Gaussian Kernel to the input data.
\begin{table}[htbp!]
\centering                                  
\begin{tabular}{|c|cc|}
\toprule
Epoch     & High pass filter (this work) & Continuum \citep{Witzel2017} \\
\hline
2004        & 14.57 & 13.51 \\
2006        & 14.04 & 13.87 \\
2009        & 14.03 & 14.38 \\
2012        & 14.76 & 15.21 \\
2013        & 14.86 & 15.22 \\
\hline
Average     & 14.45 & 14.43 \\
Std         &  0.35 &  0.69 \\
\hline
\end{tabular}
\caption{L-band magnitude comparison of G1 between high pass filtered data and the analysis of \cite{Witzel2017}. The authors of Witzel et al. did not apply any filter to the data and derived the related L-band magnitude solely from the Keck continuum observations. A dimming of the source can not be independently verified.}
\label{tab:g1_comp}
\end{table}
As shown in Table \ref{tab:g1_comp}, a consistent average L-band magnitude for G1 between this work and Witzel et al. was found until 2013. However, after this year, the NACO data suffer from confusion and crowding effects. This statement also holds for data before 2004 where G1 is confused with Sgr~A*. Additional techniques such as image deconvolution are necessary to disentangle individual objects. Although we estimate a slightly higher continuum L-band magnitude of 13.22$\pm$31 mag as listed in Table \ref{tab:single_mag_epoch_photometry_dusty_sources_lband}, we observe a reassuring decrease in the gap between the numerical values taking into account the individual uncertainties. From this short photometric discussion, we can conclude that the confusion-free Doppler-shifted Br$\gamma$ line will result in enhanced astrometric precision. For a detailed discussion of G1, we refer to Melamed et al. (in preparation).

\section{MCMC simulations}
\label{app_sec:mcmc_simulations}

For the MCMC simulations, we utilize the emcee PYTHON package described in \cite{Foreman-Mackey2013}. As a prior, we use the Keplerian fit results listed in \ref{tab:orbital_elements}. From the results of the MCMC sampler, we extract the uncertainties to reflect the variety of possible solutions. We note that the chance of a local minimum for the Keplerian solutions presented in Table \ref{tab:orbital_elements} or in this section cannot be ruled out. But using the visual confirmation as presented in Sec. \ref{sec:app_keplerian_fit_solution}, we minimize the chance of detecting an inappropriate Keplerian solution.
\begin{figure*}[htbp!]
	\centering
	\includegraphics[width=1.\textwidth]{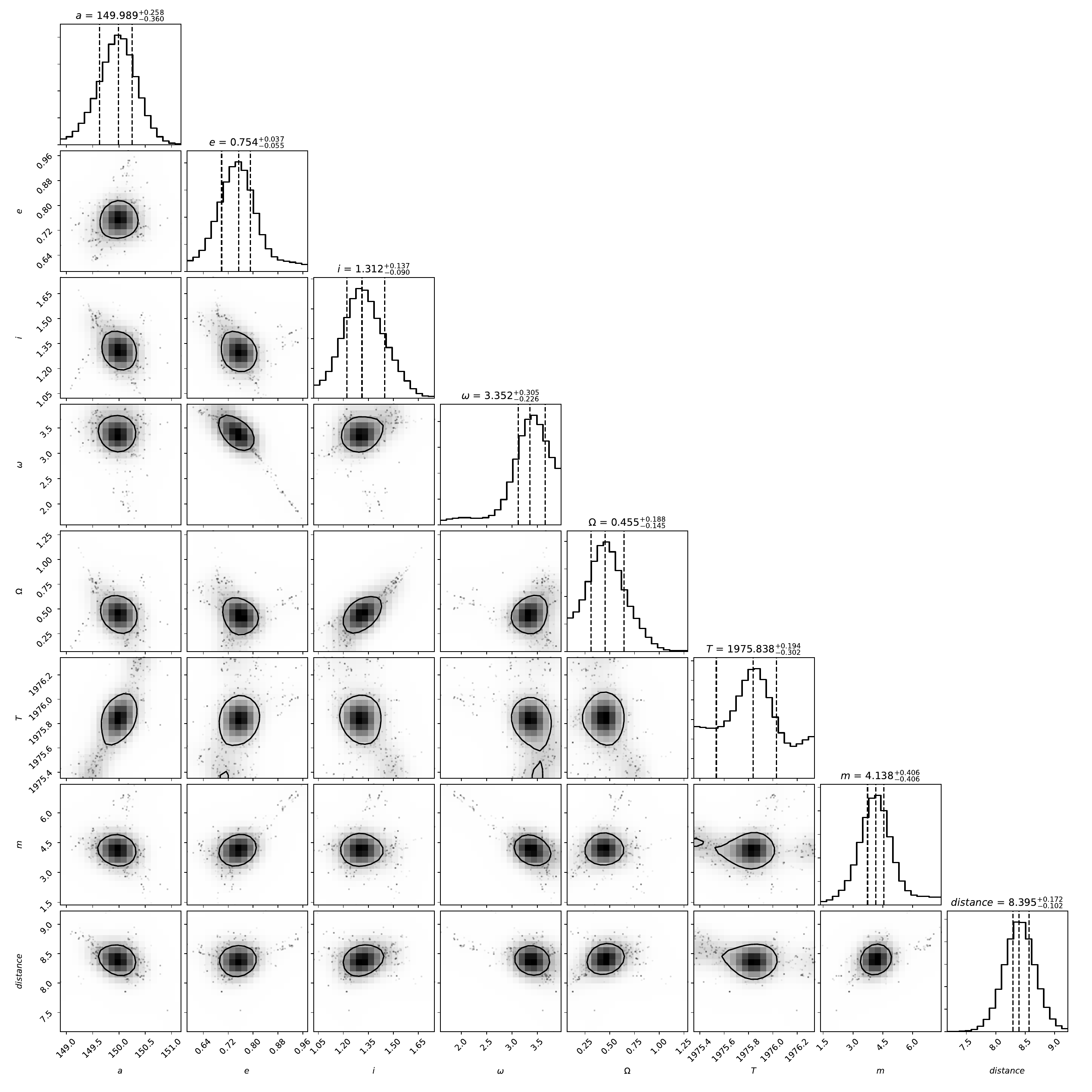}
	\caption{Monte Carlo Markov Chain simulations for the Keplerian orbits of X7. The compact shape of the possible orbital elements implies sufficiently estimated priors, i.e., the Keplerian elements listed in Table \ref{fig:all_orbits}.}
\label{fig:mcmc_x7}
\end{figure*}

\begin{figure*}[htbp!]
	\centering
	\includegraphics[width=1.\textwidth]{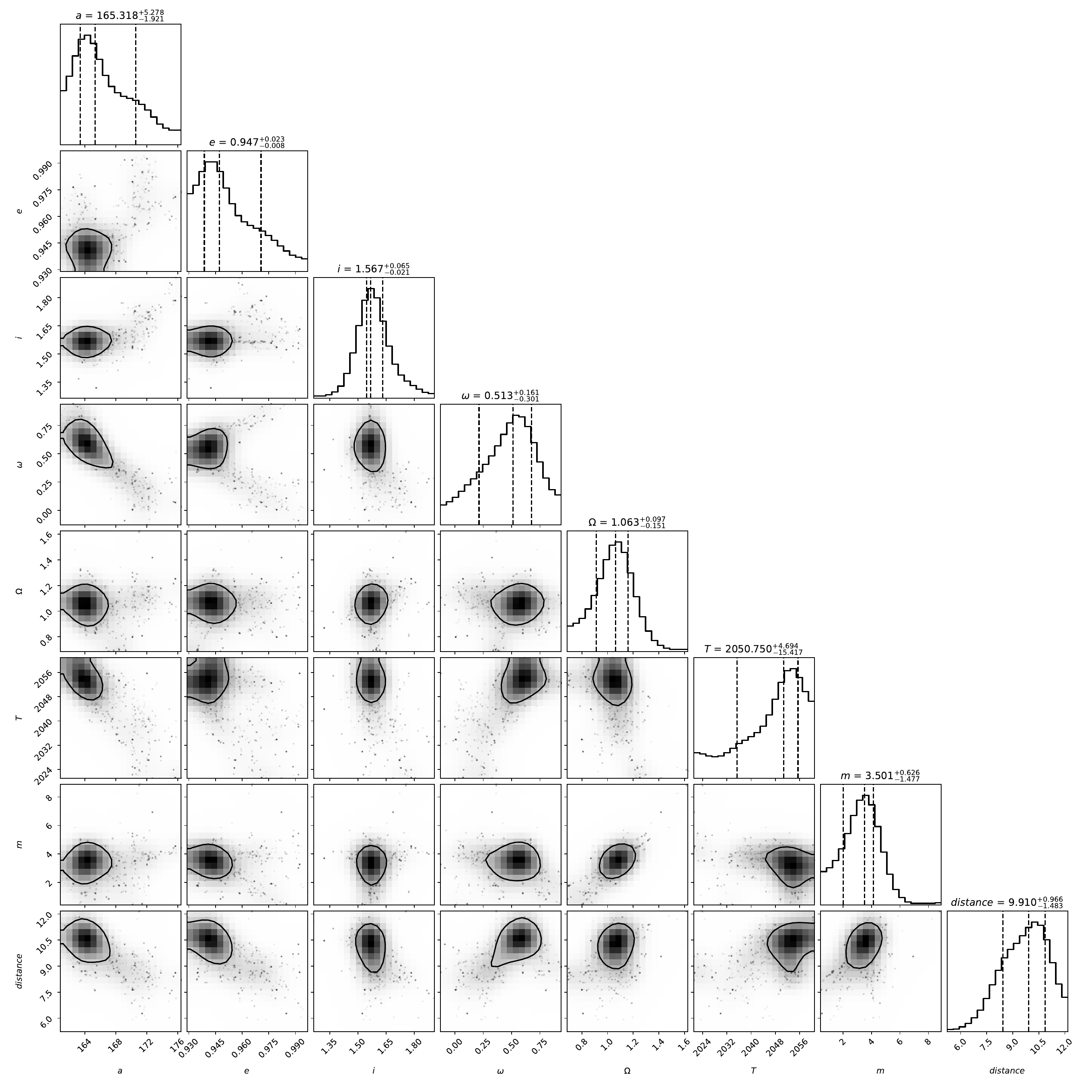}
	\caption{Model simulations of possible Keplerian elements for X7.1.}
\label{fig:mcmc_x71}
\end{figure*}

\begin{figure*}[htbp!]
	\centering
	\includegraphics[width=1.\textwidth]{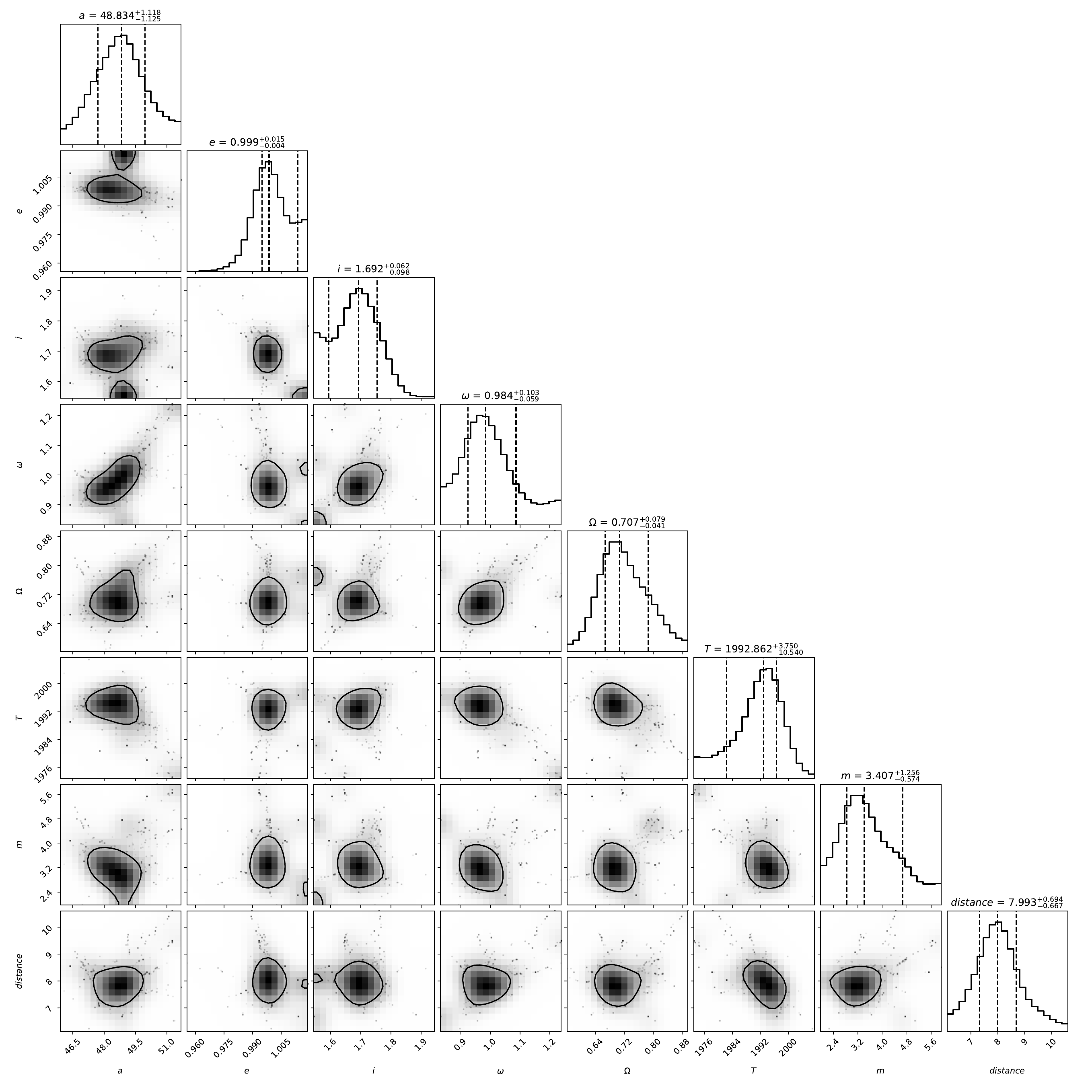}
	\caption{Possible Keplerian elements for X8. Like in Fig. \ref{fig:mcmc_x7} $\&$ \ref{fig:mcmc_x71}, we apply MCMC simulations for a sufficient coverage of possible uncertainties.}
\label{fig:mcmc_x8}
\end{figure*}

\begin{figure*}[htbp!]
	\centering
	\includegraphics[width=1.\textwidth]{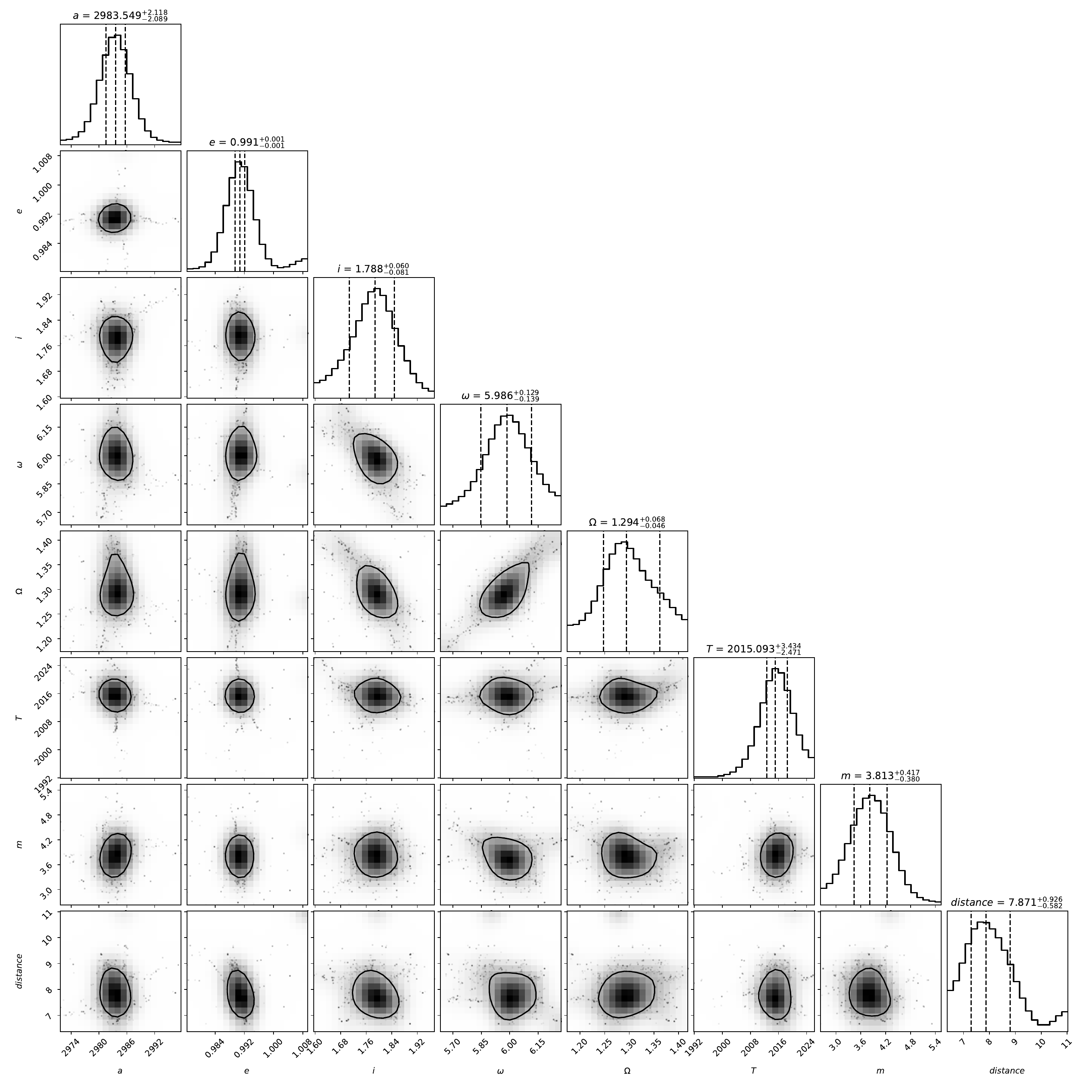}
	\caption{Post prior simulations for D5.}
\label{fig:mcmc_d5}
\end{figure*}

\begin{figure*}[htbp!]
	\centering
	\includegraphics[width=1.\textwidth]{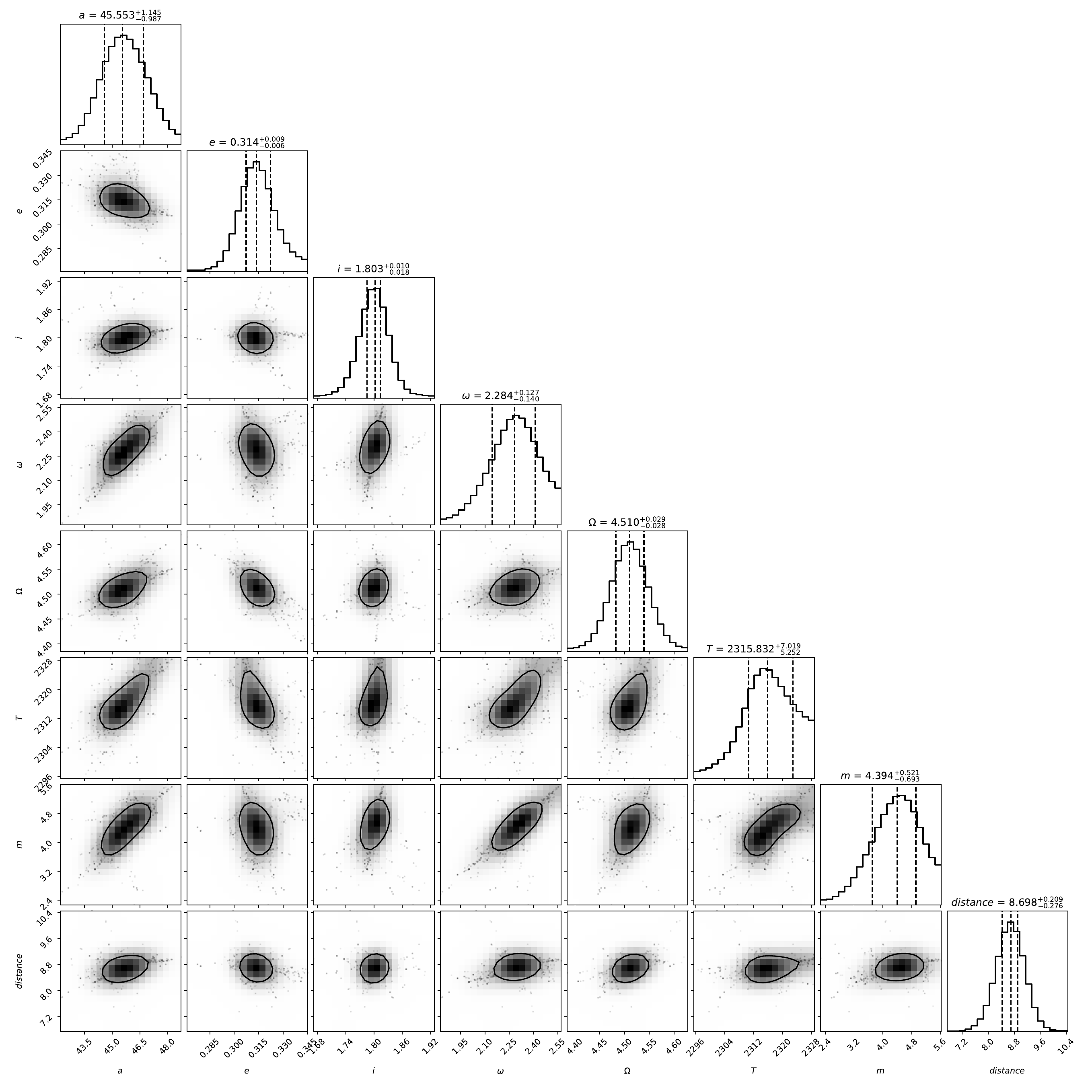}
	\caption{Results of the MCMC simulations for D9.}
\label{fig:mcmc_d9}
\end{figure*}

\end{document}